\documentclass{aa}
\usepackage[varg]{txfonts}
\usepackage{natbib}
\usepackage{float}
\usepackage{subfig}
\usepackage{rotating}
\usepackage[backref,breaklinks,colorlinks,citecolor=blue]{hyperref}

\def\aap{A\&A}

\def\apjs{ApJS}
\def\apjl{ApJL}

\title{Ejection processes in the young open cluster NGC~2264 } 
\subtitle{A study of the [OI]$\lambda$6300\AA \space emission line\thanks{Tables %C1 - C4 
\ref{table:wtts}, \ref{table:ctts}, \ref{table:oi_comp_par} and \ref{table:ctts_oi} 
are available in electronic form at the CDS via anonymous ftp to cdsarc.u-strasbg.fr 
(130.79.128.5) or via http://cdsweb.u-strasbg.fr/cgi-bin/qcat?J/A+A/}} 

\author{P. McGinnis\inst{1,2} \and C. Dougados\inst{1} \and S. H. P. Alencar\inst{2,1} 
 \and J. Bouvier\inst{1} \and S. Cabrit\inst{3} }

\institute{Univ. Grenoble Alpes, CNRS, IPAG, 38000 Grenoble, France \\ 
 \email{\href{mailto:pauline@fisica.ufmg.br}{pauline@fisica.ufmg.br}}
 \and Departamento de F\'{\i}sica -  ICEx - UFMG, Av. Ant\^onio Carlos, 6627, 30270-901
 Belo Horizonte, MG, Brazil 
 \and LERMA,  Observatoire  de  Paris,  CNRS, 61 Av. de l'Observatoire, 75014 Paris, France}
\date{Received 24 July 2017 / Accepted 21 March 2018}

\abstract{
Statistical studies of the spectral signatures of jets and winds in young stars are crucial to 
characterize outflows and understand their impact on disk and stellar evolution. 
The young, open cluster NGC~2264 contains hundreds of well-characterized classical T Tauri stars 
(CTTS), being thus an ideal site for these statistical studies. 
Its slightly older age than star forming regions studied in previous works, such as Taurus, 
allows us to investigate outflows in a different phase of CTTS evolution. 
}{
We search for correlations between the [OI]$\lambda$6300 line, a well-known tracer of jets and winds 
in young stars, and stellar, disk and accretion properties in NGC~2264, aiming to characterize the 
outflow phenomena that occur within the circumstellar environment of young stars. 
}{
We analyzed FLAMES spectra of 184 stars, detecting the [OI]$\lambda$6300 line in 108 CTTSs and two 
Herbig AeBe stars. We identified the main features of this line: a high-velocity component (HVC),  
and a broad and narrow low-velocity components (BLVC and NLVC). 
We calculated luminosities and kinematic properties of these components, then compared 
them with known stellar and accretion parameters. 
}{
The luminosity of the [OI]$\lambda$6300 line and its components correlate positively with the stellar 
and accretion luminosity. 
The HVC is only detected among systems with optically thick inner disks; the BLVC is most common 
among thick disk systems and rarer among systems with anemic disks and transition disks; and the NLVC 
is detected among systems with all types of disks, including transition disks. 
Our BLVCs present blueshifts of up to 50~km~s$^{-1}$ and widths consistent with disk winds originating 
between $\sim$0.05~au and $\sim$0.5~au from the central object, while 
the NLVCs in our sample have widths compatible with an origin between $\sim$0.5~au and $\sim$5~au, 
in agreement with previous studies in Taurus. 

A comparison of [OI]$\lambda$6300 profiles with CoRoT light curves shows that the HVC is found most 
often among sources with irregular, aperiodic photometric variability, usually associated with CTTSs 
accreting in an unstable regime. 
No stellar properties (T$_{eff}$, mass, rotation) appear to significantly influence any property of 
protosellar jets. 
We find jet velocities on average similar to those found in Taurus. 
}{
We confirm earlier findings in Taurus which favor an inner MHD disk wind as the origin of the BLVC, 
while there is no conclusive evidence that the NLVC traces photoevaporative disk winds. 
The [OI]$\lambda$6300 line profile shows signs of evolving as the disk disperses, with the HVC and 
BLVC disappearing as the inner disk becomes optically thin, in support of the scenario 
of inside-out gas dissipation in the inner disk. 
}


\titlerunning{Ejection processes in the young open cluster NGC~2264} 

\authorrunning{P. McGinnis et al.}

\begin{document}

\maketitle

\section{Introduction}\label{sec:intro} 

Outflows are a ubiquitous aspect of star formation. Powerful, highly collimated, bipolar jets are 
commonly found in association with YSOs of different evolutionary stages. The many similarities  
between jets emanating from these different sources suggest that the same physical mechanism is at 
work in different stages of star formation \citep{cabrit02}. 
Though it has been well established that these ejection phenomena are closely associated with disk 
accretion \citep[][hereafter HEG95]{cabrit90,hartigan95}, likely occurring as the byproduct of 
magnetic acceleration from the inner accretion disk \citep{ferreira06,ray07}, the exact mechanism 
which drives them is not yet well understood. 
On the other hand, spectroscopic signatures of slow winds ($v \simeq$ a few 10~km~s$^{-1}$) are also 
detected among CTTS in optical \citep[e.g.,][hereafter N14]{kwan95,natta14} as well as infrared 
forbidden lines \citep[see e.g.,][and references therein]{pascucci11,alexander14}. The link of these 
slow wind signatures with the faster jets is not well established, neither is their impact and role 
in the evolution of protoplanetary disks. Photoevaporation by high-energy radiation from the central 
object, which likely contributes significantly to the dispersion of the gas disk \citep{alexander14}, 
has been put forward as a possible origin for these winds \citep[e.g.,][]{ercolano10,ercolano16}. 

Statistical studies to test which stellar (or disk) properties scale with jet or wind properties 
are crucially needed to shed light on these outflow phenomena and their impact on disk and stellar 
evolution.
Young, accreting, low-mass stars known as classical T Tauri stars (hereafter, CTTS), are optically 
visible, thus allowing us to obtain good estimates of stellar and accretion properties. They are 
Class II objects, meaning they are associated with well-characterized Keplerian disks. These aspects 
make them ideal sites to search for correlations between jet or wind properties and stellar or disk 
properties.   

One of the strongest spectroscopic tracers of jets and winds in CTTSs is the forbidden 
[OI]$\lambda$6300 emission line. This line profile has traditionally been separated into two 
distinct features: a low-velocity component (LVC), centered close to the stellar rest frame or 
blueshifted by up to $\sim$20~km~s$^{-1}$, and a high-velocity component (HVC), blueshifted by 
$\sim$~50~-~200~km~s$^{-1}$ from the stellar rest frame (HEG95). The HVC is believed to arise in the fast, 
bipolar jets that are associated with these young objects, while the LVC seems to originate in 
slower moving winds. 
N14 found line ratios between [SII]$\lambda$4069, [OI]$\lambda$6300 and [OI]$\lambda$5578 
consistent with the LVC arising from slow ($<20$~km~s$^{-1}$), dense 
($\mathrm{n_H} > 10^8 \mathrm{cm}^{-3}$), warm (T $\sim$ 5\,000K - 10\,000K), mostly neutral gas.  
Its precise origin, however, is still debated. Possible scenarios for its 
formation are: centrifugally driven magnetohydrodynamic (MHD) disk winds \citep[HEG95,][]{kwan95}; 
thermally driven disk winds, powered by photoevaporation from X-ray \citep{ercolano10,ercolano16} 
or extreme ultra-violet \citep[EUV,][]{font04} radiation from the central object; or 
slower thermal winds from a cooler disk region heated by far-ultraviolet (FUV) radiation from the 
central object, where the [OI]$\lambda$6300,5577 emission is nonthermally excited by OH 
photodissociation \citep[e.g.,][]{gorti11}. 

In recent studies \citep[][hereafter R13 and S16, respectively]{rigliaco13,simon16}, the LVC has 
been analyzed in more detail and is found to be often composed of two components itself, a narrow 
and a broad component (NC and BC, respectively). 
By comparing not only the full widths at half maximum (FWHM) of the two components but also the 
[OI]$\lambda$5577/[OI]$\lambda$6300 line ratios, which differ between the two components in the 
sources where both were detected with high signal-to-noise ratio (S/N), S16 have concluded that 
the broad and narrow components of the LVC should originate in distinct regions of the accretion 
disk. They find that the broad component presents larger blueshifts and narrower lines with lower 
disk inclination, a trend that is expected for a disk wind. They also argue that the large 
FWHM of the broad component are consistent with emitting regions very close to the central 
object (at $\mathrm{R} \sim 0.05 - 0.5 \mathrm{au}$), within the critical radius for even 
10\,000K gas, and therefore this component cannot trace a photoevaporative wind. 
The NC seems to come from farther out in the disk (at $\mathrm{R} \sim 0.5 - 5 \mathrm{au}$). 
Its nature is less clear and the possibility that 
it could trace a photoevaporative disk wind has not been excluded. 

Statistical studies using high resolution (R$\gtrsim$ 20\,000) spectroscopic data are necessary to 
fully understand the different components of the [OI]$\lambda$6300 line. So far, S16 has been the 
only statistical work conducted in this sense, with a limited sample of 33 objects in the Taurus 
star forming region \citep[][hereafter N18, have conducted a study of the {[OI]$\lambda$6300} line 
in a sample of over 100 objects in Lupus, Chameleon and $\sigma$ Orionis, but with insufficient 
resolution to separate the broad and narrow low-velocity components]{nisini18}. It is important to 
extend such studies to a larger number of objects in other star forming regions to investigate the 
origins of the different features of the [OI]$\lambda$6300 line. 

The young, open cluster NGC~2264, located at a distance of \mbox{d $=760$ pc} \citep{sung97,gillen14}, 
has been the object of an extensive observational campaign 
entitled the Coordinated Synoptic Investigation of NGC~2264 \citep[CSI2264,][]{cody14}, in which 
a number of its low-mass members have been thoroughly characterized in terms of stellar properties 
\citep[e.g.,][]{venuti14}, accretion properties \citep[e.g.,][]{sousa16} and photometric 
variability \citep[e.g.,][]{cody14}. In this paper, we take advantage of the extensive knowledge 
available on the CTTSs belonging to NGC~2264, in order to search for correlations between the 
different components of the [OI]$\lambda$6300 line and stellar and/or disk and accretion properties. 

Our spectroscopic data, taken by the FLAMES/GIRAFFE multi-object spectrograph, mounted on the VLT 
(ESO), cover a narrower wavelength range than the previous studies mentioned (HEG95, R13, S16), 
therefore we analyze only the [OI]$\lambda$6300 line and cannot determine line ratios. However, 
our work has the advantage of its sample size, including over 100 [OI]$\lambda$6300 line profiles, 
whereas the previous studies mentioned have been limited to 30-50 objects. 
Our sample is unique in that it is the first to offer high-precision, high-cadence light curves 
from the CoRoT satellite for most of the objects studied, making it an ideal opportunity to perform 
statistical studies of outflows in young stars in conjunction with their photometric variability 
and rotational properties. Besides this, NGC~2264 is an older star forming region ($\sim$3-5 
Myr\footnote{There is, however, strong evidence for a significant age spread in this cluster 
\citep[see][]{dahm08,venuti17pop}.}) than those studied in previous works (such as Taurus and 
Lupus, of $\sim$1-3 Myr), allowing us to investigate outflows during a slightly later phase of 
CTTS evolution. 

This paper is organized in the following manner:  
in Sect. \ref{sec:obs} we describe the observations of NGC~2264 in the [OI]$\lambda$6300 line, 
taken for the present study; in Sect. \ref{sec:sample} we describe the [OI]$\lambda$6300 line 
profiles we find in our sample; in Sect. \ref{sec:oilum} we describe correlations we find between 
the different components of the [OI]$\lambda$6300 line and stellar and accretion properties; in 
Sect. \ref{sec:oicorot} we discuss the connection between the [OI]$\lambda$6300 line profile 
and the stars' photometric variability; in Sect. \ref{sec:discuss} we discuss our results and the 
possible origins of each component of the [OI]$\lambda$6300 line; and finally in Sect. 
\ref{sec:conc} we present a summary along with our conclusions.

\section{Observations and data reduction}\label{sec:obs}  

\begin{figure*}[tp]
    \centering
    \includegraphics[width=17cm]{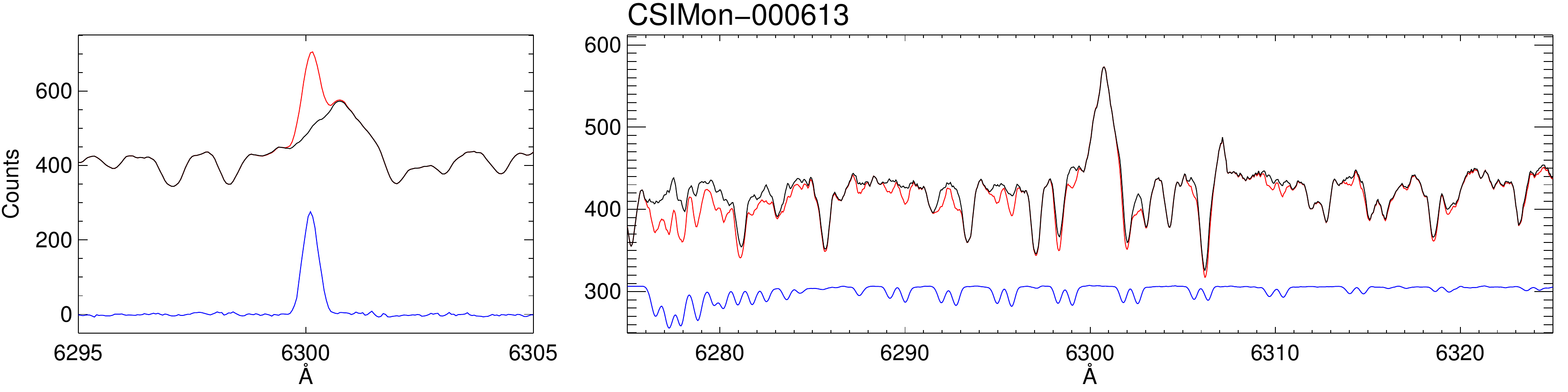}
    \caption{Example of the process to remove telluric emission and absorption lines from the 
stellar spectra. Left: the original spectrum is shown in red; the isolated contamination, taken as 
the average of 3 sky spectra that were observed simultaneously with the stellar spectrum, is shown 
below in blue; and the result after subtraction is overplotted in black. Right: the spectrum, now 
with the telluric emission line removed, is shown in red; the telluric absorption line template, 
scaled and shifted vertically to facilitate a comparison with the stellar spectrum, is shown below 
in blue; and the final decontaminated spectrum is overplotted in black.}\label{fig:skysub}
\end{figure*}

We observed a total of 487 stars in the direction of NGC~2264 with the VLT FLAMES multi-object 
spectrograph \citep{pasquini02} using the GIRAFFE/MEDUSA HR13 setup under ESO program ID 
094.C-0467(A). The data were provided by ESO after being treated for standard reduction procedures 
using ESO pipelines, which consist of bias and bad pixel removal, flat-field corrections and 
wavelength calibrations.  
The chosen setup covers the wavelength range from 6120\AA \space to 6405\AA \space with a 
spectral resolution of R$\sim$26\,500, which is very similar to the resolution used by HEG95.  
It is sufficient to disentangle the different components of the low velocity component of the 
[OI]$\lambda$6300 line (which was not originally done by HEG95). 

Of the 487 observed stars, 483 are cataloged as members of NGC~2264 according to the criteria 
described in \citet{cody14}. Of these, 237 are weak-line T Tauri stars (WTTS, nonaccreting 
low-mass stars), 182 are classified as classical T Tauri stars according to \citet{venuti14} and 
criteria described therein, and two are Herbig Ae/Be stars (hereafter HAeBe, young intermediate 
mass stars with strong emission lines). The remaining 62 have unknown spectral types and have 
not shown any indication of accretion in previous studies. 

The targets were distributed among four different fields, two of which were observed with three 
different fiber configurations in order to include all targets, giving us a total of 8 different 
configurations. Each of these configurations was observed at least once for 40 minute exposures, 
while one was observed four times in a period of $\sim$2.5 months (identified as configuration A1). 
Since the spectral feature we are interested in for the scope of this paper, the [OI]$\lambda$6300 
emission line, is not expected to vary significantly in the time between these observations 
(HEG95), we added the four spectra after having removed all contaminating effects, obtaining 
spectra of larger signal-to-noise ratio for the stars observed in this configuration. 

The [OI]$\lambda$6300 emission line is contaminated by a telluric emission line and telluric 
absorption lines that must be removed in order to recover the intrinsic line profile. For this 
reason, a number of FLAMES fibers were allocated to regions of the sky where no star was found 
and only the telluric emission spectrum could be observed. For each stellar spectrum, we used a 
$\chi^2$ minimizing technique to identify among the sky spectra observed simultaneously with it 
the three that most closely match the contamination. We shifted them in wavelength when necessary 
to best coincide with the contamination, then took the average of the three profiles and 
subtracted this from the stellar spectrum (see left panel of Fig. \ref{fig:skysub}).
This contamination was very strong on most nights, typically much stronger than in the Taurus 
spectra analyzed by S16. However, this feature is much narrower (it is unresolved in our spectra) 
than emission coming from circumstellar disks, which makes it easy to identify. 
Since it is not very variable spatially and many sky fibers were spread around the field of 
view, we were able to properly identify this contamination and subtract it, even when it was 
strongly blended with the source's [OI] emission line. 
Nonetheless, this strong feature may have overshadowed some of the weaker low-velocity emission 
in our sample, causing the [OI]$\lambda$6300 emission to go undetected in some of our sources. 

After subtracting the telluric emission line, 
we then proceeded to remove the contamination from telluric absorption lines, which affect the 
wavelength range from approximately 6275\AA \space to 6325\AA. We combined the normalized spectra 
of two A0-type stars, CSIMon-005644\footnote{The CSIMon ID is a naming scheme devised for 
the CSI2264 campaign \citep[see][]{cody14}, and comprises all cluster members, candidates and field 
stars in the NGC~2264 region observed during the campaign.} 
and \object{CSIMon-001235}, the first of which was observed four times, to 
create a template. These stars were chosen because they are hot enough that they have no photospheric 
features in this region, therefore after normalizing to the continuum only the telluric absorption 
spectrum remains. The telluric absorption spectrum is generally not very variable and this template 
proved to be a good match for all but one of our observed fields. For the final field a star observed 
simultaneously was needed. There were no A or B-type stars in this field, therefore the template was 
created using the G-type star \object{CSIMon-001126}. This star has high projected velocity ($v\sin i$), 
leading to very broad photospheric lines, which could be easily identified and fitted with synthetic 
spectra. The photospheric contribution could then be subtracted from the original spectrum in order 
to isolate the telluric absorption lines and create the template. The stellar spectra were then 
divided by the respective templates in the affected wavelength interval in order to eliminate the 
contamination (right panel of Fig. \ref{fig:skysub}). 
The left panels of Fig. \ref{fig:speccor} show the telluric corrections 
of all spectra in which a residual [OI]$\lambda$6300 emission line was detected. 

\section{The [OI]$\lambda$6300 line profiles}\label{sec:sample}

\subsection{Recovering intrinsic [OI]$\lambda$6300 emission line profiles}\label{sec:oicor} 

In order to identify forbidden [OI]$\lambda$6300 emission in the stellar spectra, it is often 
necessary to eliminate the photospheric contribution to the spectrum, thus isolating the emission 
line. Photospheric absorption lines that lie close to the [OI]$\lambda$6300 emitting region can 
affect the emission line and even bring it to levels below our detection level, especially when the 
emission is weak. We chose a sample of WTTS spectra to serve as photospheric templates to compare 
with our CTTS spectra. We selected WTTSs with spectral types in the range found for our CTTSs and 
of low $v\sin i$'s (rotational velocity projected onto our line of sight). 
Figure \ref{fig:specwtts} shows these WTTS's spectra, ordered according 
to spectral type, while Table \ref{table:wtts} (available at the CDS) 
provides some of their stellar parameters. Spectral types of both classical and weak-lined TTSs 
were taken from \citet{venuti14}, who adopted the values of \citet{dahm05}, \citet{rebull02} or 
\citet{walker56}, in that order of preference, or derived them from CFHT colors when no 
classification was available in the literature.  

\begin{figure}[t!]
    \centering
    \resizebox{\hsize}{!}{\includegraphics{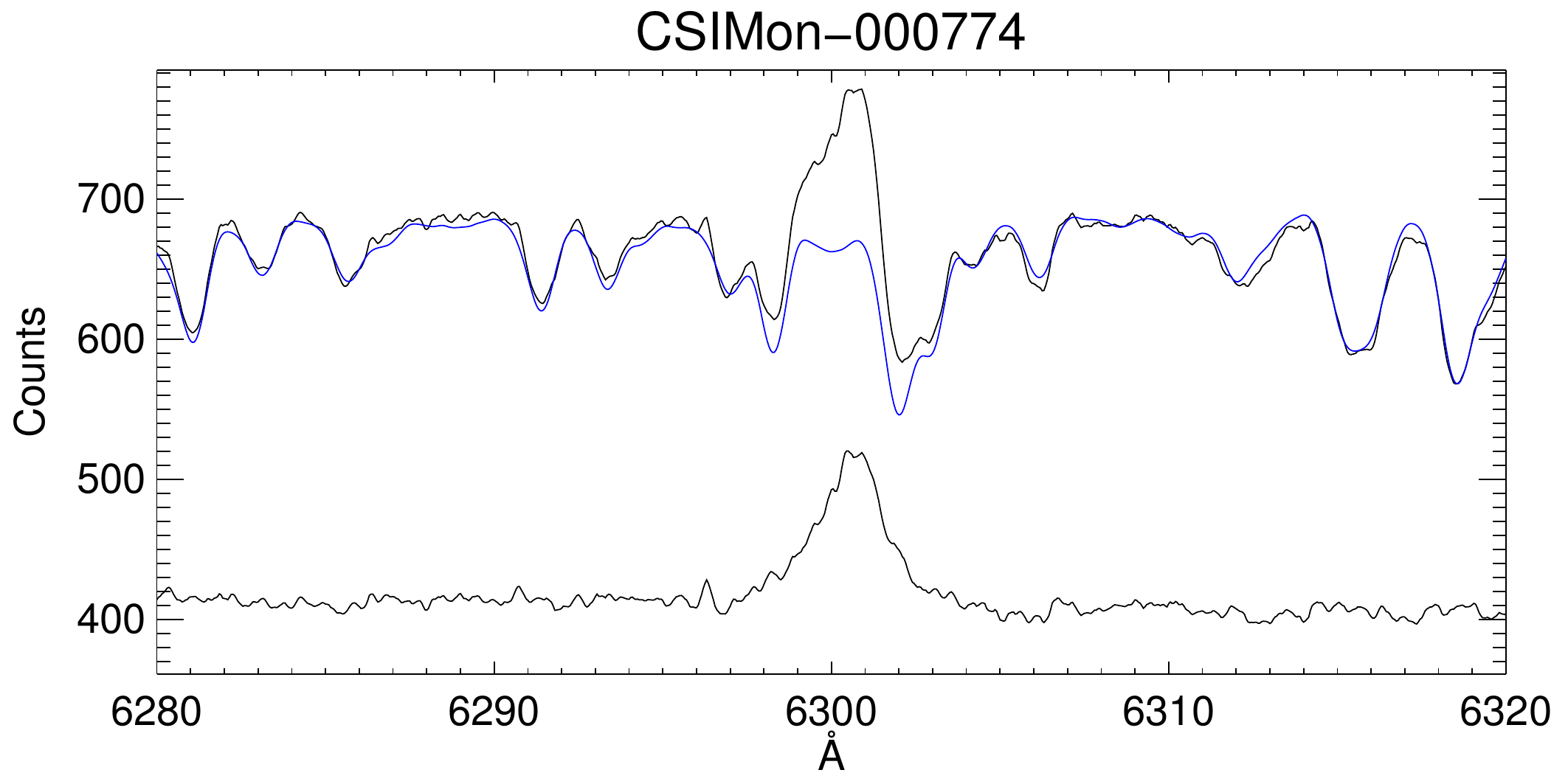}}
    \caption{Example showing the photospheric contribution being removed from a CTTS spectrum in 
order to recover the intrinsic [OI]$\lambda$6300 emission line profile. The veiled, broadened WTTS 
template is shown in blue, plotted over the CTTS spectrum in black. Below, the final product 
is shown, shifted vertically to facilitate a comparison with the original spectrum. We can see 
that the wings of this profile are recovered after removing the photospheric absorption lines. 
}\label{fig:photsub}
\end{figure}

We compared each CTTS spectrum to at least three different WTTS templates, which were broadened 
to adjust for the CTTS's rotation and had an added continuum to simulate veiling (excess accretion 
flux that affects optical and UV spectra in the presence of accretion shocks on the stellar 
surface). The templates used were of the same or nearly the same spectral type as the CTTS in 
question. However, to find the best fit, templates of different spectral types were also tested 
whenever necessary, in order to account for possible misclassifications of spectral types in the 
literature. The amount of veiling and rotational broadening were adjusted to the values that 
simultaneously minimized $\chi ^2$. The broadened, veiled template with lowest $\chi^2$ was then 
subtracted from the CTTS spectrum to recover the intrinsic [OI]$\lambda$6300 line profile (see 
Fig. \ref{fig:photsub}). 
The right panels of Fig. \ref{fig:speccor} show the photospheric 
corrections of all spectra in which a residual [OI]$\lambda$6300 emission line was detected. 

In a few cases we found that the best match for a WTTS+veiling template to the star in question was 
of a different spectral type than the one listed for the star. In three cases we found the difference 
to be by more than two sub-spectral types. We attribute this to a possible misclassification of the 
star's spectral type in the literature, possibly due to the excessive amount of veiling presented 
by these stars, which can interfere with photometric determinations of spectral type. Table 
\ref{tab:spt} provides the spectral type of the WTTS template that best reproduced their photospheric 
absorption lines. 
We also found that the stars \object{CSIMon-001149} and \object{CSIMon-001234} are double-lined 
spectroscopic binaries. Table \ref{table:ctts} (available at the CDS) 
lists stellar parameters of all CTTSs in our sample,  
such as their spectral types (from the literature and from our best fits), stellar luminosity, mass 
and radius \citep[determined uniformly for nearly all stars in our sample in][]{venuti14}, as well 
as our derived values of veiling and $v\sin i$.  

The Herbig Be star \object{CSIMon-000392} did not undergo photospheric subtraction, since there are no 
photospheric features in the vicinity of the [OI]$\lambda$6300 line in such hot stars. In this case, 
the continuum was fitted with a fourth order polynomial and subtracted from the spectrum in order to 
recover the residual profile. This was also done for the stars \object{CSIMon-000423}, 
\object{CSIMon-000632}, and \object{CSIMon-001011}, which showed such low S/N in their photospheric 
lines that no photospheric features were detected near the [OI]$\lambda$6300 line.  
The Herbig Ae star \object{CSIMon-000631} presents a shallow, broad absorption line redward of 6300\AA. 
A template of the same spectral type (A2/3) was not found for this star, since few early-type stars 
were included in our observations. Therefore, in order to remove the contaminating feature, we used 
as its photospheric template an F-type star with a similar feature close to the [OI]$\lambda$6300 
line. We fitted only the region closest to this feature, ignoring the rest of the spectrum.  

\begin{table*}[thb]
    \centering
    \caption{Spectral types from the literature and from our comparison with WTTS templates 
for the stars that showed the largest discrepancies among the two. 
    }\label{tab:spt}
    \begin{tabular}{c c c c c}
    \hline 
    \hline 
CSIMon ID & 2MASS ID & SpT (lit) & Our SpT & Veiling \\
    \hline 
\object{CSIMon-000406} & J06405968+0928438 & K3\tablefootmark{a} & M1 & 0.85 \\ 
\object{CSIMon-000474} & J06410682+0927322 & G\tablefootmark{b} & K2 & 0.15 \\
\object{CSIMon-000795} & J06411257+0952311 & G:V:e\tablefootmark{b} & K2 & 1.03 \\ 
    \hline 
    \end{tabular}
    \tablefoot{The first and second columns give the stars' IDs in the CSIMon and 2MASS catalogs, 
respectively; spectral types from the literature are given in the third column and from our comparison 
with WTTS templates in the fourth column; and the fifth column gives the value of veiling (ratio of 
excess flux to stellar flux) derived from our spectra. \\ 
    $^{(a)}$\citet{venuti14} \\ 
    $^{(b)}$\citet{dahm05}
    }
\end{table*}

After removing the photospheric contribution to the spectra, we searched for [OI]$\lambda$6300 
emission in our stars. We considered as a detection any emission with a peak above 3$\sigma$ and 
FWHM above the resolution of our spectra, which we measured as 11.3~km~s$^{-1}$. Values of $\sigma$ 
represent the noise in the residual spectra, measured from two regions of the continuum 5\AA \space 
before and after the [OI]$\lambda$6300 line. A few cases where a marginal detection was found in the 
exact region where the telluric emission line was originally located were considered as residual 
contamination from the telluric line and eliminated from the list of detections.

Among our 182 CTTSs, 108 (59\% $\pm$4\%) show emission in the [OI]$\lambda$6300 line, as well as 
both HAeBe stars in our sample. This low number is somewhat unexpected, since past studies have 
shown that most accreting T Tauri stars present emission in the forbidden [OI]$\lambda$6300 
line \citep[HEG95 and][for example, detect the {[OI]$\lambda$6300} line in 100\% of the accreting 
TTSs in their sample, while N14, S16 and N18 detect it, respectively, in 84\%, 91\%, and 77\% 
of theirs]{cabrit90}. 
As shown in Appendix \ref{sec:compl}, we believe this low detection rate may be due to incompleteness. 
Since NGC~2264 is farther than the star forming regions of these previous studies (Taurus, Lupus, 
$\sigma$ Ori) and the targets are therefore fainter, our observations may have been unable to detect 
the weakest [OI]$\lambda$6300 emission lines. Besides this, the telluric emission line in our 
spectra was often much stronger than in the spectra of S16 (this can be seen in the left panels of 
Fig. \ref{fig:speccor}), which very likely interfered with our detection, 
especially of weaker, low-velocity emission. 
To avoid any issues this incompleteness may bring about, we include upper limits of nondetections 
in our analyses whenever possible.
We discuss the completion of our sample in more detail in Appendix \ref{sec:compl}. 

All profiles shown in this paper have been shifted to the stellar rest velocity $v_{rad}$ (given 
in Table \ref{table:ctts}), which was determined during the cross-correlation with WTTS spectra.  
The stellar rest velocities of the WTTSs were found by comparing photospheric absorption lines with 
those of the synthetic spectrum of closest effective temperature to the star's, adjusted for $v\sin i$. 
Synthetic spectra were created using the program Spectroscopy Made Easy \citep[SME,][]{valenti96} 
and atomic line files extracted from the \textit{Vienna Atomic Line Database} 
\citep[VALD,][]{vald1,vald2,vald3,vald4} and were broadened to the instrumental resolution.  
Uncertainties in $v_{rad}$ are around 3~km~s$^{-1}$ for most of our sample, but they can  
reach up to $\gtrsim$10~km~s$^{-1}$ for a few spectra that present very low S/N in the photospheric 
lines ($\sim$5\% of our sample). 
Appendix \ref{sec:errors} explains how these uncertainties were estimated. 

\subsection{Characterizing [OI]$\lambda$6300 emission line profiles}\label{oiprof} 

\begin{figure}[t]
    \centering
    \resizebox{\hsize}{!}{\includegraphics{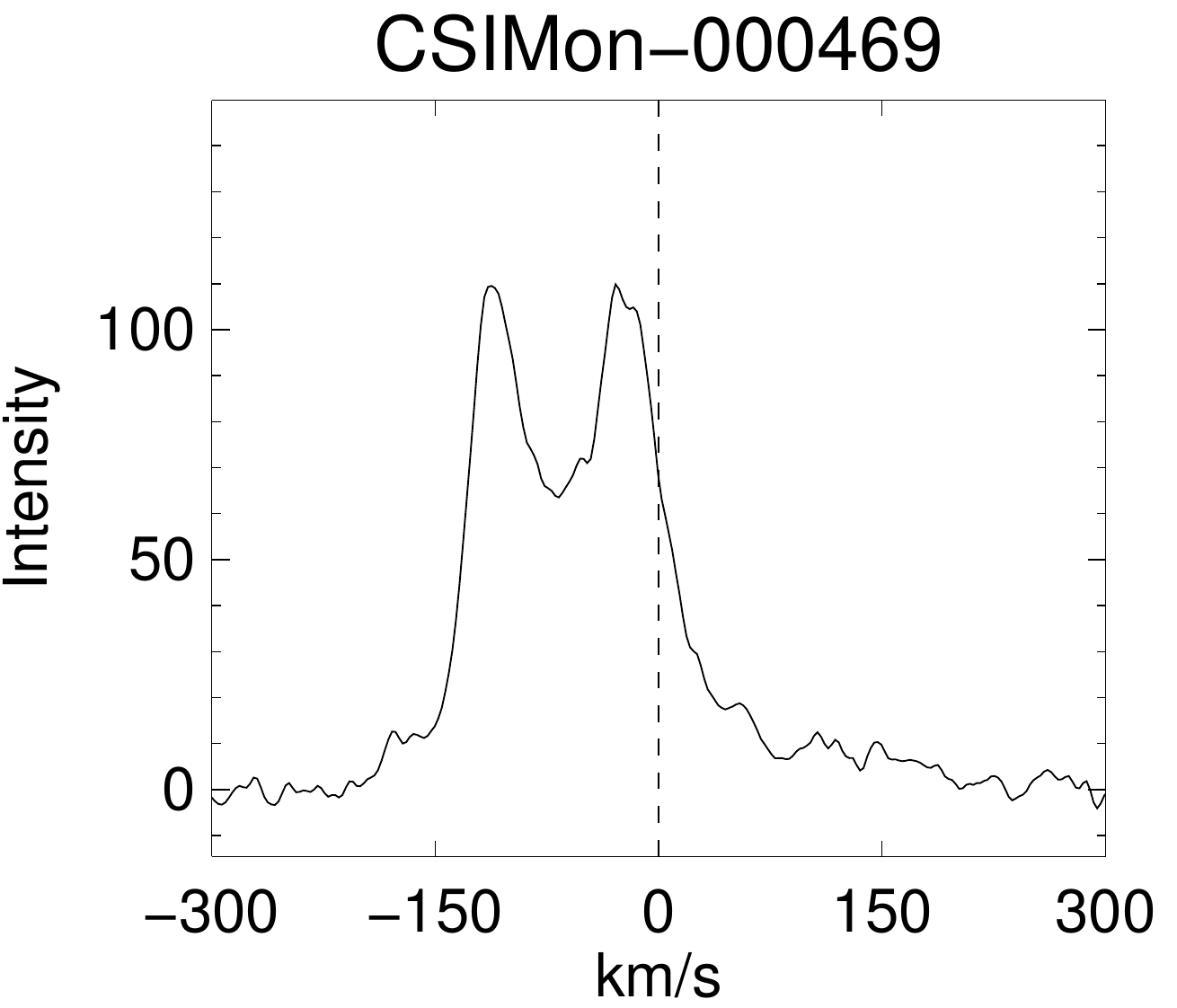}
    \includegraphics{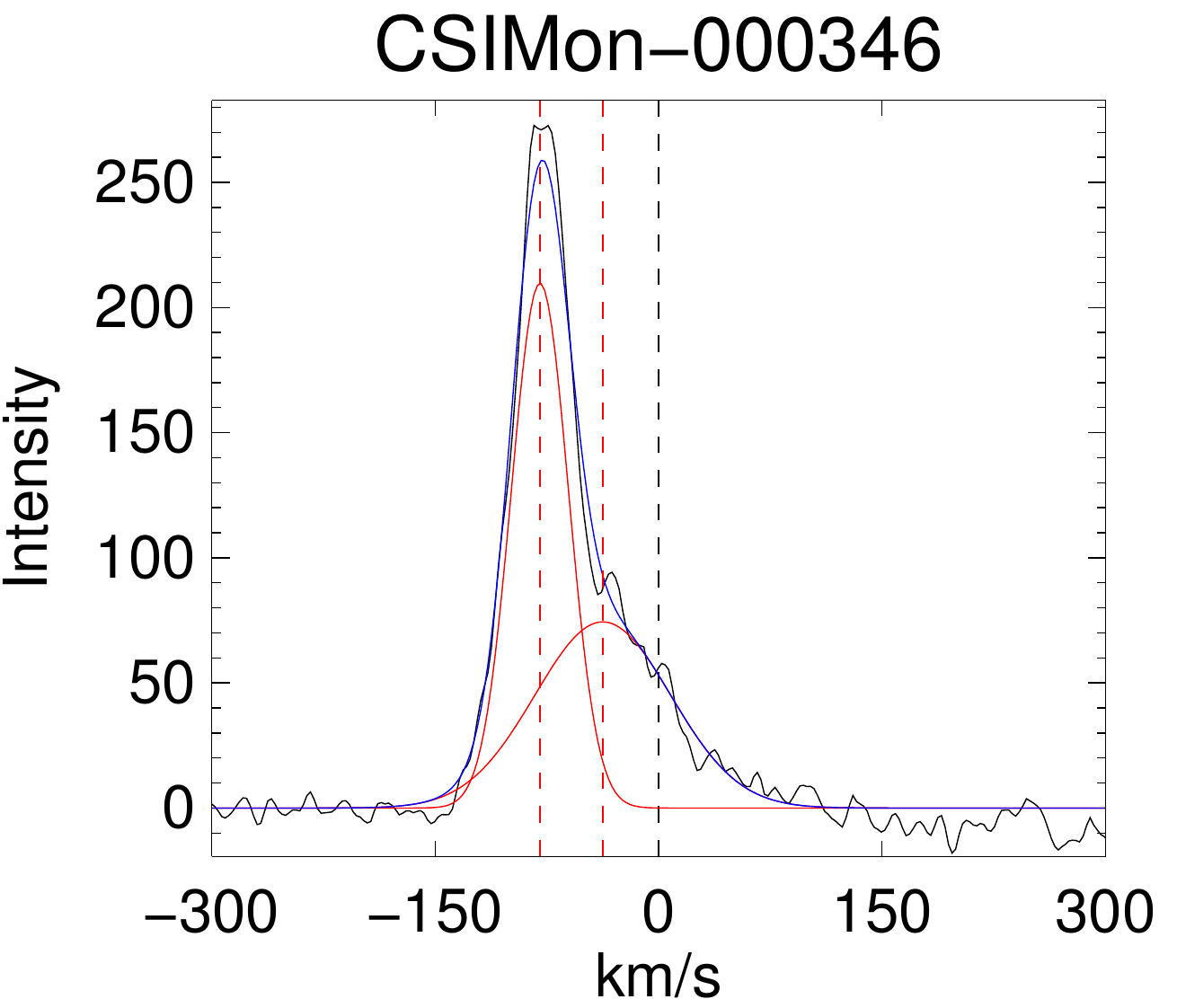}}
    \caption{Two examples of [OI]$\lambda$6300 emission line profiles found in our sample. Left: two 
distinct peaks can be identified, one close to the stellar rest velocity and the other blueshifted by 
over 100~km~s$^{-1}$. Right: Gaussian decomposition of a very asymmetric profile. Two separate Gaussians are 
shown in red, while the sum of the two is shown in blue, overplotted onto the observed profile 
(black).}\label{fig:profiles}
\end{figure}

It has been shown that the [OI]$\lambda$6300 emission line profile may present up to three 
distinct features (R13, S16): a high-velocity component (HVC), a narrow low-velocity component 
(NLVC) and a broad low-velocity component (BLVC). 
We identified, as best as we could, each of these features in our [OI]$\lambda$6300 line profiles, 
in order to investigate their origins and their connection to stellar and disk properties. 
Some profiles clearly show two or three peaks (see the left panel of Fig. \ref{fig:profiles} for an 
example), though many profiles have only one clear peak but are asymmetric or very broad at the base. 
We therefore performed a Gaussian decomposition of all profiles in order to find which ones consist 
of various components and estimate the centroid velocity and width of each component, as well as 
their uncertainties (derived in Appendix \ref{sec:errors}). To determine how many Gaussians are 
necessary to reproduce an emission profile, reduced $\chi ^2$ values were compared for fits with up 
to four Gaussians, and the number that resulted in $\chi ^2$ closest to 1 was taken, as long as all 
components had peak intensity above 2$\sigma$. The right panel of Fig. \ref{fig:profiles} shows 
an example of this Gaussian decomposition (the Gaussian decomposition of all profiles is shown in 
Fig. \ref{fig:gausdecomp}). 

In order to classify the different components found through these Gaussian fits, we initially used 
the definition of NLVC, BLVC and HVC provided by S16: all components of \mbox{$|v_c| \geq 30$ km~s$^{-1}$} 
were classified as HVCs, those of \mbox{$|v_c| < 30$ km~s$^{-1}$} with \mbox{FWHM $> 40$ km~s$^{-1}$} were classified 
as BLVCs, and those of \mbox{$|v_c| \leq 30$ km~s$^{-1}$} with \mbox{FWHM $\leq 40$ km~s$^{-1}$} were classified as 
NLVCs. However, when using these criteria to separate the components in our sample, we found a few 
inconsistencies. 
There are ten systems in which, rather than a NLVC and a BLVC, we identified instead
two BLVCs, i.e., two components of \mbox{$|v_c| \leq 30$ km~s$^{-1}$} and \mbox{FWHM $> 40$ km~s$^{-1}$}. 
No source in S16 presented more than one NLVC or BLVC. This suggests that the threshold 
between narrow and broad LVCs may be slightly different in our sample than in S16's. 
Besides this, among the profiles with blended HVCs, we found three components identified as HVCs that 
presented relatively low centroid velocities \mbox{(30 km~s$^{-1}$ $< |v_c| <$ 50 km~s$^{-1}$)} and were very broad, 
showing \mbox{FWHM $>$ 165 km~s$^{-1}$} (and up to \mbox{FWHM $\sim$ 250 km~s$^{-1}$}). This is broader than 
the spectrally resolved HVCs in our sample, all of which show \mbox{FWHM $<$ 165 km~s$^{-1}$}.  
Hence, it is possible that these are in fact BLVCs extending to slightly larger blueshifts (up to 
50~km~s$^{-1}$) than probed by the S16 sample.  

These issues suggested the need to derive new criteria to classify the different components based 
on our own sample, rather than rely on previously derived definitions. This would account for 
intrinsic differences that may exist between the different samples. This is also the 
first time that the [OI]$\lambda$6300 line is studied in over 100 profiles at high 
spectral resolution\footnote{N18 have studied a sample of similar size, but with lower resolution,
which did not allow them to disentangle the narrow and broad low-velocity components.},
allowing for a statistical analysis of the distribution of FWHM and centroid velocity of the 
different components of the [OI]$\lambda$6300 line. We perform this analysis in the following 
section in an attempt to properly identify the different components in our sample. 
It is important to note, however, that the limits between the different components will always be 
somewhat arbitrary when analyzing only one emission line, especially since there is some overlap 
between the different components (this is because of the different geometries in which the systems 
are viewed, which affect both the widths and centroid velocities of the emission). 

\subsection{Classification of the different components of the [OI]$\lambda$6300 line for the NGC~2264 sample} 

\begin{figure*}[t!]
    \centering
    \includegraphics[width=17cm]{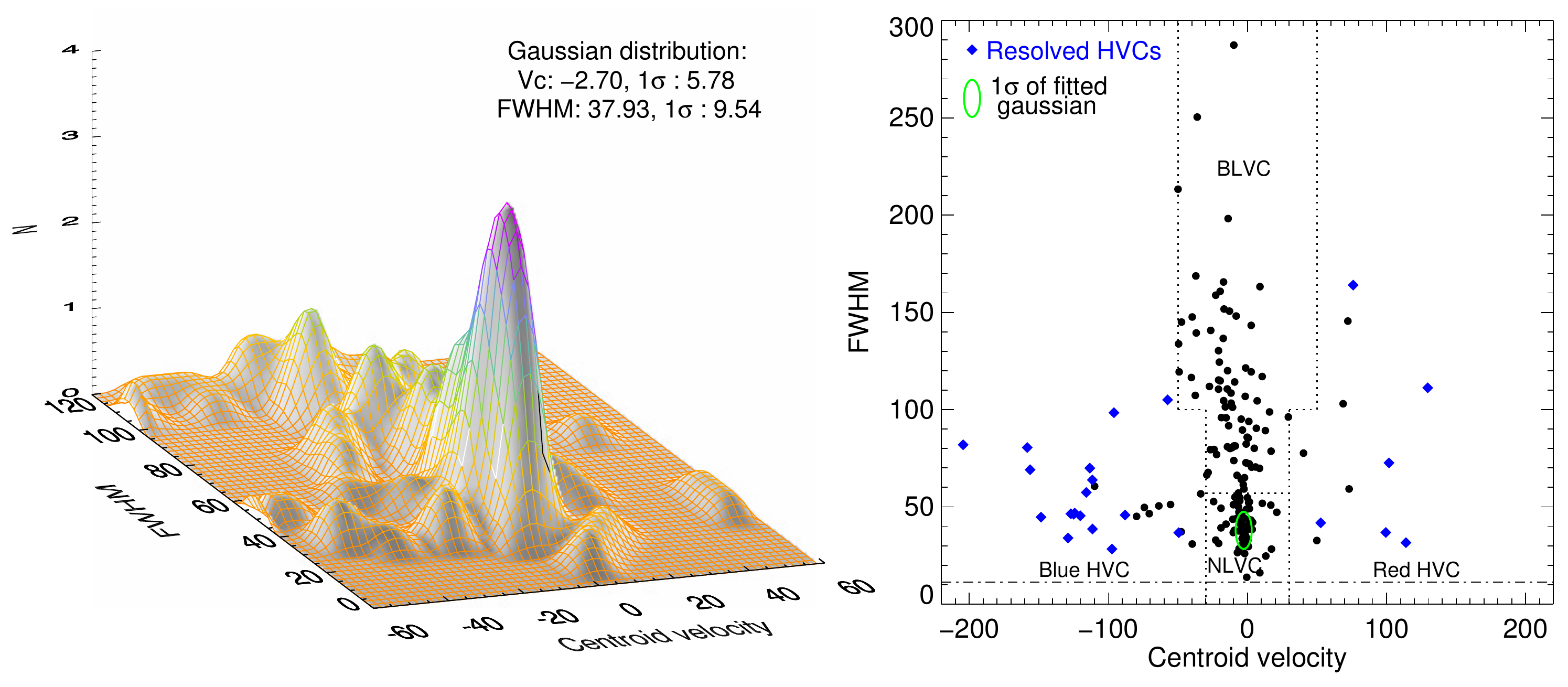}
    \caption{Right: distribution of FWHM and centroid velocity of all components from all profiles 
derived from the Gaussian fits. Dotted lines indicate the separation between different components - 
narrow and broad LVC, red- and blueshifted HVC, according to our analysis 
(though there is some overlap between the NLVC and BLVC). The dashed-dotted line represents 
our instrumental profile, below which no components are found. Blue diamonds represent 
high-velocity components whose peaks are resolved, and not blended with the low-velocity emission. 
Left: surface plot of the central region of the distribution shown in 
the right panel, where the z axis represents the number of Gaussian components in square bins of 
5~km~s$^{-1}$ $\times$ 5~km~s$^{-1}$. The primary peak corresponds to the locus of the narrow LVC, while the 
secondary peak represents a concentration of broad LVCs. 
    }\label{fig:vel_comp1}
\end{figure*}

The right panel of Fig. \ref{fig:vel_comp1} shows the distribution of 
FWHM and centroid velocity of all of the components derived from all of the Gaussian 
fits, and on the left panel a surface plot of the same distribution, with the z axis representing 
the number of Gaussian components in square bins of 5~km~s$^{-1}$ $\times$ 5~km~s$^{-1}$. There is evidently a very 
high concentration of components at low, slightly blueshifted centroid velocities and low FWHM. We 
interpret this as the locus of the narrow LVCs studied by R13 and S16. 
A Gaussian fit around the surface plot (left panel of Fig. \ref{fig:vel_comp1}) has a 1$\sigma$ 
width represented in the right panel as a green ellipse. If this truly represents the locus of 
the narrow component of the LVC, then this component should typically have $v_c=-2.7\pm5.8$~km~s$^{-1}$ 
and FWHM $=37.9\pm9.5$~km~s$^{-1}$ (where the 1$\sigma$ uncertainty has been taken as the standard 
deviation of the distribution). If we consider a 3$\sigma$ threshold, then the narrow LVCs should have 
$-20.0$~km~s$^{-1}$~$< v_c < 14.6$~km~s$^{-1}$ and FWHM~$< 66.6$~km~s$^{-1}$. None of the components 
are narrower than 11.3~km~s$^{-1}$, since this is the resolution of our spectra and therefore this 
width represents the instrumental profile. 

\begin{figure}[t]
    \centering
    \resizebox{\hsize}{!}{\includegraphics{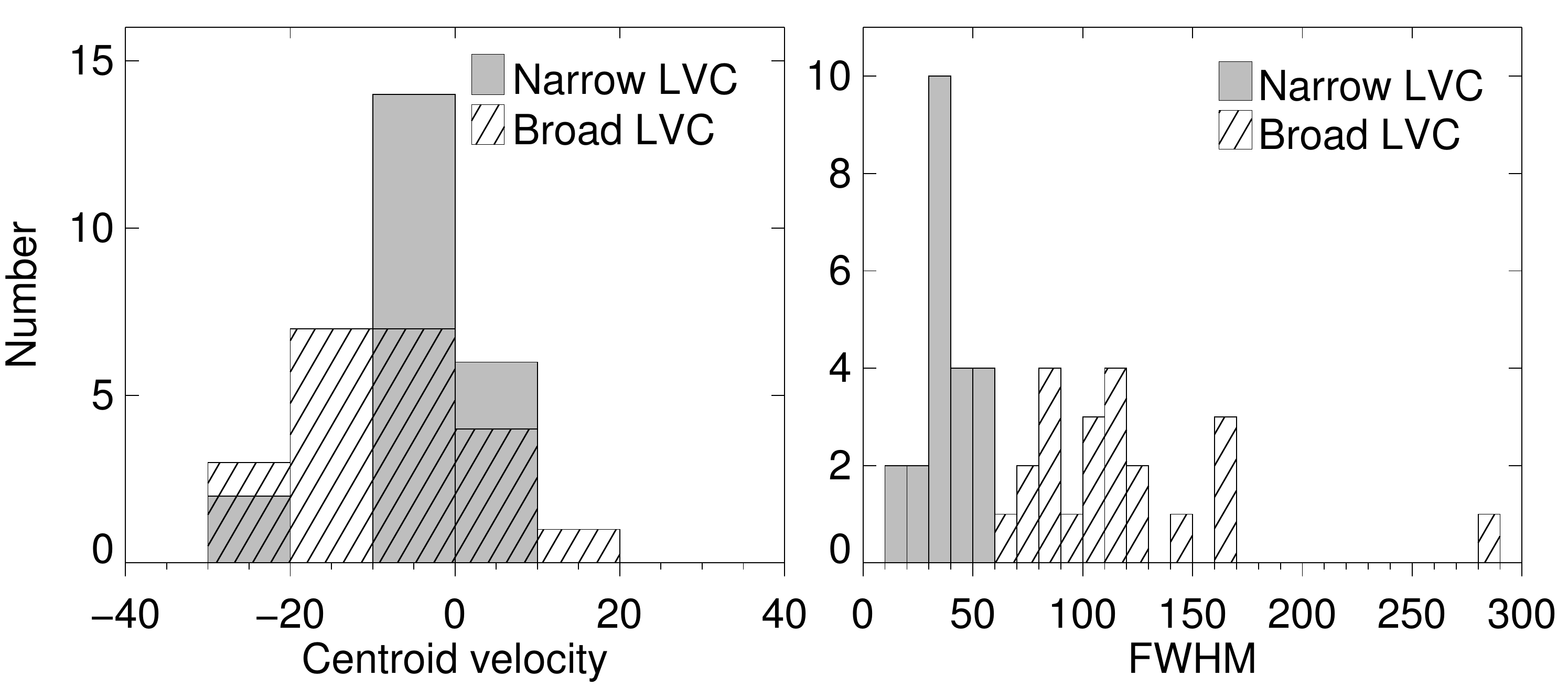}} 
    \caption{Distribution of centroid velocity (left) and FWHM (right) of the components from our 
Gaussian fits for a group of stars that present two LVCs ($|v_c| < 30$~km~s$^{-1}$). Shaded bins represent 
the narrowest of the two components while striped bins represent the broadest.
    }\label{fig:gaus_hist1}
\end{figure}

The cutoff in centroid velocity between LVCs and HVCs varies in the literature from 30~km~s$^{-1}$ (S16) to 
60~km~s$^{-1}$ (HEG95) and is always somewhat arbitrary. In our sample, the lowest centroid velocity of a HVC 
with a spectrally resolved peak is -48.3~km~s$^{-1}$, but this does not necessarily represent the cutoff 
between low and high velocity components, only our ability to resolve separate peaks. 
In an attempt to better characterize the different components and search for the best value at which 
to separate the HVC from the NLVC and BLVC, we analyzed the distribution of centroid velocity and FWHM 
of the different components that are present within the same profile. 
We began by selecting a sample of 21 stars whose profiles consisted of two Gaussian components 
with centroid velocity below 30~km~s$^{-1}$, in which one was clearly broader than the other\footnote{We exclude from 
this sample two systems which have photometric variability attributed to occultation by the inner 
disk \citep{mcginnis15}, since they are likely observed at high inclination and therefore projection 
effects may be causing a HVC to have $|v_c| < 30$~km~s$^{-1}$.}. 
These profiles can help to better distinguish between the narrow and broad components of the LVC. 
Figure \ref{fig:gaus_hist1} shows the distribution of centroid velocity and FWHM of the LVCs of these 
21 stars, with shaded bins representing the narrowest of a profile's two components, and 
striped bins representing the broadest. We see that the NLVC is generally close to stellar 
rest velocity or slightly blueshifted, while the centroid velocity of the BLVC varies more. 

\begin{figure}[t]
    \centering
    \resizebox{\hsize}{!}{\includegraphics{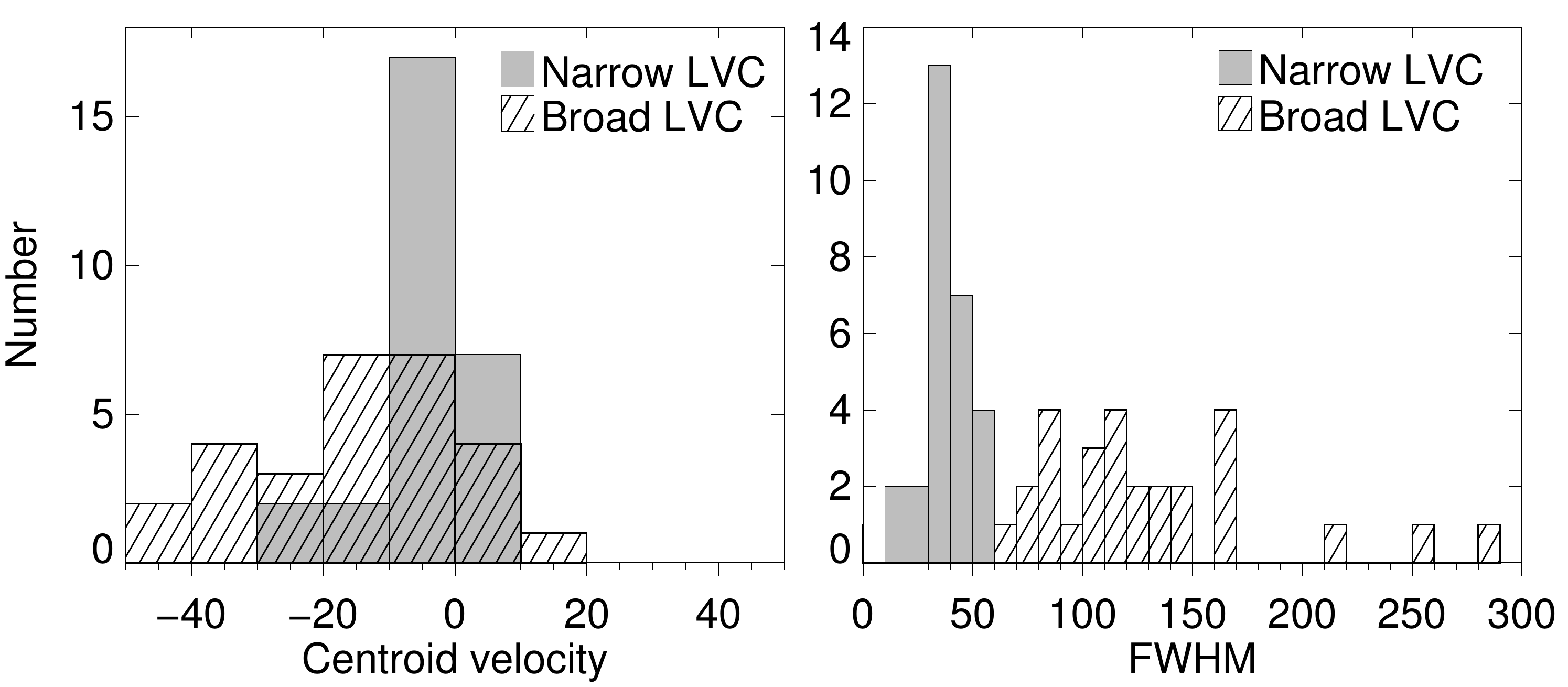}} 
    \caption{Same as Fig. \ref{fig:gaus_hist1}, but taking components with centroid velocities 
less than 50~km~s$^{-1}$.}\label{fig:gaus_hist2}
\end{figure}

We then repeated Fig. \ref{fig:gaus_hist1}, this time for a group of 28 stars that showed two components 
of centroid velocity \mbox{$|v_c| < 50$~km~s$^{-1}$} where, once again, one was clearly broader than the other 
(Fig. \ref{fig:gaus_hist2}). The narrower of the two components continues to show the same behavior, 
having centroid velocities between -24.3~km~s$^{-1}$ and 8.9~km~s$^{-1}$ in both Figs. \ref{fig:gaus_hist1} and 
\ref{fig:gaus_hist2}, consistent with the value of \mbox{$|v_c| < 30$ km~s$^{-1}$} used by S16 and with our 
characterization of the NLVC from Fig. \ref{fig:vel_comp1}.  
The broader component, however, extends to more blueshifted velocities in Fig. \ref{fig:gaus_hist2}, 
up to $\sim$50~km~s$^{-1}$. The right panel of Fig. \ref{fig:vel_comp1} also shows that there are a number 
of components with blueshifts of up to 50~km~s$^{-1}$ that are broader than the clear HVCs (those that present 
\mbox{$|v_c| > 50$~km~s$^{-1}$}), whose FWHMs typically fall below $\sim$100~km~s$^{-1}$. These components are therefore 
likely BLVCs, showing that this component can indeed present blueshifts of up to $\sim$50~km~s$^{-1}$, unlike 
the NLVC. 

The FWHM of the NLVC reaches 59.2~km~s$^{-1}$, close to the lower limit of 66.2~km~s$^{-1}$ of the broader 
component. This would suggest that the threshold between these two components is around 60~km~s$^{-1}$. However, 
S16 found at least two [OI]$\lambda$6300 line profiles with two low-velocity components, in which the 
broadest of the two had FWHM~$< 60$~km~s$^{-1}$. This shows that there is a small overlap between the two 
components, probably due to projection effects because of the different geometries of the individual 
star-disk systems. Therefore we take 57~km~s$^{-1}$ as our cutoff between NLVC and BLVC, which corresponds to 
$2\sigma$ of a normal distribution around the primary peak seen in Fig. \ref{fig:vel_comp1}, in the FWHM 
axis. More than 95\% of the NLVCs in Figs. \ref{fig:gaus_hist1} and \ref{fig:gaus_hist2} have FWHM below 
this value, in accordance with the $2\sigma$ level of a normal distribution. Using this value rather than 
$3\sigma$ will likely minimize the amount of BLVCs that are misclassified as NLVCs, though some 
misclassification will be inevitable, given that there is some overlap between the components. 

\begin{table*}[t!]
 \centering
  \caption{Number of profiles that present each of the different components (narrow LVC, broad LVC, redshifted 
  HVC and blueshifted HVC) or a combination of two or more components. 
  }\label{tab:compon}
 \begin{small}
  \begin{tabular}{ c c c l c c c c }
\hline 
\hline 
\multicolumn{3}{ c }{\textbf{Only LVC}} & \multicolumn{5}{ c }{\textbf{Have HVC}} \\ 
\hline
  Only NLVC  &   Only BLVC  &   NLVC+BLVC  &              & NLVC+HVC    & BLVC+HVC   & NLVC+BLVC+HVC & Only HVC \\ \cline{1-3}\cline{5-8}
%\hline                                                                                         
             &              &              &              & 7 profiles  & 15 profiles & 8 profiles & 1 profile   \\ \cline{5-8}
             &              &              &              & \multicolumn{4}{ c }{Of which the HVC are:} \\ \cline{5-8}
 27 profiles & 30 profiles  & 22 profiles  & Blueshifted: & 3 profiles  & 12 profiles & 5 profiles & 0 profiles  \\ 
             &              &              & Redshifted:  & 1 profiles  & 2 profile   & 1 profile  & 0 profiles  \\ 
             &              &              & Both:        & 3 profiles  & 1 profile   & 2 profiles & 1 profile   \\ 
\hline                                                                                         
\multicolumn{8}{ c }{Total number of stars with \space\space NLVCs: 64; \space\space BLVCs: 75; \space\space HVCs: 31}    \\ 
\hline
  \end{tabular}
 \end{small}
\end{table*}

\begin{table*}[t!]
    \centering
    \caption{Information about the centroid velocities ($v_c$) and FWHM of the different 
components of the [OI]$\lambda$6300 emission line. 
    }\label{tab:vel_cmp}
    \begin{tabular}{l c c c c c c}
\hline 
\hline 
                          &       &  Min  &  Max  &  Mean & Median & 1$\sigma$ \\ 
\hline 
\textbf{NLVC:}            & $v_c$ &  -29.2 &  21.1 &   -3.5 &   -2.7 &   9.1 \\ 
Detected in 64 profiles   & FWHM  &  13.8 &  66.4  &  41.6  &   41.1 &  10.6 \\ 
\hline 
\textbf{BLVC:}            & $v_c$ &  -49.9 &  29.4 &  -12.4 &  -11.8 &  16.5 \\ 
Detected in 75 profiles   & FWHM  &   61.4 & 287.4 &  112.5 &  104.5 &  41.7 \\ 
\hline 
\textbf{Blueshifted HVC:} & $v_c$ & -204.4 & -33.6 & -100.3 & -109.9 &  41.3 \\   
Detected in 27 profiles   & FWHM  &   28.3 & 105.0 &   54.3 &   46.8 &  19.3 \\ 
\hline 
\textbf{Redshifted HVC:}  & $v_c$ &   40.3 & 129.6 &   79.8 &   73.1 &  28.2 \\ 
Detected in 11 profiles   & FWHM  &   31.6 & 164.0 &   79.6 &   72.6 &  46.1 \\ 
\hline 
    \end{tabular} 
    \tablefoot{All values are in units of km~s$^{-1}$. The instrumental profile is 11.3~km~s$^{-1}$.  
 In the final column, $1\sigma$ is the standard deviation of the distributions of $v_c$ and FWHM. 
    }
\end{table*}

Figures \ref{fig:gaus_hist1} and \ref{fig:gaus_hist2} suggest that the cutoff in centroid velocity 
between the LVC and the HVC should be around 50~km~s$^{-1}$. However, for systems seen at high inclinations, 
the [OI]$\lambda$6300 emission coming from a high-velocity jet would be observed at lower projected 
velocities. If, for instance, a system is observed at an inclination of $80^{\circ}$, emission from 
a jet traveling at 200~km~s$^{-1}$ would be centered at 35~km~s$^{-1}$ ($v_{jet}*\cos i$). We appear to have a 
few cases like this in our sample. For example, the star \object{CSIMon-000228} shows a broad 
component centered close to rest velocity and a narrower component (FWHM $= 56.7$~km~s$^{-1}$) centered 
at -33.6~km~s$^{-1}$, which is more blueshifted than a typical narrow LVC. We do not have direct measurements 
of inclination for the stars in this cluster, but we can deduce from its rotation properties that 
this star is observed at a very high inclination. This component is therefore consistent with a HVC. 

In the end, the following criteria were used to classify each component. 
Firstly, all components with $|v_c| < 30$~km~s$^{-1}$ were classified as LVCs. When more than one 
component from an emission profile satisfied this condition, the broadest was classified as a 
broad LVC and the narrowest as a narrow LVC. When there was only one component, it was classified 
as narrow if FWHM $<$ 57~km~s$^{-1}$ and broad if FWHM $\geq$ 57~km~s$^{-1}$ (from the analysis of Figs. 
\ref{fig:vel_comp1}, \ref{fig:gaus_hist1} and \ref{fig:gaus_hist2}). 
There were no cases of more than two components in this velocity bin. 
Secondly, all components with $|v_c| > 50$~km~s$^{-1}$ were classified as HVCs.\footnote{There were 
two cases (stars \object{CSIMon-000260} and \object{CSIMon-000392}) in which two blueshifted HVCs 
were found. Multiple HVCs have been detected in other studies as well (e.g., S16).} 
Finally, the components with centroid velocity between 30~km~s$^{-1}$ and 50~km~s$^{-1}$ were 
classified as HVCs when their FWHM was less than 100~km~s$^{-1}$ and as BLVCs if 
FWHM~$\geq$~100~km~s$^{-1}$. 
This is justifiable by the fact that the FWHM of the HVC measures the internal velocity dispersion 
within the jet, and is not expected to exceed 100~km~s$^{-1}$ unless there is a considerable velocity 
variability within the beam. 
Furthermore, using this criterion, there were no cases in which more than one BLVC was identified in 
the same source. 
However, we must note that the choice of where to establish a threshold in FWHM between HVC and BLVC 
is based only on the distribution seen in Fig. \ref{fig:vel_comp1} and is thus somewhat arbitrary. 
Therefore there may be some misclassification in this velocity bin. 

Of the 182 CTTSs and 2 HAeBe stars observed in NGC~2264, we detected emission in the 
[OI]$\lambda$6300 line of both HAeBe stars and 108 CTTSs, all but one of which presented a LVC 
(the only exception being \object{CSIMon-001249}, which shows both a red and blueshifted HVC; 
a LVC is likely present, but blended within the two HVCs). 
Among them, we found a total of 64 narrow LVCs and 75 broad LVCs, with 30 systems presenting both 
of these components. A HVC, however, was only found in 31 profiles (28\% of the total detections). This 
represents only 17\% of all CTTSs and HAeBe stars observed, a much lower detection rate than was 
expected\footnote{This may be due to a sensitivity issue, as discussed in Appendix \ref{sec:compl}.}. 
We performed K-S tests between the distributions of FWHM and $v_c$ of the different components 
and all proved to be statistically different, with a $\lesssim 1\%$ probability that the distributions 
draw from the same population. 

Table \ref{tab:compon} presents a summary of the different types of profiles found in our sample, 
showing how many detected profiles present each of the different components and how many present 
a combination of two or more of these components. The different components are 
characterized in terms of their centroid velocities ($v_c$) and full width at half maximum (FWHM) 
in Table \ref{tab:vel_cmp}. 

Both the BLVC and the NLVC appear to be more blueshifted whenever a HVC is present in the system. 
For the BLVC, the average $v_c$ is -9.3~km~s$^{-1}$ (with a standard deviation of 
\mbox{$\sigma =$ 15.7 km~s$^{-1}$}) and -19.2~km~s$^{-1}$ (\mbox{$\sigma =$ 16.4 km~s$^{-1}$}) among 
systems without and with a detected HVC, respectively. For the NLVC, the respective numbers are 
-1.5~km~s$^{-1}$ (\mbox{$\sigma =$ 8.7 km~s$^{-1}$}) and -9.9~km~s$^{-1}$ (\mbox{$\sigma =$ 7.7 km~s$^{-1}$}). 
For the NLVC, a KS test shows that the distributions of the centroid velocities 
of the two samples (stars that present a HVC and stars that do not) are statistically different, 
with a probability of deriving from the same population of less than 1\%. There is no statistically 
significant difference in the distributions of FWHM among the two samples for either component. 
This difference in centroid velocities may be caused by a geometrical effect, since it has been 
previously shown that systems observed at lower inclinations, which are more favorable to detect 
a HVC, tend to present BLVCs of larger blueshifts (S16). 
On the other hand, this effect could simply be due to the difficulty of separating the 
low-velocity components when a HVC is blended in the profile. In these cases, the LVCs may have 
been artificially shifted toward the blue in the Gaussian decomposition. 

\subsection{Comparison of our sample of LVCs with previous studies}\label{sec:oicomp} 

\begin{figure*}[t!]
    \centering
    \includegraphics[width=17cm]{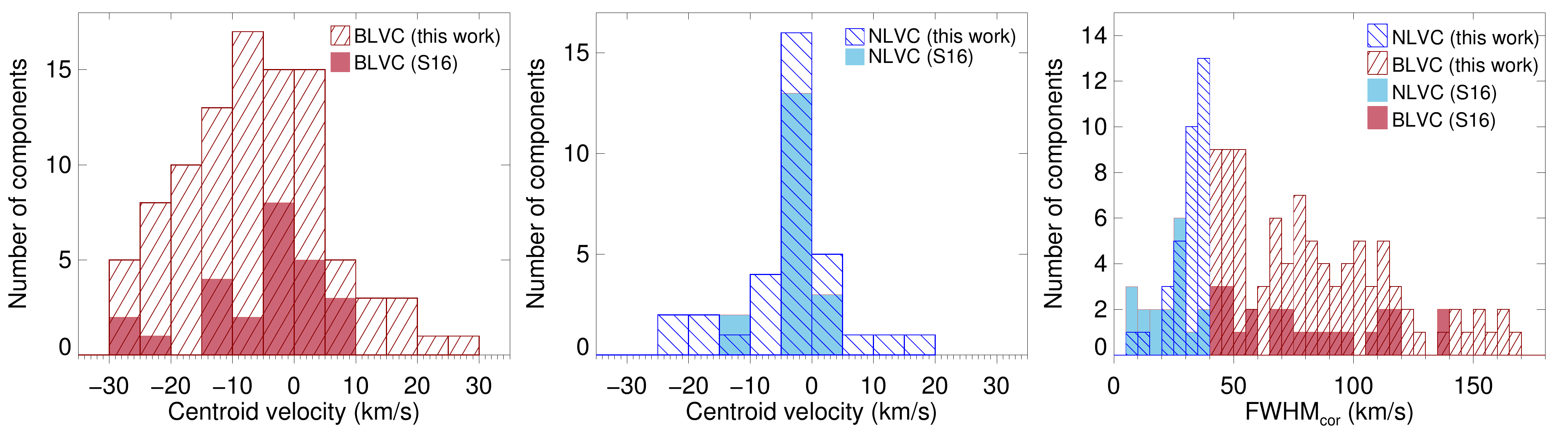}
    \caption{Comparison of the centroid velocities and FWHM of the low-velocity components of 
the [OI]$\lambda$6300 line in our sample and in S16. Striped red bins represent our broad 
LVCs, shaded red bins represent the BLVCs of S16, striped blue bins represent our narrow LVCs, 
and shaded blue bins represent the NLVCs of S16. 
For this figure, we use the definition of S16 to classify our components, in order to compare the 
two samples (i.e., BLVCs have \mbox{$|v_c| < 30$~km~s$^{-1}$} and \mbox{FWHM $> 40$~km~s$^{-1}$} and NLVCs have 
\mbox{$|v_c| < 30$~km~s$^{-1}$} and \mbox{FWHM $\leq 40$~km~s$^{-1}$}). 
In the right panel, FWHM have been deconvolved from the instrumental profiles in both samples, in 
order to compare without the influence of the different spectral resolutions. 
}\label{fig:oicomp}
\end{figure*}

In this section, we summarize the differences between the narrow and broad low-velocity components 
of the [OI]$\lambda$6300 line in our sample and in the sample of S16, the only previous study in 
which the distinction between these two components has been made for a large number of profiles. 
Figure \ref{fig:oicomp} shows the distributions of centroid velocity and FWHM for both the 
narrow and broad components of the LVCs in our sample and in that of S16. 
In this section, we use the widths corrected for the instrumental profiles (referred to as 
FWHM$_{cor}$, to avoid confusion with the uncorrected widths), in order to accurately compare the 
two samples of different spectral resolutions. 
Besides this, throughout this section we use S16's definition of NLVC and BLVC for our components as 
well, in order to compare the two samples regardless of the difference in definitions used to classify 
the components (i.e., BLVCs have \mbox{$|v_c| < 30$~km~s$^{-1}$} and \mbox{FWHM$_{cor}$ $> 40$~km~s$^{-1}$} 
and NLVCs have \mbox{$|v_c| < 30$~km~s$^{-1}$} and \mbox{FWHM$_{cor}$ $\leq 40$~km~s$^{-1}$}).  
Using these definitions, we find a total of 29 NLVCs and 100 BLVCs in our sample, while S16 had 
found 18 NLVCs and 25 BLVCs in the Taurus sample.

Table \ref{tab:cmps16} summarizes the comparison between the two samples. This table gives the 
mean and standard deviation ($\sigma$) of the distributions of centroid velocity and FWHM$_{cor}$ 
of each component in both samples. We also give the results of KS tests performed between the 
distributions. 

The left and middle panels of Fig. \ref{fig:oicomp} show that, when using the same criteria to 
classify the low-velocity components as S16, both the NLVCs and BLVCs show very similar distributions 
of centroid velocity between the two samples. A KS test between the centroid velocities of the BLVCs 
in our sample and in the sample of S16 gives a high probability (P~$=20$\%) that the distributions 
derive from the same population, with a very similar result (P~$=24$\%) for the NLVC. 
Even if we include the BLVCs in our sample that extend beyond $|v_c|=30$~km~s$^{-1}$ up to 
$|v_c|=50$~km~s$^{-1}$, we still find a probability of 8\% that the samples derive from the same 
population, which is high enough to suggest that the difference in the range of $|v_c|$'s we observe 
may not only be due to the different cutoffs chosen between LVC and HVC, but possibly to our larger 
number statistics, which may have allowed us to sample the tail of the most blueshifted components.  

In the right panel of Fig. \ref{fig:oicomp}, we can see that the distribution of FWHM of the BLVC is 
also very similar among the two samples, with a KS test giving a 91\% probability that they derive from 
the same population. Our values extend further than those of S16 (the figure is clipped at 180~km~s$^{-1}$ 
for clarity, but there is one component in our sample whose FWHM reaches nearly 300~km~s$^{-1}$), but the 
result of the KS test shows that this is simply a matter of our larger number statistics, which allowed us 
to better sample the tail of the distribution. As for the NLVC, we see that the distributions 
of FWHM differ considerably, with S16 finding many more components narrower than 25~km~s$^{-1}$ than we do, 
despite our larger sample size. It is possible that the contamination of the telluric emission line, 
which is much stronger in our observations than in those of S16 (see the left panels of Fig. 
\ref{fig:speccor} and Fig. 28 of S16), may have made it very difficult for us to recover 
the narrowest [OI]$\lambda$6300 line profiles. The narrow LVCs are generally very close to stellar rest 
velocity (as illustrated by the middle panel of Fig. \ref{fig:oicomp}), which means that they closely 
coincide with the telluric emission line, and many profiles with only a very narrow LVC may have been 
entirely engulfed by this contamination. This would also help to explain our low detection rate 
(discussed in Sect. \ref{sec:oicor} and Appendix \ref{sec:compl}) in comparison with previous studies 
of the [OI]$\lambda$6300 line among accreting T Tauri stars. 
The higher spectral resolution of S16 (almost twice ours) will also have made it much easier for them 
to properly remove this contamination. In light of this, we compare the two samples of NLVCs only for 
components of FWHM~$>$~25~km~s$^{-1}$. In this case, we find a probability of 30\% that the distributions 
of FWHM of the two samples derive from the same population, and a probability of 59\% for 
the centroid velocity of this component.

\begin{table}[t]
    \centering
    \caption{Comparison between the [OI]$\lambda$6300 LVC in NGC~2264 (our sample) and in Taurus (S16)
    }\label{tab:cmps16}
    \begin{tabular}{c c c c c c} % c}
    \hline 
    \hline 
    & \multicolumn{2}{ c }{This paper} & \multicolumn{2}{ c }{S16} & KS test \\ 
    \hline 
               & Mean  & $\sigma$ & Mean & $\sigma$ & P \\ 
    \hline 
\textbf{NLVC:} &       &          &      &          &       \\ 
 $v_c$         & -3.4  &   8.2    & -2.5 &    4.1   & 59\%  \\   
 FWHM$_{cor}$  & 30.2  &   9.5    & 22.5 &    9.3   & 30\%  \\   
    \hline 
\textbf{BLVC:} &       &          &      &          &       \\  
 $v_c$         & -6.4  &  11.7    & -5.0 &    9.8   & 20\%  \\   
 FWHM$_{cor}$  & 87.2  &  40.9    & 79.9 &   30.5   & 91\%  \\   
    \hline 
    \end{tabular}
    \tablefoot{
The first four columns show the mean and standard deviation ($\sigma$) of the distributions of centroid 
velocities ($v_c$) and FWHM for the NLVCs and BLVCs of the [OI]$\lambda$6300 line in our sample and in 
the sample of Taurus from S16. The last column shows the results of KS tests between the two samples, 
restricted to components of FWHM~$> 25$~km~s$^{-1}$. 
P is the probability that the distributions derive from the same population.  \\  
FWHM$_{cor}$ represents the FWHM corrected for the instrumental broadening of \mbox{FWHM$_{inst} =$ 11.3~km~s$^{-1}$}, 
for our sample, and \mbox{FWHM$_{inst} =$ 7.0~km~s$^{-1}$}, for S16's sample. \\ 
The definitions of S16 are used to classify NLVCs and BLVCs in both samples, in order to compare them
regardless of the different definitions. 
    }
\end{table}

This study of NGC~2264 and the study of Taurus conducted by S16 probe similar ranges 
of mass, stellar luminosity and mass accretion rates. However, the average values of 
all three of these parameters are larger in our sample, when considering only systems
 with positive [OI]$\lambda$6300 detections, than in the sample of S16. Our sample with 
[OI] detection has an average mass of 0.88$\pm$0.51 M$_{\odot}$ (ranging from 0.14M$_{\odot}$ 
to 3M$_{\odot}$), compared to 0.67$\pm$0.30 M$_{\odot}$ for S16; an average stellar 
luminosity of $\log{\mathrm{L}_*}/{\mathrm{L}_{\odot}} = -0.08\pm0.45$, compared to 
S16's $\log{\mathrm{L}_*}/{\mathrm{L}_{\odot}} = -0.36\pm0.37$; and an average 
accretion luminosity of $\log{\mathrm{L}_{acc}}/{\mathrm{L}_{\odot}} = -1.05\pm0.75$, 
compared to $\log{\mathrm{L}_{acc}}/{\mathrm{L}_{\odot}} = -1.69\pm0.76$ for S16. 
The values of X-ray luminosities are very similar in the two samples: 
$\log{\mathrm{L}_X}/{\mathrm{L}_{\odot}} = -3.55\pm0.54$ in NGC~2664, versus 
$\log{\mathrm{L}_X}/{\mathrm{L}_{\odot}} = -3.66\pm0.59$ in Taurus. 

\section{[OI]$\lambda$6300 line luminosities}\label{sec:oilum} 

\subsection{Recovering equivalent widths and line luminosities}\label{sec:oilumcal} 

Equivalent widths (EW) were measured for each [OI]$\lambda$6300 line profile and its components. 
In order to estimate uncertainties, we added a random distribution of noise at the same level of 
the flux uncertainty and calculated the resulting EW. This was done 200 times, then the mean and 
standard deviation were taken as our final value of EW and its uncertainty. We estimated the EW 
of the narrow and broad components of the LVC from their Gaussian fits. For the HVC, we took the 
line's total value, measured by integrating above the continuum over the full profile, minus the 
EWs of all other components. This is to account for the fact that the HVC usually deviates from 
a Gaussian profile more than the narrow and broad components of the LVC and therefore using the 
value derived from the Gaussian fit may result in a larger error. 
For the 7 stars that present both a redshifted and a blueshifted HVC, we first estimated 
the EW of each of these components from their Gaussian fits. We then made a second estimate by 
subtracting the values of all other Gaussian components, including that of the other HVC, from 
the total [OI]$\lambda$6300 line's EW. 
This slightly overestimates each HVC's EW because the residuals from both components as they 
deviate from a Gaussian profile are added to each component's measure of EW. Therefore to 
minimize this effect, we consider this value to be an upper limit and the estimate from the 
Gaussian fit to be a lower limit, and take the average of the two as the final value.
The equivalent widths, as well as the centroid velocities and FWHM, of each component 
in our sample are given in Table \ref{table:oi_comp_par} (available at the CDS). 

Our spectra are not flux calibrated, therefore in order to retrieve [OI]$\lambda$6300 line 
luminosities we estimated the continuum flux around 6300\AA \space using the average $r$-band 
magnitude from the CFHT Megacam 2012 campaign \citep[described in ][]{venuti14}. To account for 
a star's variability, we took the amplitude of variability of its CoRoT light curve(s) as the 
error bar. For the brightest stars that saturated in the CFHT observations, we used the CoRoT 
$R$ magnitudes, which were converted from the mean CoRoT flux using the photometric zero-point 
of 26.74, as determined by \citet{cody14}. A comparison of the CFHT $r$ and CoRoT $R$ magnitudes 
in our sample showed that the two coincide very well among the brightest stars ($R \lesssim 14$) 
and therefore we introduced very little error by using the CoRoT magnitudes for the stars with 
no CFHT photometry. Magnitudes were corrected for extinction (see below), then converted to 
flux using a distance to the cluster of d~$=760$ pc \citep{sung97,gillen14}. The continuum flux 
was then multiplied by the measured equivalent widths in order to retrieve the luminosities of 
the [OI]$\lambda$6300 line's individual components, which were then added together in order to 
retrieve the total [OI]$\lambda$6300 line luminosity. The values found for each component and 
for the full [OI]$\lambda$6300 line are given in Table \ref{table:ctts_oi} (available at the CDS). 

To correct for extinction we used the individual values derived for each star by 
\citet{venuti14}. When no value was available, we used the average extinction toward the 
cluster of $A_R=0.40$, estimated from the stars in our sample with measured extinction. 
This was done when calculating the luminosity of the [OI]$\lambda$6300 line's LVCs. The HVC, 
however, is believed to arise from jets far enough above the disk mid-plane that it is likely 
only affected by interstellar extinction in our line-of-sight toward the cluster, and not by 
the circumstellar disk material of individual objects. Therefore, when calculating the flux 
in the [OI]$\lambda$6300 line's blueshifted HVC, we used the average extinction toward the 
cluster for all stars rather than their individual values. 
However, interstellar extinction is not necessarily uniform across the entire cluster, so 
using the average value may introduce additional uncertainties in the HVC luminosities. 
Therefore, we took into account the differences between the average value of extinction and 
individually determined values when calculating the error bars in the [OI]$\lambda$6300 HVC 
luminosity. Most of the stars in NGC~2264 present low extinction that is very similar to the 
average value of $A_R=0.40$, meaning that this did not result in a large addition to the  
uncertainty, except for a few stars which seem to be in more embedded regions. 

When calculating the luminosity of the redshifted HVCs, the individually derived values of 
extinction were used, just as for the LVCs. Even so, it is likely that their luminosities were 
often underestimated, because this emission suffers additional extinction when traversing the 
circumstellar disk, which should intersect our line-of-sight to the receding part of the 
protostellar jet. In fact, when a star presents both redshifted and blueshifted high velocity 
components, the luminosity of the former is usually lower than that of the latter. 
Therefore care must be taken when analyzing luminosities of the redshifted HVC. 

We also calculated H$\alpha$ line luminosities from equivalent widths and the same estimated 
continuum flux as for the [OI]$\lambda$6300 line, using $r$-band or CoRoT magnitudes corrected 
for individual extinction values. For the range of spectral types considered in this paper 
there should not be a considerable difference between the continuum flux at 6300\AA \space 
and at 6563\AA. H$\alpha$ equivalent widths were taken from previous FLAMES campaigns 
\citep{sousa16} or from the literature \citep{rebull02,dahm05}. 

\subsection{Correlations with stellar and accretion properties}\label{sec:oilumcor} 

We analyzed the relation between the luminosity of the [OI]$\lambda$6300 line and its 
individual components (NLVC, BLVC, and HVC) and different properties of the star, disk and 
accretion (these properties are given in Table \ref{table:ctts} 
and the relations can be seen in Fig. \ref{fig:oi_lum_lbol}).
In order to evaluate possible correlations between these properties and the [OI]$\lambda$6300 
emission line and its different components, we used the ASURV (Astronomy SURVival) package 
\citep{lavalley92}, which implements the methods described in \citet{isobe86}. We chose these 
methods because they allow us to perform linear regression and correlation tests while taking 
into account the upper limits of our nondetections and thus diminishing possible observational 
biases. 

\paragraph{Stellar luminosity}

Stellar bolometric luminosities were determined in a self-consistent way for nearly all 
stars in our sample by \citet{venuti14} and were determined by us following the procedures 
described therein for the few remaining stars.  
We find a strong correlation between the stellar luminosity and all components, especially 
the low velocity components, something that is also seen by N14\footnote{The studies 
performed by N14 and N18 were conducted using X-shooter, which has a resolution of only 
R=8,800 in the [OI]$\lambda$6300 line. These studies were therefore unable to separate the 
broad and narrow components of the LVC in their sample and analyzed the LVC as a whole.}, 
S16 and N18\footnotemark[8]. A Kendall $\tau$ test gives a probability of less than 0.01\% 
that these correlations do not exist. 
        
\paragraph{Accretion luminosity}

Accretion luminosities were calculated from UV excess luminosity by \citet{venuti14} when 
available, and H$\alpha$ luminosity otherwise \citep[following the relation proposed by][]{fang09}. 
All components of the [OI]$\lambda$6300 line correlate well with the accretion luminosity L$_{acc}$, 
or similarly with the mass accretion rate $\dot{M}_{acc}$, as has been noted in many previous studies 
\citep[][HEG95, R13, N14, S16, N18]{cabrit90}. 
These correlations remain even when we normalize both L$_{acc}$ and the luminosity of the 
components of the [OI]$\lambda$6300 line by the stellar luminosity L$_*$, showing that these 
correlations are not driven by an underlying correlation with L$_*$. 

A Kendall $\tau$ test shows that the probability that L$_{acc}$ and the luminosities of the different 
components of the [OI]$\lambda$6300 line are not correlated is less than 0.008\% for all components. 
However, for the HVC, if we examine only positive detections we no longer find a clear correlation 
between L$_{acc}$ and L$_{HVC}$. 
This correlation has been found in other studies (e.g., N18) and many of the nondetections in our 
sample belong to sources of low accretion luminosities, meaning that they may in fact have weak 
[OI]$\lambda$6300 HVC emission which was below our detection threshold. Therefore it is reasonable 
to assume that this lack of correlation between the HVC luminosity and accretion luminosity when 
considering only positive detections should be due to an observational bias. This demonstrates the 
importance of including upper limits in this analysis. 

Using the parametric EM algorithm available in ASURV, we find the following relations 
for each component of the [OI]$\lambda$6300 line: 

\vspace{-0.5cm}
\begin{equation}
\log \mathrm{L}_{NLVC} = -4.64(\pm0.14) + 0.84(\pm0.11) \log \mathrm{L}_{acc} 
\end{equation}

\vspace{-0.5cm}
\begin{equation}
\log \mathrm{L}_{BLVC} = -4.52(\pm0.11) + 0.57(\pm0.07) \log \mathrm{L}_{acc} 
\end{equation}

\vspace{-0.5cm}
\begin{equation}\label{eq:hvc}
\log \mathrm{L}_{HVC} = -4.78(\pm0.20) + 0.89(\pm0.15) \log \mathrm{L}_{acc} .
\end{equation}

\begin{figure}[tb]
    \centering
    \resizebox{\hsize}{!}{\includegraphics{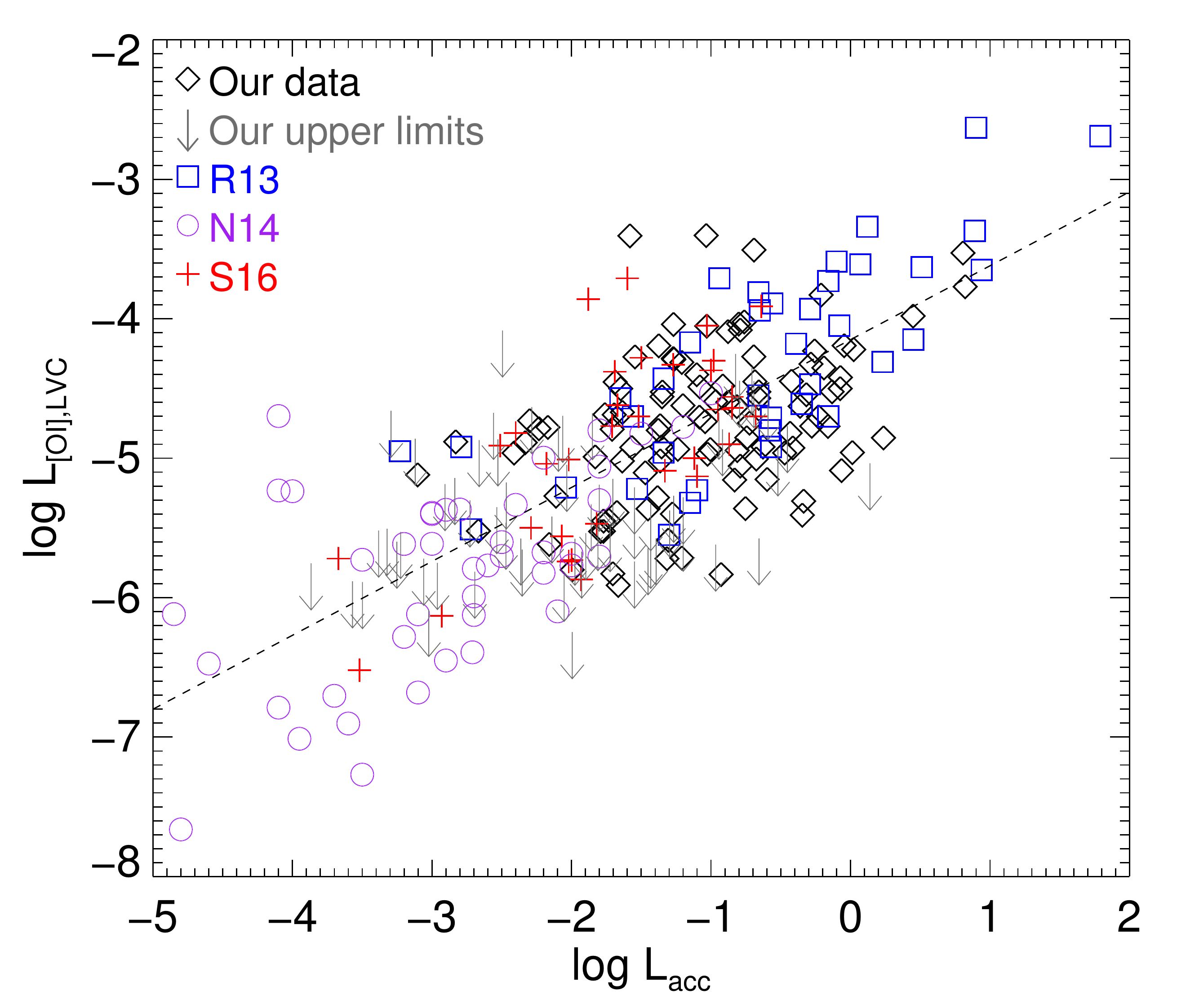}} 
    \caption{[OI]$\lambda$6300 LVC luminosity versus accretion luminosity. 
Black diamonds represent our detections, while gray arrows represent our $3\sigma$ upper limits 
to the LVC luminosities when no emission in [OI]$\lambda$6300 was detected. Blue squares, purple 
circles and red crosses represent the values given in R13, N14 and S16 for their samples, respectively. 
The dashed lines shows a linear fit to the combination of all four samples, including our upper limits. 
    }\label{fig:lvccmp}
\end{figure}

\begin{figure*}[t!]
    \centering
    \includegraphics[width=17cm]{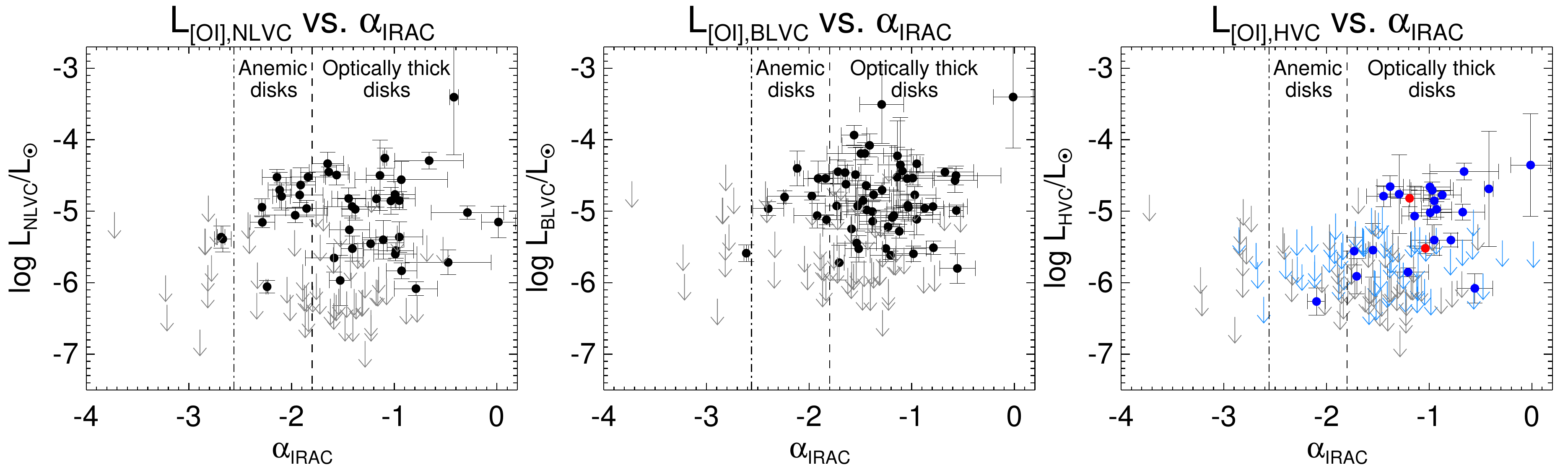} 
    \caption{Luminosity of each [OI]$\lambda$6300 component versus the $\alpha_{IRAC}$ index for 
all accreting systems. 
Gray arrows represent upper limits for [OI]$\lambda$6300 line luminosities when no emission in 
[OI]$\lambda$6300 was detected. In the HVC plots (right column): light blue arrows represent upper 
limits for HVC luminosities when only low-velocity [OI]$\lambda$6300 emission was detected; blue 
filled circles represent blueshifted HVCs while red filled circles represent redshifted HVCs 
(systems that did not present a blueshifted component). 
Dashed lines represent the boundary between systems with optically thick inner disks (right) 
and systems with anemic disks (middle), while dashed-dotted lines represent the boundary between 
systems with anemic disks and naked photospheres, according to $\alpha_{IRAC}$ (left). 
    }\label{fig:oi_lum_airac}
\end{figure*}

We can compare our correlations of the LVC with those of R13, N14 and S16. Figure \ref{fig:lvccmp} 
shows the luminosity of the full (narrow+broad) LVC of the [OI]$\lambda$6300 line versus accretion 
luminosity for all four samples. We consider 
the full LVC rather than the broad or narrow components individually in order to compare with R13, 
who only decomposed their LVC into separate components for two stars, and N14, who did not have 
enough spectral resolution to perform this decomposition. We see that all of the samples agree 
well in both relations, allowing us to establish a well-defined link between the [OI]$\lambda$6300 
line's LVC luminosity and the accretion luminosity over 7 orders of magnitude. Using once again 
the EM method, we find the following for the combination of all four samples: 

\vspace{-0.5cm}
\begin{equation}\label{eq:lvclacc}
\log \mathrm{L}_{LVC,all} = -4.15(\pm0.06) + 0.53(\pm0.04) \log \mathrm{L}_{acc}
\end{equation}

%\noindent
Equation \ref{eq:lvclacc} is in agreement with the relation found for only our sample, 
using the same method: 
        
\vspace{-0.5cm}
\begin{equation}
\log \mathrm{L}_{LVC,ours} = -4.44(\pm0.11) + 0.64(\pm0.07) \log \mathrm{L}_{acc} . 
\end{equation}

\noindent
The relation between L$_{LVC}$ and L$_{acc}$ is also in agreement with those found by R13, 
S16 and N18, though our slope is not as steep as the one found by N14 for their data. 
As for the HVC, our slope of L$_{HVC}$ vs. L$_{acc}$ (Eq. \ref{eq:hvc}) agrees with that of N18, 
though our values of L$_{HVC}$ in general lie below theirs for the same L$_{acc}$. 

In Fig. \ref{fig:lvccmp}, we can see that our upper limits in [OI]$\lambda$6300 LVC 
luminosity are within the same range as the luminosities measured by N14 and S16 in their 
samples, particularly among the stars of low mass accretion rates (L$_{acc} \lesssim 10^{-2} 
\mathrm{L}_{\odot}$). This strongly supports the idea that our low detection rate, especially 
among these stars of low accretion luminosity, is likely not due to a physical effect but 
instead to a sensitivity limit. Many of the stars in our sample where no emission was detected 
in [OI]$\lambda$6300 could in fact present [OI]$\lambda$6300 emission that was unfortunately 
below our detection threshold. 
        
\paragraph{X-ray luminosity}

We also searched for correlations between the luminosity of the [OI]$\lambda$6300 line and its 
components and X-ray luminosities\footnote{Measured in the 0.5-8.0 keV band.} 
\citep[provided by Ettore Flaccomio, private communication; some of these values are published 
in][]{guarcello17}. The X-ray luminosities do not correlate with the HVC luminosity. However, 
we find a positive correlation with the luminosity of the LVC, a different result than the one 
obtained by S16, who find a 10.8\% likelihood that no such correlation exists in their sample. 
In fact, when we separate the contributions of the narrow and broad components, we find that 
there is only a 0.05\% chance that the X-ray luminosity (L$_X$) does not correlate with the broad 
LVC luminosity, and an even smaller one ($<$~0.001\%) that there is no correlation with the 
narrow LVC. However, since our data also show a correlation between the X-ray luminosity and the 
stellar luminosity, it is unclear to us whether the relationships found between L$_X$ 
and L$_{NLVC,BLVC}$ are real or whether they are driven by an underlying correlation 
with the stellar luminosity. 
In an attempt to remove the possible influence of the stellar luminosity (L$_*$), we 
normalized the X-ray luminosities and the [OI]$\lambda$6300 line's NLVC and BLVC 
luminosities by L$_*$. We find no correlation between these two quantities, even spanning 
a range of around 2 orders of magnitude in L$_X/\mathrm{L_*}$ . Therefore it 
is likely that the correlations we find between L$_X$ and L$_{NLVC}$ or 
L$_{BLVC}$ are not intrinsic, but are driven by the stellar luminosity.  

\paragraph{UV luminosity}

We also find a correlation between both the NLVC and BLVC luminosities and the luminosity of the 
CFHT $u$-band, with a very low ($< 0.001\%$) probability that it is false. However, since the 
$u$-band flux shows a strong positive correlation with stellar luminosity, it is unclear if the 
relation we find is real and not driven by the stellar luminosity, as was the case for L$_X$. 
Again, no correlations are found when the [OI]$\lambda$6300 line luminosities and UV luminosity 
are normalized by stellar luminosity (again spanning around 2 orders of magnitude in 
L$_{UV}/\mathrm{L_*}$), meaning that this relation is very likely driven by the 
stellar luminosity and is not an intrinsic one. Still, it is interesting to note that the NLVC 
correlates best with the X-ray luminosity, while the BLVC correlates better with the UV flux. 

\paragraph{The $\alpha_{IRAC}$ index}

Figure \ref{fig:oi_lum_airac} shows the relation between the luminosities of each of the 
[OI]$\lambda$6300 components and the $\alpha_{IRAC}$ index, which indicates the amount of dust 
present in the inner circumstellar disk. This index corresponds to the slope of a star's spectral 
energy distribution between 3.6$\mu$m and 8$\mu$m and can be used to classify a system's inner 
($\sim0.1$~au) disk structure \citep{lada06}. 
The dashed line in Fig. \ref{fig:oi_lum_airac} represents the boundary between, on the right, systems 
with optically thick inner disks and, in the middle, systems with anemic disks (optically thin inner 
disks), while the dashed-dotted line represents the boundary between systems with anemic disks and, 
on the left, systems that would be classified as naked photospheres according to the $\alpha_{IRAC}$ 
index. It is clear from the right panel that in order to possess high velocity [OI]$\lambda$6300 
emission, a system must be optically thick in this region of the disk. All systems showing this 
emission have optically thick inner disks, with the exception of one star (\object{CSIMon-000250}) 
that has an anemic disk and presents a faint [OI]$\lambda$6300 HVC. The NLVC is present in a number 
of anemic disk systems and even some systems classified as naked photospheres according to 
$\alpha_{IRAC}$, while the BLVC seems to show an intermediate behavior, having been observed in some 
anemic disk systems but not as many as the LVC.
These plots point to an evolution of the [OI]$\lambda$6300 line profile with the evolution of the 
disk. Systems with primordial disks often have a BLVC, or a combination of NLVC and BLVC, associated 
with a HVC. As the disk evolves the HVC disappears, followed by the BLVC, leaving only a NLVC 
among systems with optically thin inner disks. This is in line with what is discussed by 
\citet{ercolano17} and observed by S16, who show that systems with transition disks (those that 
present a dust cavity in their inner disk) present almost exclusively a NLVC in their 
[OI]$\lambda$6300 line profile. 
 
\paragraph{Infrared excess}

\begin{figure}[tb]
    \centering
    \resizebox{\hsize}{!}{\includegraphics{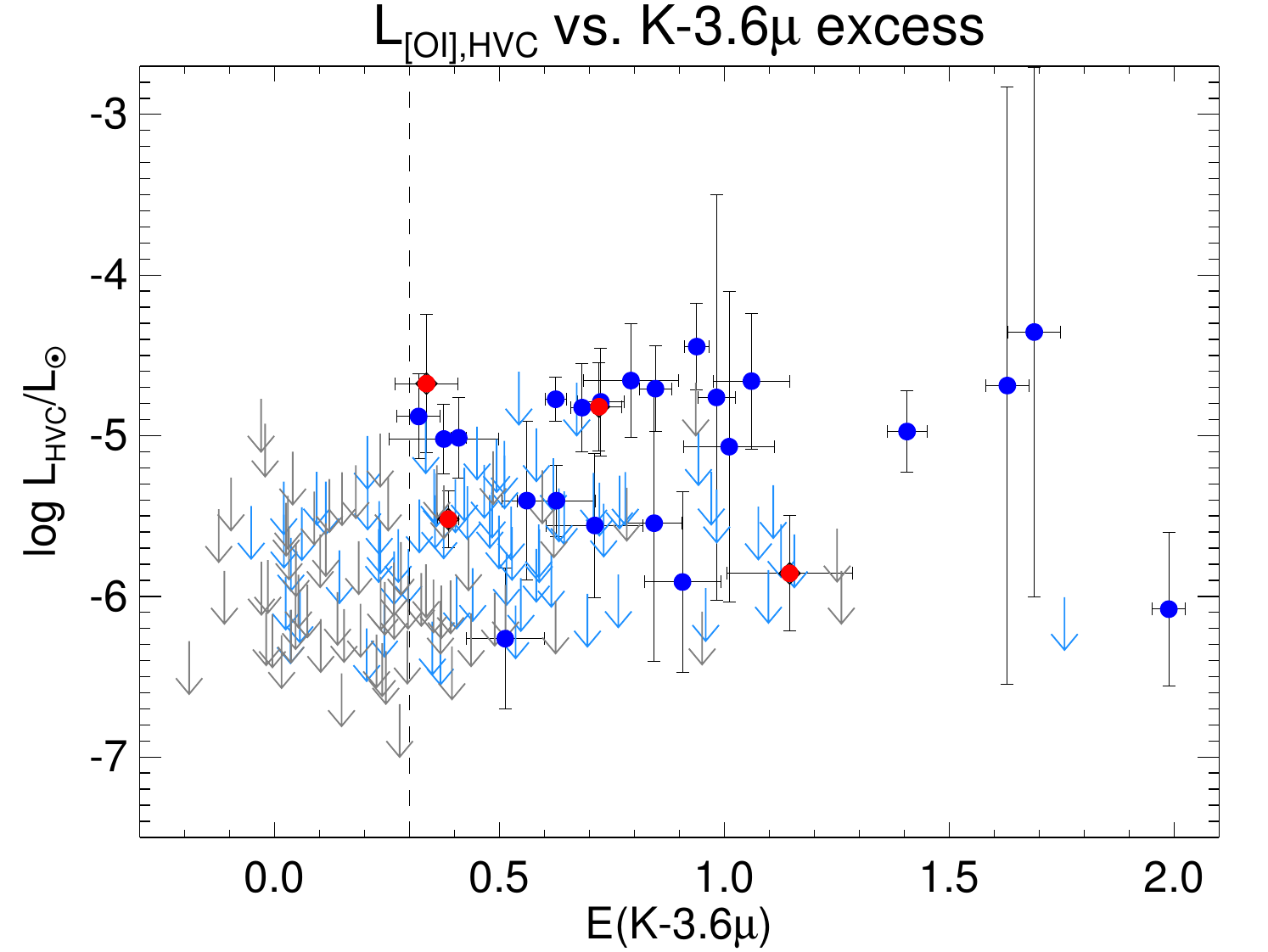}} 
    \caption{[OI]$\lambda$6300 HVC luminosity versus K-IRAC$_{3.6\mu}$ excess. 
The dashed line marks the threshold at K-IRAC$_{3.6\mu} = 0.3$, 
below which the HVC is not detected. 
}\label{fig:oi_lum_ir}
\end{figure}

The $\mathrm{K}-3.6\mu$m excess probes a region of the disk that is even closer to the star than 
$\alpha_{IRAC}$. Figure \ref{fig:oi_lum_ir} shows an even clearer threshold for HVC emission: all 
systems showing this emission, whether bright or faint, have $\mathrm{E(K}-3.6\mu \mathrm{m}) > 0.3$, 
the same conclusion reached by HEG95. This indicates that in order for a high velocity jet to be 
launched, the inner disk region must be optically thick. There is however no correlation 
between the amount of infrared excess and the jet's luminosity or velocity. 
 
\section{Photometric variability and the [OI]$\lambda$6300 line}\label{sec:oicorot}

\begin{figure*}[t!]
    \centering
    \includegraphics[width=17cm]{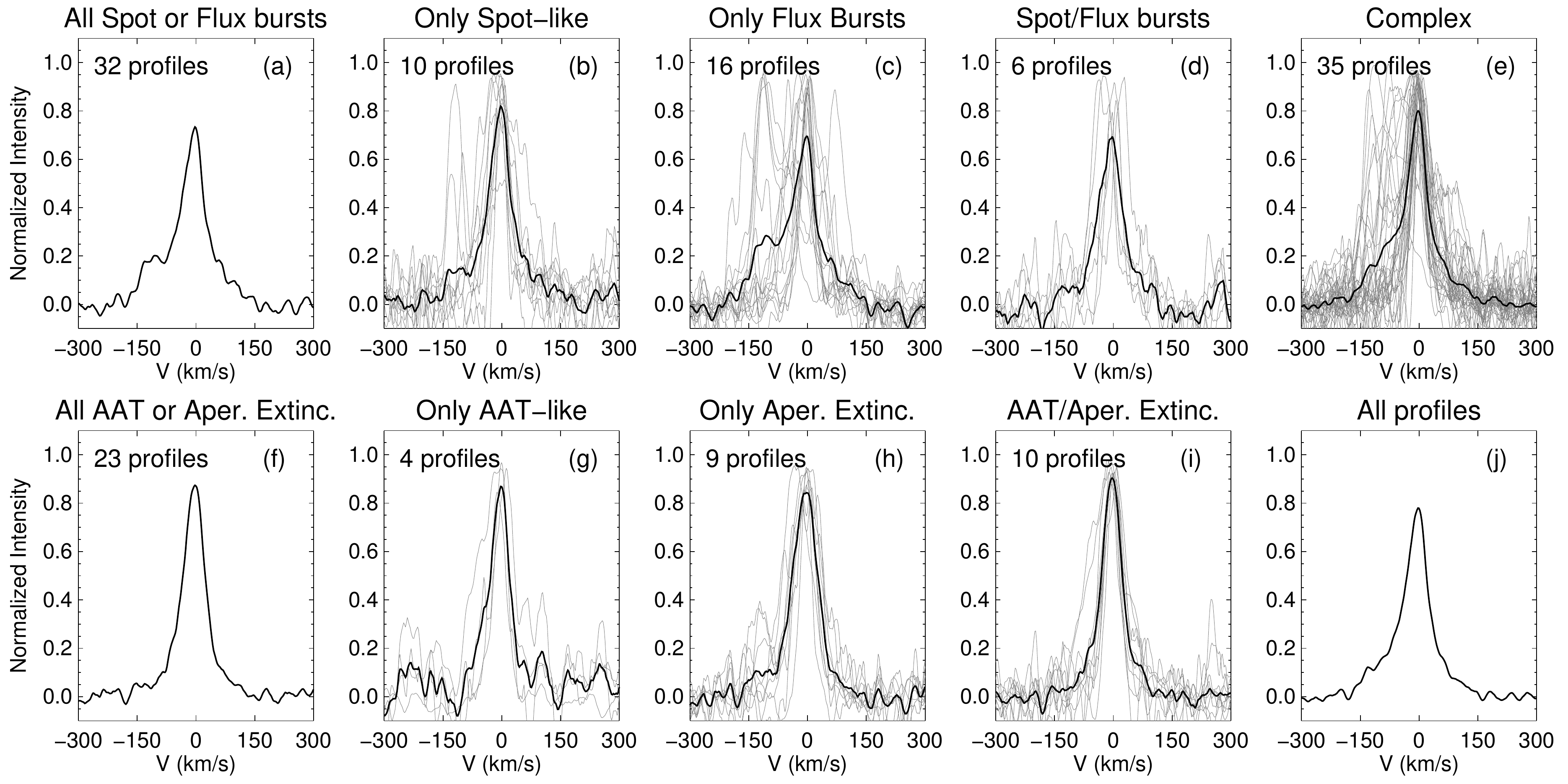}
    \caption{[OI]$\lambda$6300 line profiles for different types of photometric variability. 
Profiles of individual stars are shown in gray while the average in each group is shown in a 
thick black. 
From left to right, top to bottom: 
(a) stars that presented either spot-like or flux burst-dominated light curves in either CoRoT 
observation (likely observed at medium to low inclinations);
(b) stars with spot-like light curves in both CoRoT observations; 
(c) stars with light curves dominated by stochastic flux bursts in both CoRoT observations; 
(d) stars that altered between spot-like and flux bursts between 2008 and 2011; 
(e) stars with complex (difficult to classify) light curves; 
(f) stars that presented either AA Tau-like or aperiodic extinction-dominated light curves in either 
CoRoT observation (likely observed at high inclinations); 
(g) stars that presented only AA Tau-like light curves in both CoRoT observations; 
(h) stars that presented only aperiodic extinction-dominated light curves in both CoRoT observations; 
(i) stars that alternated between AA Tau-like and aperiodic extinction between 2008 and 2011; 
and finally (j) an average of all [OI]$\lambda$6300 line profiles. 
Each profile was normalized by its maximum intensity before the average was taken.
}\label{fig:lctype1}
\end{figure*}

In this section, we take advantage of the highly detailed light curves made available by the 
CoRoT satellite for many members of NGC~2264 and for the first time investigate properties of 
the [OI]$\lambda$6300 line profile in relation to a source's photometric behavior. This 
cluster was observed by CoRoT for 23 days in 2008 and again between Dec 1 2011 and Jan 9 2012, 
resulting in high precision light curves of hundreds of cluster members \citep{alencar10,cody14}. 
Of the 108 CTTSs in our sample that present [OI]$\lambda$6300 emission, 91 were observed by the 
CoRoT satellite in either of these two observing runs. The light curves of T Tauri stars have 
been classified in previous papers according to their morphology into the following groups: 
spot-like light curves, characterized by a very regular, periodic variability that has been 
attributed to rotational modulation of stable cold spots on the stellar surface \citep{alencar10}; 
light curves dominated by irregular flux bursts attributed to abrupt events of increased accretion 
\citep{cody14,stauffer14}; AA Tau-like light curves \citep{mcginnis15}, which resemble that of the 
CTTS AA Tau in \citet{bouvier99}, attributed to quasi-periodic occultation events of the stellar 
photosphere by an inner disk warp; and light curves dominated by aperiodic extinction events 
attributed to the occultation of the photosphere by irregularly distributed material in the disk 
\citep{mcginnis15}. However, many light curves present a more complex behavior not easily 
attributed to one of these groups, which may or may not have an underlying periodicity. They are 
likely the product of a combination of physical mechanisms, though a portion of these light 
curves has been attributed mainly to time-variable mass accretion producing transient hot spots 
on the stellar surface \citep{stauffer16}. 

It is reasonable to assume that a relatively high inclination is necessary to observe either 
AA Tau-like occultations or aperiodic extinction events, since optically thick material from 
the disk must intersect our line-of-sight in order to produce the occulting events. On the 
other hand, spot-like variability and light curves dominated by flux bursts are probably 
observed in systems with medium to low inclinations, since extinction events are absent from 
these light curves, meaning there is no optically thick disk material intersecting our 
line-of-sight. The complex, difficult to classify, light curves probably represent an
 intermediate case, where we observe both extinction events and spots on the stellar surface. 
We can therefore sometimes use the light curve classification to infer system inclinations, 
which is very useful since we do not have direct inclination measurements for this cluster.

In Fig. \ref{fig:lctype1}, 
comparing the average profile of stars with AA Tau-like or aperiodic extinction-dominated light 
curves (bottom left) with that of stars with spot-like or flux burst-dominated light curves (top 
left), we see that the low velocity components of the two profiles are very similar, whereas the 
high velocity components are not. The average profile of the stars with spot-like or flux 
burst-dominated light curves has a blueshifted HVC with a clearly separated peak, which is 
consistent with systems observed close to pole-on possessing fast bipolar jets, since the velocity 
of these jets projected in our line-of-sight would be close to their terminal velocities and the 
resulting high velocity [OI]$\lambda$6300 emission would be blueshifted enough to be easily 
distinguished from the [OI]$\lambda$6300 emission line's LVC.  
The average [OI]$\lambda$6300 line profile for the stars with AA Tau-like or 
aperiodic extinction-dominated light curves presents a typical LVC with a somewhat broad base 
and only a modest extended blue wing. This is consistent with a scenario where these systems are 
observed nearly edge-on, and therefore even if there were fast jets present, they would be nearly 
perpendicular to our line-of-sight and the projected velocity would be so low that we would barely 
be able to distinguish the HVC they produce from a LVC. 

\begin{figure}[t!]
    \centering
    \resizebox{\hsize}{!}{\includegraphics{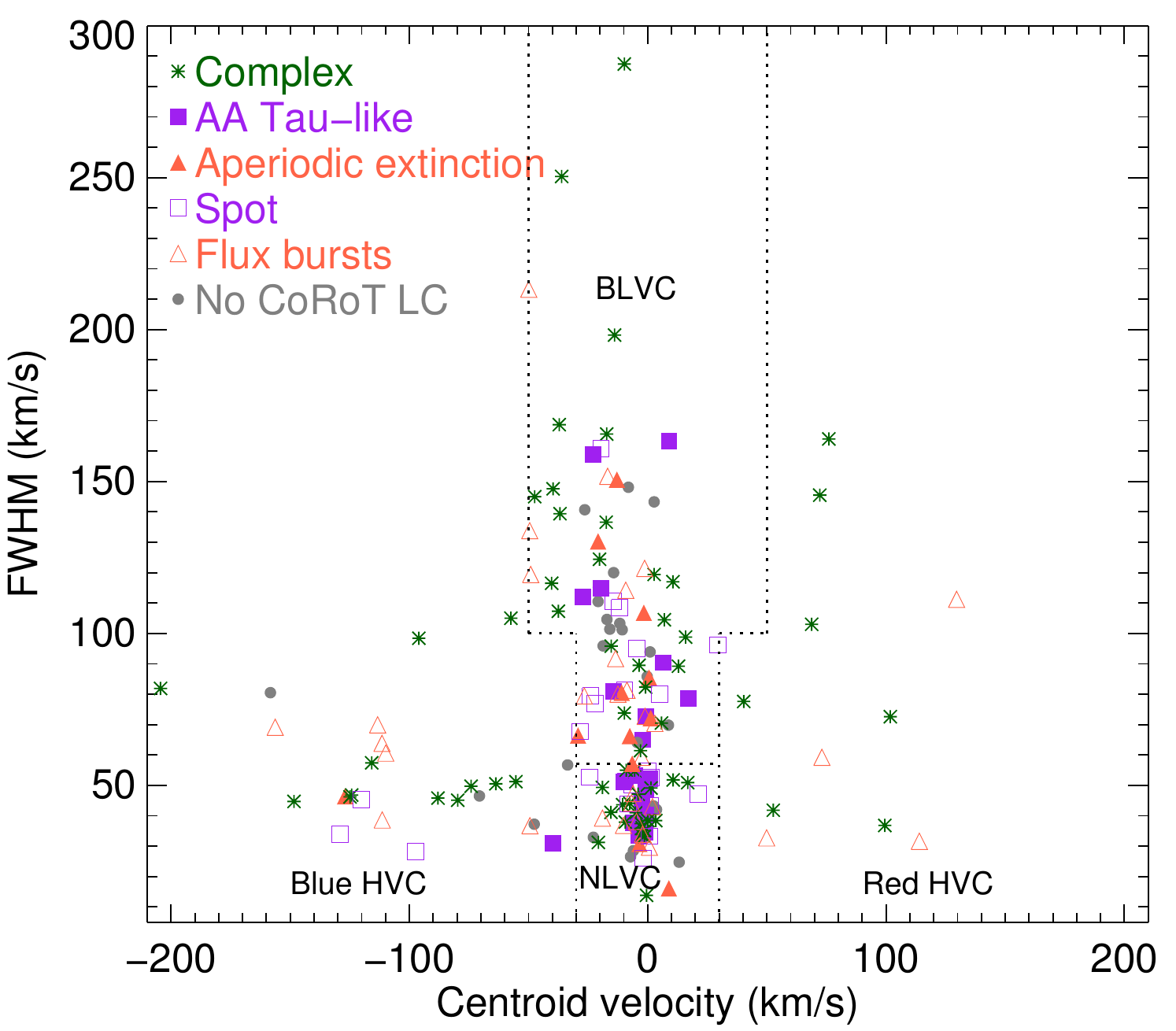}} 
    \caption{Distribution of centroid velocities and FWHM of all components in our sample, 
with different colors and symbols representing different types of photometric variability. 
Purple symbols represent systems that showed a periodic behavior in one or both CoRoT observing 
runs, with filled squares corresponding to AA Tau-like systems and open squares to those with 
spot-like behavior. Orange symbols represent systems that were nonperiodic in both 2008 and 2011, 
with filled triangles corresponding to systems whose light curves are dominated by aperiodic 
extinction and open triangles to those dominated by flux bursts. 
Green asterisks represent light curves that were classified as complex. 
Dashed lines indicate approximately the separation between different components - narrow and 
broad LVC, red- and blueshifted HVC. 
}\label{fig:lctype3}
\end{figure}

The light curve classification can help indicate in which way a CTTS is accreting matter. 
Magnetohydrodynamic (MHD) simulations \citep[e.g.,][]{romanova08,kulkarni08} have shown that it 
is possible for a star to accrete matter from the disk via a stable accretion regime, where one 
major accretion funnel flow and accretion shock are present on each hemisphere of the star-disk 
system; or via an unstable accretion regime, where a series of accretion funnel flows (or 
accretion "tongues") appear randomly around the star; or even in a sort of boundary between the 
two, such as the "ordered unstable" regime discussed in \citet{blinova15}. AA Tau-like photometric 
variability has been associated with a stable accretion regime \citep{bouvier07, romanova13, 
mcginnis15}, in which an inner disk warp may be produced as a consequence of the intense 
interaction between the inner disk and an inclined, strongly bipolar stellar magnetic field. If a 
system undergoing stable accretion is observed at a high enough inclination, its light curve would 
then present quasi-periodic flux dips due to this warp occulting the stellar photosphere as the 
system rotates. If this system were observed at a lower inclination its light curve would likely 
be dominated by rotational modulation of stable spots on the stellar surface. On the other hand, 
\citet{kurosawa13} show that a system undergoing unstable accretion would probably show stochastic 
photometric variability due to irregularly distributed hot spots on the stellar surface. 
\citet{mcginnis15} proposed that if these systems are observed at sufficiently high 
inclinations, the light curves could be dominated by irregular obscuration of the photosphere 
from material lifted above the disk mid-plane in the accretion tongues. Therefore, we might 
assume that CTTSs that present periodic or quasi-periodic light curves (spot-like and AA Tau-like) 
are accreting in the stable regime, whereas those that present aperiodic light curves (dominated 
by flux bursts or extinction events) are accreting in the unstable regime. 

It is important to remember that a CTTS's light curve morphology is not a permanent 
characteristic of that system, but rather a reflection of the system's current state, and as 
that system progresses through different phases of accretion and of its evolution, its light 
curve morphology will change accordingly. \citet{mcginnis15} and \citet{sousa16} show that 
many of the CTTSs in NGC~2264 that were observed by CoRoT in both epochs suffered a transition 
from one light curve type in 2008 to a different one in 2011. These transitions were usually 
between spot-like and flux burst-dominated light curves or between AA Tau-like and aperiodic 
extinction-dominated light curves, and were attributed to a change in these systems' accretion 
regimes between the two observing runs. According to \citet{kurosawa13}, some of the most 
important factors to determine in which regime a CTTS will accrete are mass accretion rate and 
the strength and topology of the stellar magnetic field. Systems undergoing a stable accretion 
regime tend to have lower mass accretion rates and more ordered magnetic fields, with stronger 
bipolar components, than those undergoing an unstable accretion regime. Therefore if any of 
these factors change over time, the CTTS may transition from one regime to another. 

Figure \ref{fig:lctype1} shows (in black) the average [OI]$\lambda$6300 line profiles of the 
different groups of our CTTSs that present [OI]$\lambda$6300 emission, separated according to 
their light curve morphology in both CoRoT epochs and according to whether they maintained a 
constant photometric behavior or transitioned between different classifications. Each profile 
was normalized by its maximum intensity before the average was taken. The individual normalized 
profiles are also shown (in gray), in order to highlight the diversity of the [OI]$\lambda$6300 
line profiles in each group. 
We can see that stars that present AA Tau-like light curves in one or both CoRoT epochs show a 
narrow profile, with only a broad base or a modest blue wing, while stars that present aperiodic 
extinction in both CoRoT epochs show a much broader profile, with a more extended blue wing. In 
a similar fashion, stars with light curves dominated by flux bursts in both CoRoT epochs have a 
very broad profile and a clear blueshifted HVC with a distinct peak, whereas the stars with 
spot-like light curves in one epoch or another present a much weaker HVC, whether in the form 
of an extended blue wing or clearly separated from the LVC. This may be explained by the 
fact that the stars with very irregular, aperiodic photometric variability are often the 
ones with the highest mass accretion rates, especially those whose light curves are dominated 
by flux bursts \citep{sousa16}. 
These higher mass accretion rates would be connected to more intense jets and therefore 
stronger [OI]$\lambda$6300 emission line HVCs. 

\begin{figure}[t!]
    \centering
    \resizebox{\hsize}{!}{\includegraphics{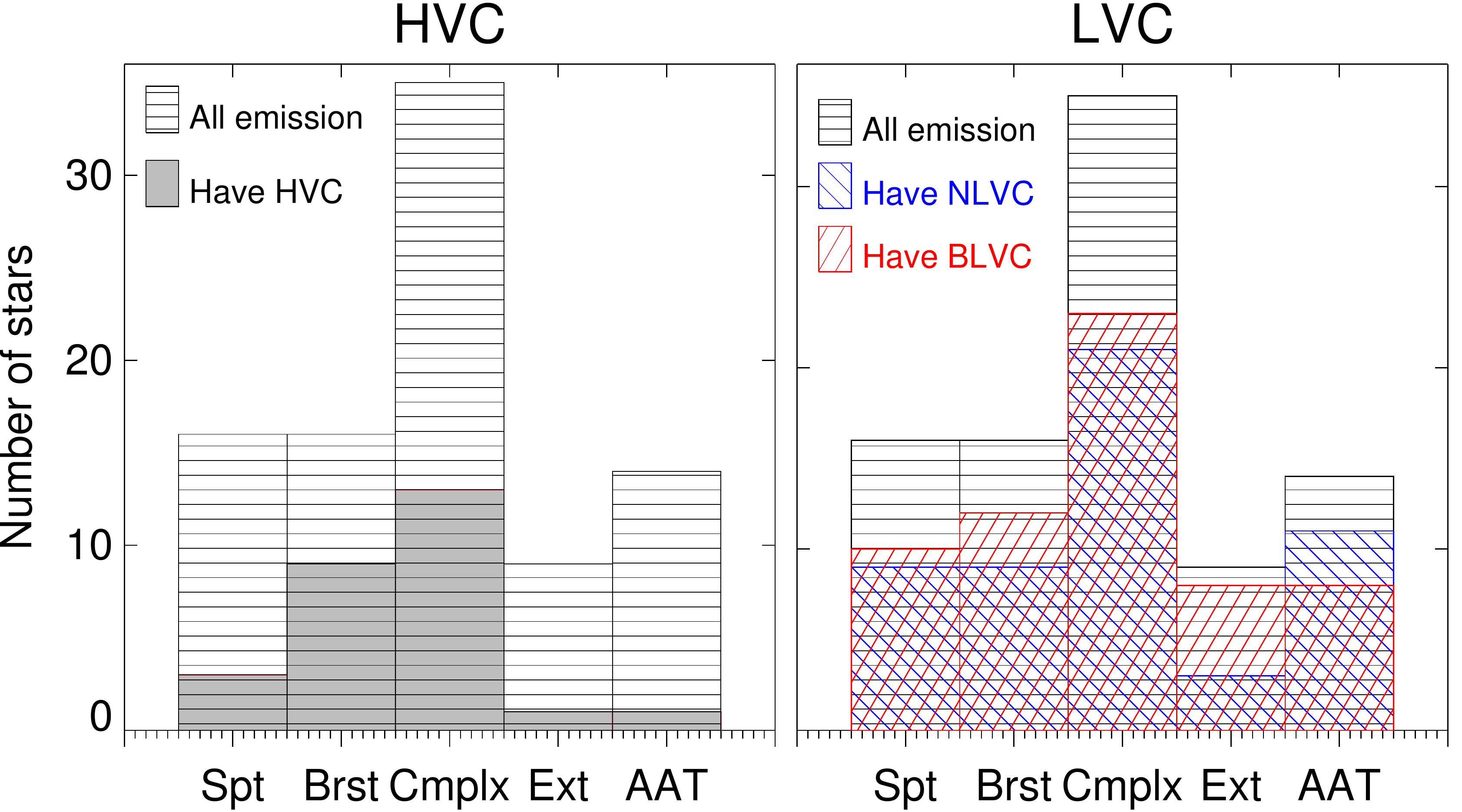}} 
    \caption{Histogram of light curve types for the systems where [OI]$\lambda$6300 
emission was detected (black striped bins), for those that present a HVC (left panel, shaded 
gray bins), for those that present a NLVC (right panel, blue), and for those that present a 
BLVC (right panel, red). 
The ordinate represents the number of systems that present each light curve type. From left 
to right, the bins represent systems with spot-like light curves in one or both CoRoT runs; 
systems with accretion burst-dominated light curves in both CoRoT runs; systems with complex 
light curves in both CoRoT runs; systems with aperiodic extinction-dominated light curves in 
both CoRoT runs; and systems that showed an AA Tau-like light curve in one or both CoRoT runs. 
}\label{fig:lctypehvc}
\end{figure}

Figure \ref{fig:lctype3} shows the distribution of centroid velocities and FWHM of all 
components in our sample, with different colors and symbols representing different types of 
photometric variability. 
This figure helps to illustrate how the HVC is much more often associated with systems that 
show irregular, nonperiodic variability (green and orange symbols), while systems that show 
periodic variability (purple symbols) usually present only low-velocity [OI]$\lambda$6300 
emission. This becomes even clearer in the left panel of Fig. \ref{fig:lctypehvc}, which shows 
a histogram of light curve types for all systems in which the [OI]$\lambda$6300 line was 
detected (striped black bins) and for those that present a HVC (shaded gray bins). We can see 
that most (81$\substack{+5\\-10}$\%) of the systems with a HVC are associated with either 
light curves dominated by flux bursts or complex variability (compared with 57$\pm$5\% for the 
full [OI]$\lambda$6300 sample). 
A few systems are dominated by spot-like light curves, which are generally viewed at low 
inclinations (close to face-on), favorable to detect a HVC. Only very few objects with a HVC 
present AA Tau-like light curves or photometric variability attributed to aperiodic extinction 
events, which can be explained by the high system inclinations necessary to observe these 
phenomena, making it difficult to deconvolve the HVC from the LVC. Besides this geometrical 
effect, Fig. \ref{fig:lctypehvc} shows us that the [OI]$\lambda$6300 line's HVC is indeed most 
often associated with irregular photometric behavior. Among the few systems 
with periodic (AA Tau-like or spot-like) light curves in which a HVC is found, this component 
is on average weaker than among the systems with complex or flux burst-dominated light curves. 
The periodic systems also present on average lower mass accretion rates and infrared excess 
than the other systems where a HVC is found. 
It seems that the occurrence of large-scale jets is not only more common among the 
sources accreting in the unstable regime, but is also stronger in these systems. 
The right panel of Fig. \ref{fig:lctypehvc} shows that the LVC does not present a clear trend 
with photometric variability.  

\section{Discussion}\label{sec:discuss} 

\subsection{Where does the low-velocity emission come from?}\label{sec:lvc} 

Emission originating in the disk is subject to broadening by Keplerian rotation, therefore if 
the narrow and broad components of the LVC come from disk winds, their widths can indicate at 
which disk radii they originate\footnote{Assuming that no other mechanism strongly affects the 
line broadening besides Keplerian rotation and the instrumental profile, which can easily be 
removed.}. 
Since it is the rotation velocity projected in our line-of-sight that determines the observed 
width of a line, then the FWHM of the NLVC and BLVC should show strong trends with the system 
inclination, as was found by S16 for their sample. 
If Keplerian rotation is the only major source of line broadening for both the NLVC and BLVC, 
then there should be a correlation between their FWHM among systems where both were detected, 
since the system inclination should affect both components in a similar manner (S16). 
We find this correlation in our sample, as S16 find in theirs, with a Kendall $\tau$ test 
giving only a 0.2\% probability of false correlation. 
Nevertheless, there is a significant spread in this relation, meaning that other factors besides 
the system inclination should be influencing the FWHM of one or both of these components. 

In order to further investigate the region in the disk where the [OI]$\lambda$6300 emission 
originates, we must derive system inclinations. We do not have direct measurements of the 
inclinations of our systems, but we can estimate them from stellar rotation properties, using 
the relation $v\sin i = (2 \pi R_*/P_{rot}) \sin i$ (where $v\sin i$ is the stellar rotation 
velocity projected in our line of sight, $R_*$ is the stellar radius and $P_{rot}$ the stellar 
rotation period). This method is subject to a number of uncertainties and the resulting 
inclinations are very inaccurate \citep[see, e.g.,][for a comparison between inclinations 
derived from rotation properties and those measured directly]{appenzeller13}. Nevertheless, 
though direct imaging of disks should give much more accurate system inclinations, the 
inclinations provided are of the outer disk (beyond 10~au). In-depth studies of T Tauri and 
Herbig Ae/Be disks are beginning to show that a misalignment between the inner and outer disk 
is not a rare phenomenon \citep[e.g.,][]{marino15,loomis17,min17}. Using stellar rotation 
properties ensures that we are estimating the inclination of the inner disk, where 
[OI]$\lambda$6300 emission is believed to arise, though the accuracy is far from ideal. 

To estimate inclinations, we used the periods obtained from the 2011 CoRoT light curves given 
in \citet{venuti17} or the 2008 CoRoT light curves given in \citet{alencar10}. Stellar radii 
were taken from \citet{venuti14}, who compared locations on the HR diagram to pre-main sequence 
model grids of \citet{siess00}. Values of $v\sin i$ were measured directly in our FLAMES 
spectra by comparing to synthetic spectra (described in Sect. \ref{sec:oicor})
of the closest effective temperature to the star's, $\log g=4$, and broadened to the 
instrumental resolution of $R=26\,500$. An extra continuum was added to simulate veiling and 
the synthetic spectrum was rotationally broadened, with both veiling and $v\sin i$ adjusted 
simultaneously to minimize $\chi ^2$. 

\begin{figure}[t!]
    \centering
    \resizebox{\hsize}{!}{\includegraphics{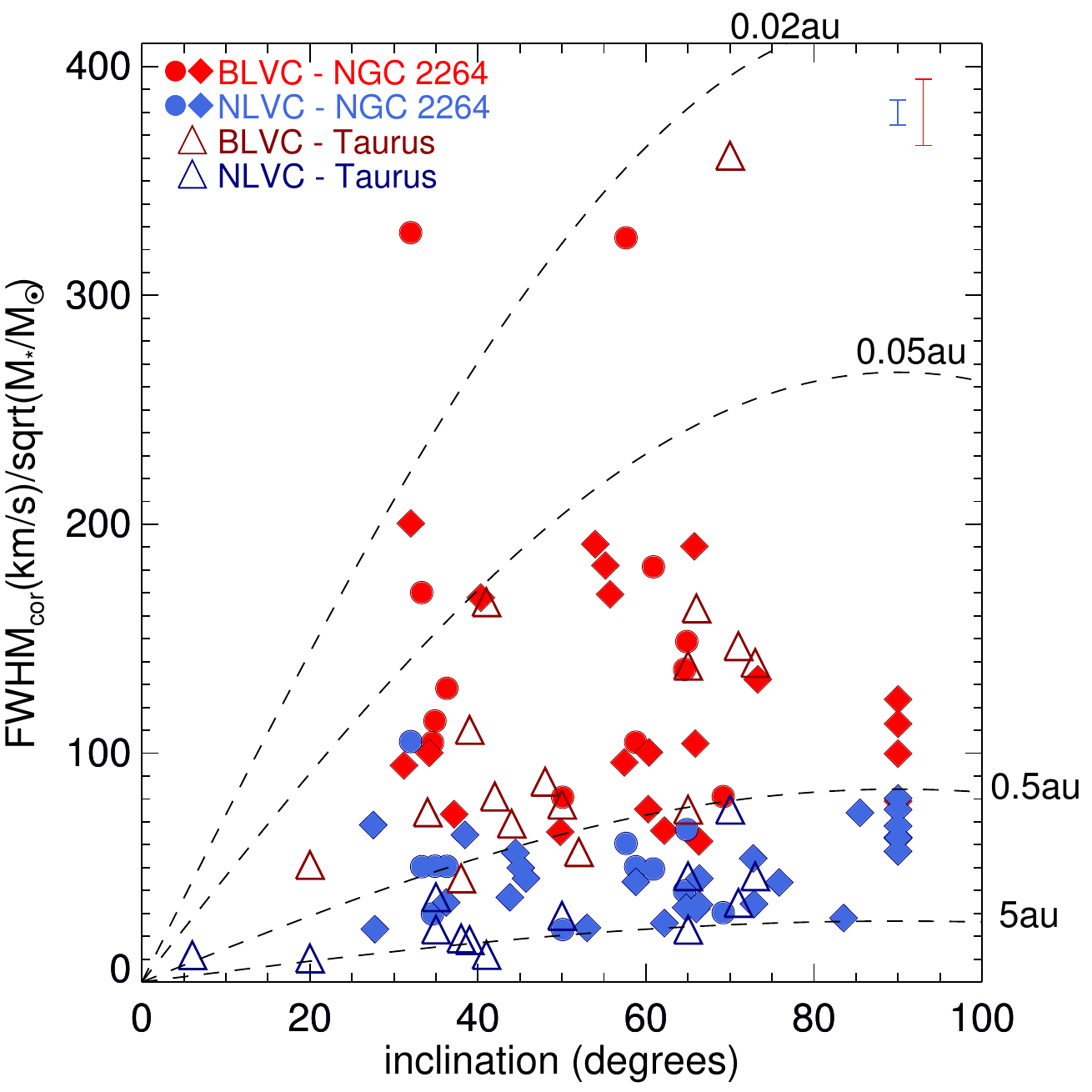}} 
    \caption{FWHM$_{cor}$$^{10}$ normalized by the square root of the stellar mass vs. inclination. Filled 
red symbols represent our broad LVCs, while filled blue symbols represent our narrow LVCs. Circles 
show cases where two LVCs were identified and diamonds show cases where only a BLVC or a NLVC
was identified. 
Open symbols represent the data from S16, dark red for the BLVC and dark blue for the NLVC. 
Black dashed lines show line width as a function of inclination for Keplerian rotation at four different 
disk radii: 0.02~au, 0.05~au, 0.5~au, and 5~au (assuming that $\mathrm{FWHM} = 2v_{Kep}(r)$). 
Typical error bars on the ordinate are shown in the top left corner in blue and red for our NLVCs and 
BLVCs, respectively. 
}\label{fig:gaus_inc_fwhm}
\end{figure}

Figure \ref{fig:gaus_inc_fwhm} shows the FWHM$_{cor}$/$\sqrt{\mathrm{M_*}}$\footnote{FWHM$_{cor}$ 
represents the FWHM corrected to remove the instrumental profile and $M_*$ is the stellar mass.} 
of the broad and narrow LVCs versus system inclinations, which describes the relation expected for 
Keplerian broadening and therefore indicates the region in the disk where each component originates 
(if the line width is dominated by Keplerian rotation, then the deprojected half width at half 
maximum corresponds to \mbox{$v_{Kep}(r_0) = \sqrt{GM_*/r_0}$}, where $r_0$ is the emitting radius, 
so \mbox{$\mathrm{FWHM}_{cor}/\sqrt{M_*} \propto (r_0)^{-1/2}*\sin i$}). 
There is a smooth transition between the two components in this plot, with some overlap between them 
(though it is possible that the overlap may be caused partly by inaccurate inclination 
estimates). The sample of S16 also shows a smooth transition. 
In Fig. \ref{fig:gaus_inc_fwhm} we plot the data from S16 as open symbols in order to compare our 
results with theirs. We see that our data lie in the same region as theirs, though 
we do not find linear correlations between the FWHM$_{cor}/\sqrt{\mathrm{M_*}}$ of either component 
and inclinations as S16 did, possibly due to our large uncertainties in inclination. 

Figure \ref{fig:gaus_inc_fwhm} suggests that most of the NLVCs in NGC~2264 originate between 
$\sim0.5$~au and $\sim5$~au, while most of the BLVCs seem to come from a region between $\sim0.05$~au 
and $\sim0.5$~au, in agreement with what S16 find in Taurus. For a small number of BLVCs, we infer 
launching radii $<$~0.05~au, however the large uncertainty in inclination makes at least two of them 
compatible with 0.05~au. For the remaining two systems, their truncation radii could be located at 
$\sim$~0.02~au if their stellar magnetic fields are on the order of 100~G, according to Equation 6 of 
\citet{bessolaz08}. This is somewhat low for CTTSs, but not implausible. In any case, by disregarding 
any broadening mechanisms besides instrumental and Doppler broadening due to Keplerian rotation, we 
may be somewhat underestimating these inferred launching radii. 

\subsection{A comparison between the NLVC and photoevaporative disk wind models}\label{sec:nlvc} 

\citet{ercolano10} propose that the LVC of the [OI]$\lambda$6300 emission line may be 
produced in an X-ray driven photoevaporative disk wind. R13 argue against this, based on 
a number of reasons, such as the fact that these models could not reproduce the observed 
widths of the [OI]$\lambda$6300 LVCs. 
However, this was before the LVC was found to be composed of two components, one narrower than 
the other. In the work of S16, the broad component is shown to be consistent with an origin in 
an MHD disk wind, while the NLVC is investigated in the context of a photoevaporative disk wind, 
though they do not find enough evidence to confirm this as its origin. In this and the following 
section, we discuss whether the data from NGC~2264 supports or conflicts with this scenario. 

It is interesting that in our sample a weak correlation is found between the luminosity of the 
[OI]$\lambda$6300 LVC and the X-ray luminosity, while both S16 and R13 find no such correlation. 
In Sect. \ref{sec:oilumcor} we have noted that the correlation with X-ray luminosity is strongest 
for the NLVC. Even if it is driven by an underlying correlation with the stellar luminosity, as 
was suggested in Sect. \ref{sec:oilumcor}, this 
alone would not exclude the possibility that a photoevaporative disk wind may be the origin of 
the narrow component of the [OI]$\lambda$6300 line's LVC. \citet{ercolano16} argue that 
a correlation between L$_X$ and L$_{[OI],LVC}$ is not necessarily expected, since the observed 
L$_X$ traces mostly hard X-rays, while it is the soft X-rays that would be the main responsible 
for heating the gas in the [OI]$\lambda$6300 emitting region, as they are strongly absorbed and 
thus not detectable by X-ray observations. 

The launching radii of the NLVC implied by Fig. \ref{fig:gaus_inc_fwhm} are compatible with the 
critical radius expected for photoevaporative winds \citep{alexander14}. This is confirmed by the 
updated photoevaporative disk wind model of \citet{ercolano16}. 
However, this model is still unable to reproduce the widths of the observed profiles. 
Using a very similar spectral resolution to ours, \citet{ercolano16} produce [OI]$\lambda$6300 
low-velocity profiles of \mbox{14.36~km~s$^{-1}$ $<$ FWHM $<$ 32.89~km~s$^{-1}$} for a star of 
0.7~M$_{\odot}$, much narrower than most of the values we observe 
(more than half of our sample of NLVCs have FWHM larger than their upper limit). 

\begin{figure}[t!]
    \centering
    \resizebox{\hsize}{!}{\includegraphics{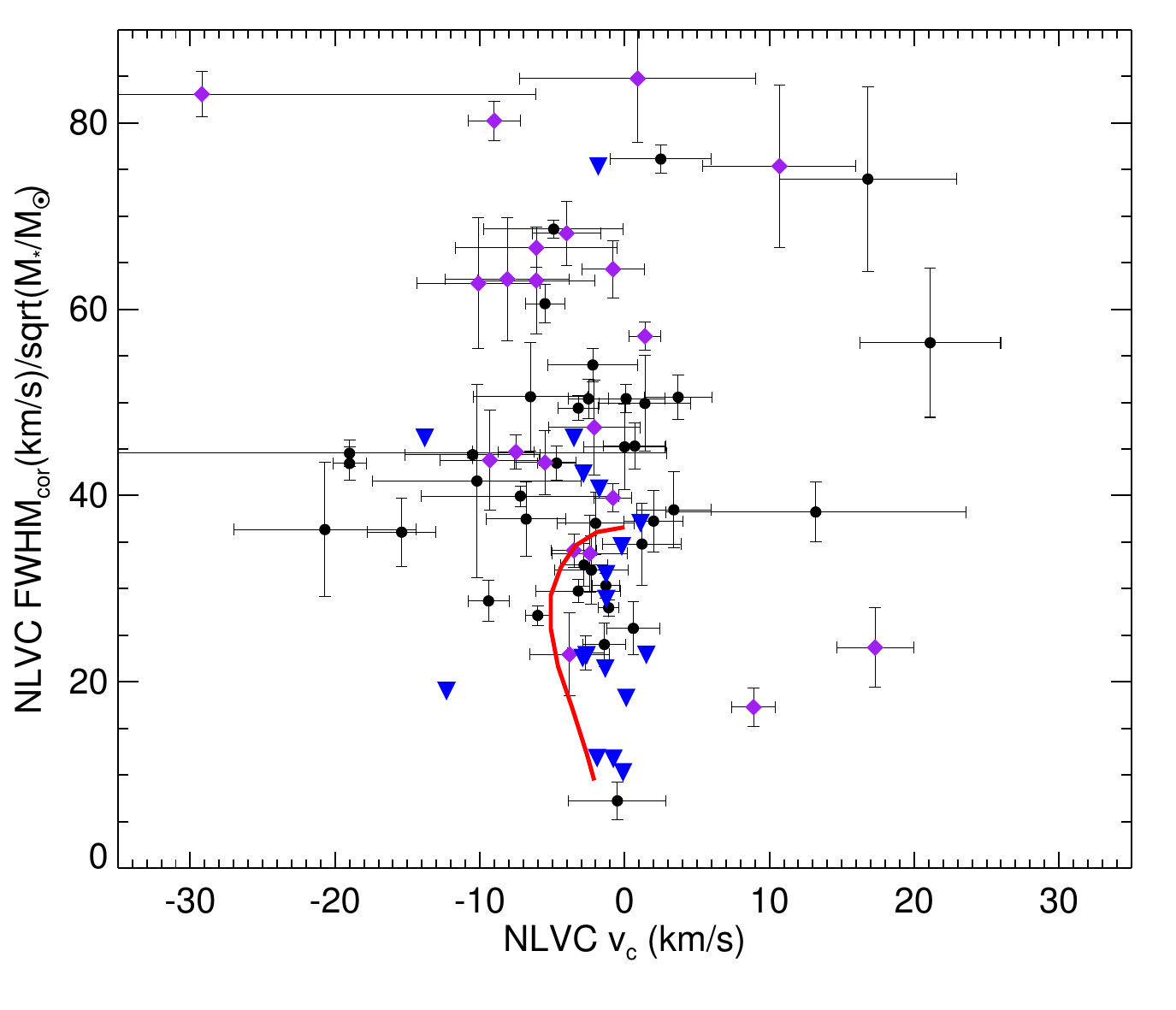}}  
    \caption{Distribution of FWHM$_{cor}/\sqrt{M_*}$$^{10}$ versus centroid velocities for the 
narrow LVCs in our sample (black dots) and in S16 (blue triangle), compared with theoretical expectations 
from \citet{ercolano16} for 
a star of M$_* = 0.7 M_{\odot}$, L$_X = 2 \times 10^{30}$ erg/s, and L$_{acc}=\mathrm{L}_{bol}$, 
varying the system inclination from $i=0^{\circ}$ to $i=90^{\circ}$ (thick red line). 
Purple diamonds represent systems observed at high inclinations, therefore there could be 
misclassification of the components in their [OI]$\lambda$6300 line profile due to projection effects. 
}\label{fig:nlvc_eo16}
\end{figure}

Figure \ref{fig:nlvc_eo16} shows the relation between the FWHM$_{cor}/\sqrt{M_*}$\footnotemark[10] 
and centroid velocities of our observed NLVCs, along with the expected relation according 
to \citet{ercolano16} for a photoevaporating disk wind model around a star of 
M$_* = 0.7 M_{\odot}$, L$_X = 2 \times 10^{30}$ erg/s, and L$_{acc}=\mathrm{L}_{bol}$, varying the 
system inclination from $i=0^{\circ}$ to $i=90^{\circ}$ (thick red line). 
As was noted in Sect. \ref{sec:oicomp}, we may be incomplete at FWHM~$<25$~km~s$^{-1}$, so we add the 
sample of S16 to this figure to better populate the lower region of this diagram (represented in 
Fig. \ref{fig:nlvc_eo16} as blue triangles). We highlight the systems that are likely observed at 
high inclinations with regard to our line-of-sight (based on $\sin i \gtrsim 1$ or extinction-dominated 
CoRoT light curves), since some of these low-velocity components may in fact be misclassified HVCs 
projected at low ($v_c<30$~km~s$^{-1}$) velocities (represented in Fig. \ref{fig:nlvc_eo16} as purple diamonds). 
We see that the model does not agree well with the bulk of the distribution of FWHM$_{cor}/\sqrt{M_*}$, 
even if we ignore the components that may be misclassified. 
As for the centroid velocities, in total 43/64 (67\%) of our values fall, within the 1$\sigma$ 
uncertainties, in the range of \mbox{-5.1~km~s$^{-1}$ $\leq v_c \leq 0$~km~s$^{-1}$} that is expected 
from this model, and 58/64 (91\%) fall in this range within 3$\sigma$. 
Hence, most of the centroid velocities of our NLVCs are in agreement with theoretical 
expectations for photoevaporative disk winds, 
however we cannot exclude that a more efficient acceleration mechanism may be required 
for at least some of these systems \citep[e.g., magneto-thermal disk winds,][]{bethune17}.  
In total, if we exclude the systems observed at high inclinations, we still find that nearly half 
of our sample of NLVCs (27/64) have a combination of FWHM and centroid velocity that cannot be 
reproduced by this model. 

\subsection{The BLVC as the base of high-velocity jets}\label{sec:blvc} 

The high blueshifts (up to nearly 50~km~s$^{-1}$) we find for the broad LVCs in our sample are 
compatible with disk winds of up to 50~km~s$^{-1}$ (assuming random system inclinations, the largest 
blueshifts likely arise from systems viewed closer to face-on, meaning that their centroid 
velocities may represent the maximum velocities of these disk winds). 
Figure \ref{fig:gaus_inc_fwhm} points to an origin in the innermost part of the disk, less 
than one au from the star, too close for a photoevaporative wind to arise. 
In this region of the disk the interaction with the stellar magnetosphere is strongest, meaning 
this could represent emission coming from a magnetized inner disk wind, as was proposed by S16. 
The strong correlation we find between its luminosity and the accretion luminosity (stronger 
than the one found for all other components), besides the positive correlation S16 find 
between the FWHM of the BLVC and the accretion luminosity, suggest that accretion plays an 
important role in launching this wind.  

The emitting region of the BLVC deduced from Fig. \ref{fig:gaus_inc_fwhm} is consistent with 
estimates of the launching region of protostellar jets (see, e.g., \citealt{tabone17}, who 
estimate from jet rotation a launching range of 0.05 to 0.3~au for the HH 212 SiO jet, and 
\citealt{anderson03,pesenti04}, who infer that the launching radius for DG Tau can extend up to 
$\sim$3au), leading us to speculate that the BLVC may be tracing the base of the same wind that 
is accelerating into a protostellar jet. The blueshifts of up to $\sim$50~km~s$^{-1}$ found for 
the BLVC would then represent the intermediate propagation velocity of a wind that is accelerating 
from 0~km~s$^{-1}$ to typical velocities of protostellar jets. We find that most 
(74$\substack{+6\\-9}$\%) of the systems where a HVC was detected also present a BLVC. If they 
are indeed connected, then we would expect to find a BLVC in all systems with a HVC. As discussed 
in Appendix \ref{sec:compl}, our sample is likely subject to a sensitivity issue and may be incomplete.
Therefore it is possible that these systems indeed present a BLVC that was unfortunately below 
our detection threshold. 

If the BLVC represents the base of the jet traced by the HVC, then we may expect there to be 
a correlation between parameters of these two components among the systems that present both. 
However, we do not necessarily expect there to be a direct correlation between the FWHM 
of the two, since the HVC is dominated by jet emission on much larger scales from the source 
($\gtrsim 30$~au), where collimation has been achieved. Its FWHM is therefore likely dominated 
by turbulent and/or thermal processes, leading to a much narrower profile. 

A relationship we could investigate is between the emitting region of the BLVC and the 
expected launching radius of the jet associated with the HVC. 
For a jet launched via a magneto-centrifugal wind from a Keplerian disk, \citet{blandford82} 
argue that the maximum velocity of the jet should scale with the Keplerian velocity of the 
launching radius ($r_0$) as $v_{jet}=v_{Kep}(r_0)*\sqrt{(2\lambda - 3)}$, where $\lambda$ 
is the magnetic lever arm and $v_{Kep}(r_0)=\sqrt{(GM_*/r_0)}$. 
If Keplerian rotation is the main responsible for the broadening of the BLVC, then half of 
its FWHM, deconvolved from the instrumental profile, should correspond to the Keplerian 
velocity at the emitting radius projected in our line-of-sight. Therefore, if the BLVC 
originates in a disk wind that is the base of the HVC-emitting jet, then its FWHM would be 
equivalent to $2 v_{Kep}(r_0) * \sin i$ (where $i$ is the system inclination), and this 
quantity should correlate with the HVC jet velocity.  

We search for a correlation between the centroid velocities of the HVC (which correspond to 
the average jet velocities projected in our line-of-sight) and the BLVCs' FWHM 
deconvolved from the instrumental profile, but do not find a convincing one. We must keep in 
mind, however, that the system inclination would strongly interfere with any correlation that 
may exist, since the deprojected velocities are perpendicular to each other. Therefore this 
does not exclude the possibility that the BLVC may be connected to the base of the large scale 
jets traced by the HVC. Ideally we would search for a correlation between the deprojected 
velocities, but unfortunately we only have reliable estimates of system inclination for seven 
systems in which both a HVC and a BLVC were detected, which subjects this result to very low 
number statistics, and the uncertainties in inclination are large enough to mask any correlation. 
If there is indeed a connection between the two components, then the relation between the 
deprojected velocities of these seven systems gives an average value of $\lambda \sim 4.5$. 
In order to confirm this scenario, however, it is necessary to analyze a statistically 
significant sample of systems that present both components and have accurate measurements 
of system inclinations. 

\subsection{Evolution of the LVC with the evolution of the disk}\label{sec:lvc3}

\begin{figure}[t!]
    \centering
    \resizebox{\hsize}{!}{\includegraphics{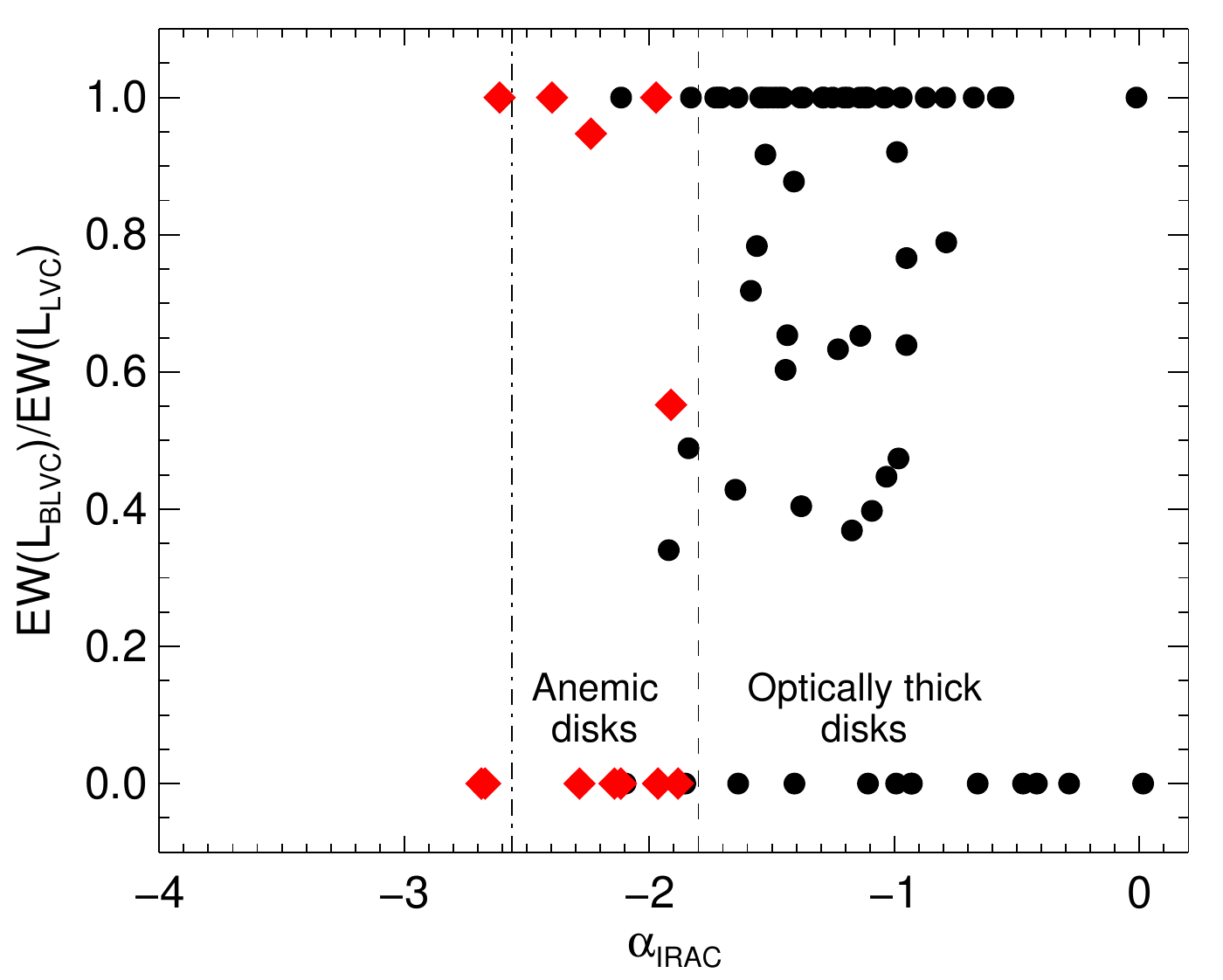}} 
    \caption{Ratio of the [OI]$\lambda$6300 BLVC equivalent width to the equivalent width of 
the total LVC versus the $\alpha_{IRAC}$ index. Transition disk sources are shown as red filled 
diamonds, while all other sources are shown as black filled circles.  
 }\label{fig:tdairac}
\end{figure}

In our study, we have seen that the BLVC is much more common than the NLVC among 
systems with thick disks and rarer than the NLVC among systems with a thin inner disk 
(left and middle panels of Fig. \ref{fig:oi_lum_airac}). 
The overall detection rate of the BLVC among accreting TTSs in our sample decreases from 
50$\pm$5\% in the optically thick range of the $\alpha_{IRAC}$ index, to 28$\substack{+10\\-7}$\% 
in the range considered to have anemic disks, to only 6$\substack{+11\\-2}$\% in the range that 
would be classified as a naked photosphere based only on $\alpha_{IRAC}$. On the other hand, the 
NLVC has a lower detection rate among systems with optically thick disks (29$\pm$4\%) than among 
those with anemic disks (41$\pm$9\%), and its detection rate decreases again to 
12$\substack{+12\\-4}$\% in the most optically thin range.  
This is in agreement with S16, who show that the NLVC is often present among systems with 
transition disks, whereas the BLVC rarely is. 
In our sample, the NLVC is also more common among transition disk systems than the 
BLVC. Among the 13 transition disk systems (Alana Sousa, private communication) with a detected 
[OI]$\lambda$6300 line, seven show only a NLVC, three show both a NLVC and a BLVC, 
and only three show a BLVC alone. 
This can be seen in Fig. \ref{fig:tdairac}, which shows the ratio of the broad component's 
equivalent width to the equivalent width of the total LVC versus $\alpha_{IRAC}$.  

These results support the scenario proposed by S16 and discussed by \citet{ercolano17} of an 
evolution of the [OI]$\lambda$6300 line profile with the evolution of the disk. In this scenario, 
the broad LVC disappears as the inner disk evolves and, as an inner hole is formed, the narrow 
LVC tends to dominate the [OI]$\lambda$6300 line profile. 
Since it has been shown that the BLVC originates closer to the central object than the NLVC (Fig. 
\ref{fig:gaus_inc_fwhm}), this points to an inside-out gas dissipation of the inner disk. 
Even among the transition disk sources that still present a BLVC, their inferred launching regions 
are all consistent with originating beyond 0.2~au from the central object. This is close to the 
upper limit for the BLVC and therefore consistent with an inside-out clearing of the inner disk.  

\citet{ercolano10} predict that, if the [OI]$\lambda$6300 emission comes from a photoevaporative 
disk wind, then this emission would be approximately twice as luminous among sources with 
transition disks 
than among sources with primordial disks. If the NLVC traces such a photoevaporative wind and the 
BLVC traces an inner MHD disk wind, then we could expect to detect much stronger NLVCs than BLVCs 
as the systems evolve. 
However, in our sample alone, we do not see a tendency of the NLVC luminosity with the 
clearing of the inner dust disk (left panel of Fig. \ref{fig:oi_lum_airac}). 

\subsection{The high-velocity component: testing launching models}\label{sec:hvc}

In order to investigate which properties of the star-disk system may be connected to the mechanism 
that drives protostellar jets, we search for trends between these properties and the 
[OI]$\lambda$6300 HVC.  
However, the expected correlations for models of jet launching via stellar or magnetospheric 
winds often involve not one stellar parameter, but a complex combination of them. To properly 
explore this issue, it is necessary to delve into the theoretical background of the possible jet 
launching mechanisms, which is not the scope of this paper. 
We therefore discuss here only the general trends found in our sample, and reserve a more 
thorough analysis of the HVC origin from a statistical viewpoint to a forthcoming paper, in which 
we will also include results from other star forming regions in order to increase the size of our 
sample. 

\begin{figure*}[t!]
    \centering
    \includegraphics[width=8.5cm]{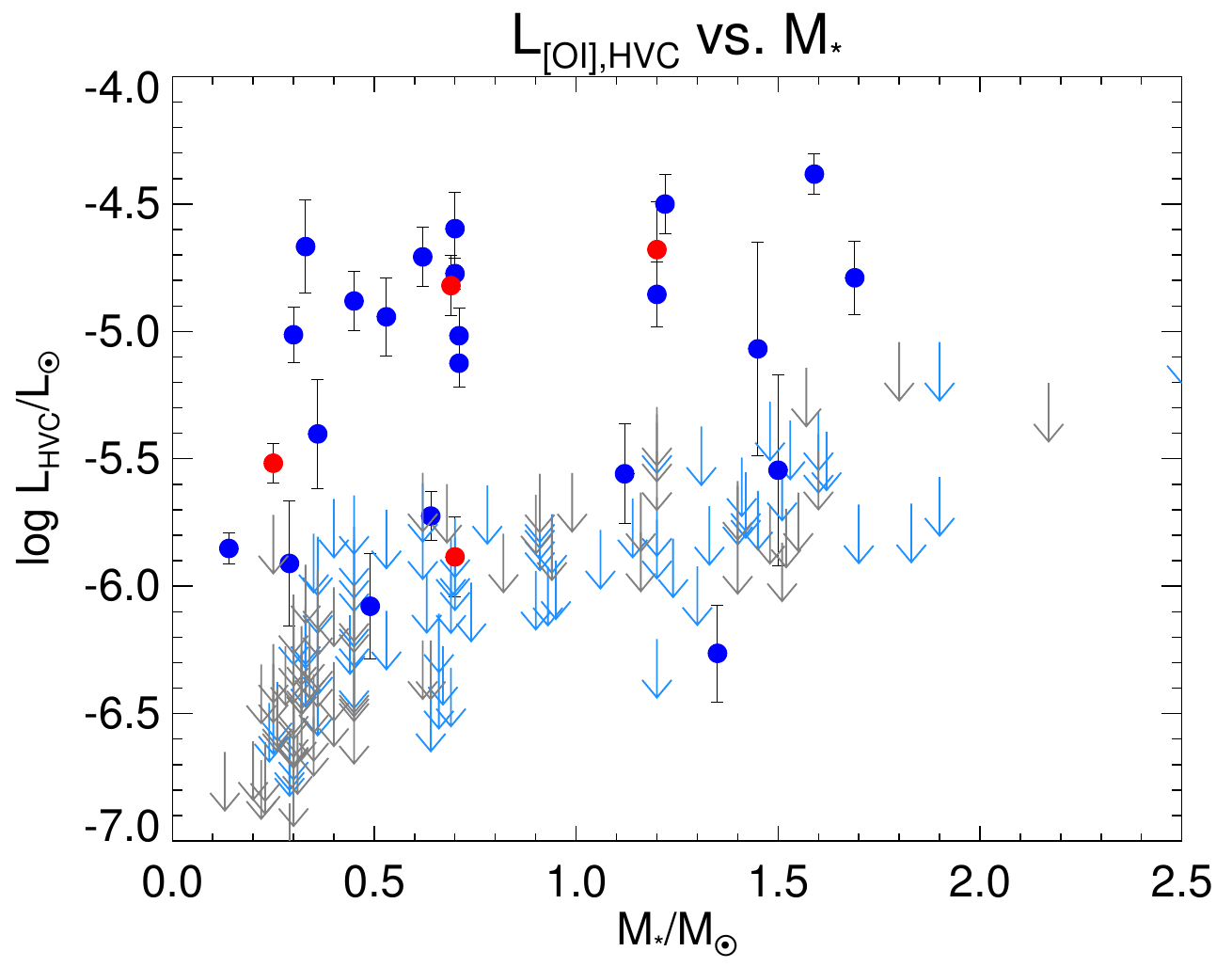} 
    \includegraphics[width=8.5cm]{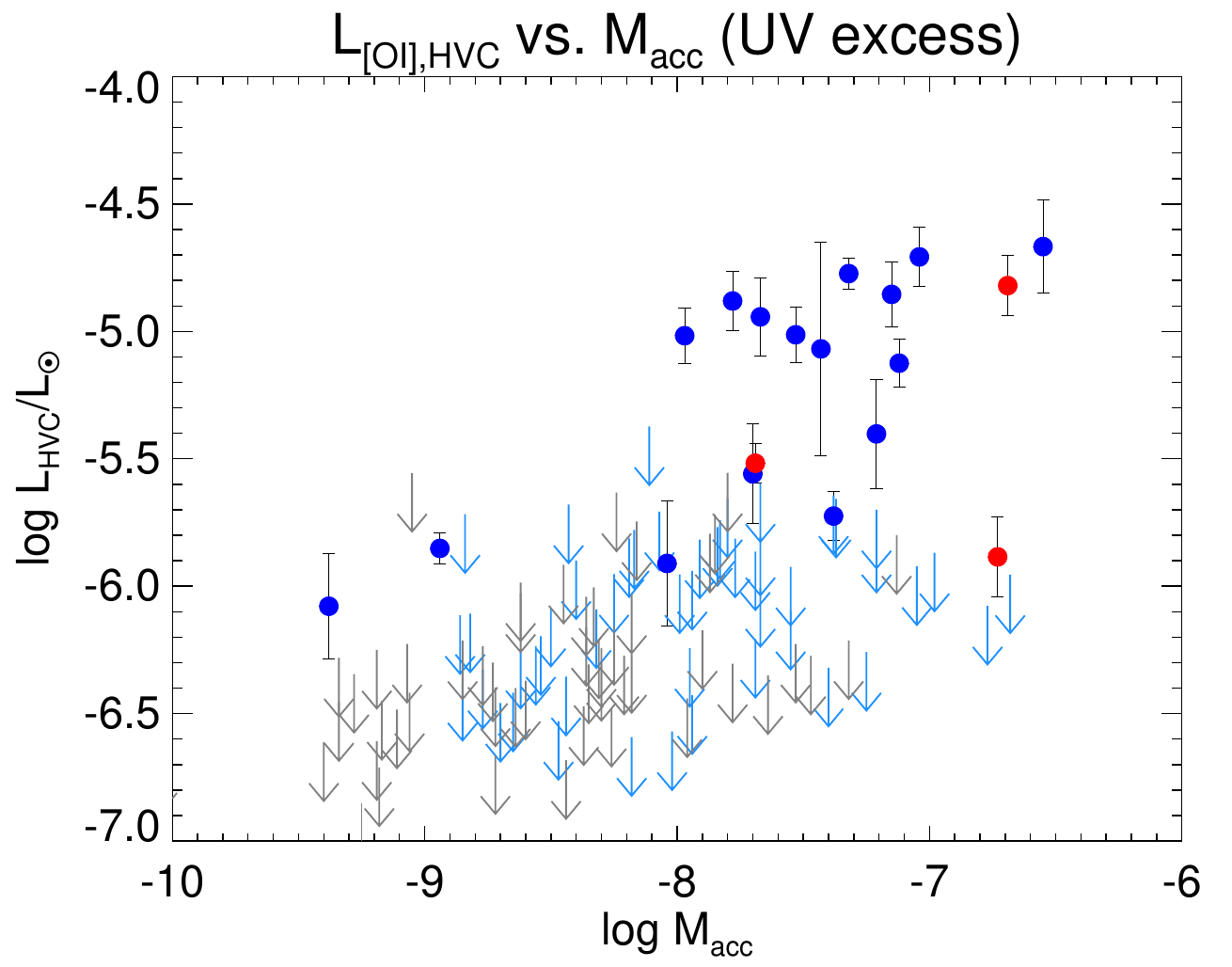} 
    \caption{[OI]$\lambda$6300 HVC luminosity versus stellar mass (left) and mass accretion rate 
derived from the UV excess (right). 
Gray arrows represent upper limits for [OI]$\lambda$6300 line luminosities when no emission in 
[OI]$\lambda$6300 was detected, while light blue arrows represent upper limits for HVC luminosities 
when only low-velocity [OI]$\lambda$6300 emission was detected. Blue filled circles represent 
blueshifted HVCs, while red filled circles represent redshifted HVCs (systems that did not present 
a blueshifted component). Luminosities of redshifted HVCs should be considered as lower limits, 
since the redshifted part of the jet suffers additional extinction from the circumstellar disk. 
}\label{fig:hvc_macc}
\end{figure*}

We first analyzed the relationship between the luminosity of the HVC of the [OI]$\lambda$6300 
line and stellar and accretion properties. Figure \ref{fig:hvc_macc} shows the HVC luminosity 
versus stellar mass, on the left, and versus mass accretion rates, on the right. 
We plot in this Fig. only the mass accretion rates derived from the UV excess and not the 
values derived from the H$\alpha$ equivalent widths, in order to compare a more self-consistent 
sample\footnote{Determining mass accretion rates using these two different methods leads to slightly 
different values \citep[see e.g.,][]{sousa16}. UV excess is a more direct measurement of accretion, 
since the H$\alpha$ line can have contributions from a wind and from accretion shocks, and does not 
necessarily originate only in accretion columns as is normally assumed when estimating 
$\dot{M}_{acc}$, whereas the UV excess originates only in the accretion shocks.}.   
We find no direct correlation between the HVC luminosity and stellar mass. The left panel of Fig. 
\ref{fig:hvc_macc} shows an apparently bimodal behavior: some of the stars have [OI]$\lambda$6300 
HVC luminosities that fall in the same range as many of the upper limits of nondetections, while 
another group shows higher luminosities that are nearly constant among the entire mass range, with 
$\log \mathrm{L}_{HVC}/\mathrm{L}_{\odot} \sim -4.7$. The right panel, on the other hand, shows 
that there is a relation between the [OI]$\lambda$6300 HVC and mass accretion, as has already been 
established here through the correlation between HVC luminosity and accretion luminosity in Sect. 
\ref{sec:oilumcor} and in other papers \citep[][HEG95]{cabrit90}. However, there is not a clear 
threshold between the two groups identified in the left panel. 
It is possible that these groups represent two slightly different populations that coexist in this 
star forming region, as other studies have found evidence of multiple substructures and 
subpopulations in NGC~2264 \citep{venuti17pop}.
However, we find no significant difference between the ages, radial velocities, or spatial 
distribution of the two groups we observe. We are, nonetheless, subject to low number statistics, 
therefore we cannot necessarily rule out this possibility.  

In order to relate jet velocities with stellar and accretion parameters, we calculated the 
deprojected jet velocities from the HVC centroid velocities and from estimates of inclinations 
derived from stellar rotation properties in Sect. \ref{sec:lvc}. 
We assume that the centroid velocity of the [OI]$\lambda$6300 emission line's HVC represents 
the average propagation velocity of the jet responsible for the emission, projected in our 
line-of-sight. 
Since the jet is perpendicular to the accretion disk, then the deprojected jet velocity ($v_{jet}$) 
is the HVC centroid velocity divided by the cosine of the system inclination ($v_{c,HVC}/\cos i$). 
We recall that deriving system inclinations from stellar rotation properties is subject to large 
uncertainties, therefore our values of jet velocity are also very imprecise. 
We find deprojected jet velocities mostly in the range $100 \mathrm{km~s^{-1}} < v_{jet} < 200$~km~s$^{-1}$. 
Within the uncertainties they all agree with the average value of $v_{jet} = 140(\pm76)$~km~s$^{-1}$. 
It is therefore unclear if the large spread we observe in $v_{jet}$ is due to an intrinsic spread 
in jet velocities across our sample or to the large uncertainties in deriving inclinations. 
We can also estimate the average jet velocity independently of inclination estimates by taking 
the median and standard deviation HVC peak velocity of $v_c=97.4(\pm38.8)$~km~s$^{-1}$ (found when 
considering both red and blueshifted HVCs), and deprojecting this value considering a median 
system inclination of 60$^{\circ}$. This gives us $v_{jet} = 195(\pm78)$~km~s$^{-1}$ for our sample, 
slightly larger than our first estimate and in agreement with the average jet velocity of 
$v_{jet} = 196(\pm16)$~km~s$^{-1}$ found in Taurus by \citet{appenzeller13} from resolved [NII] lines. 

We searched for correlations between the deprojected jet velocities and stellar properties such 
as effective temperature, mass and rotation rate, and found none. 
It is important to note, nevertheless, that our very large error bars may be masking possible trends. 
\citet{appenzeller13}, who used direct measurements of inclination for a number of stars in Taurus, 
also searched for correlations between jet velocities and various stellar properties (escape velocities, 
photospheric radii, rotation periods, magnetic field data, spectral veiling, mass accretion rates and 
mass loss rates) and found no convincing correlations, except with the mass loss rate. 

\begin{figure}[t!]
    \centering
    \resizebox{\hsize}{!}{\includegraphics{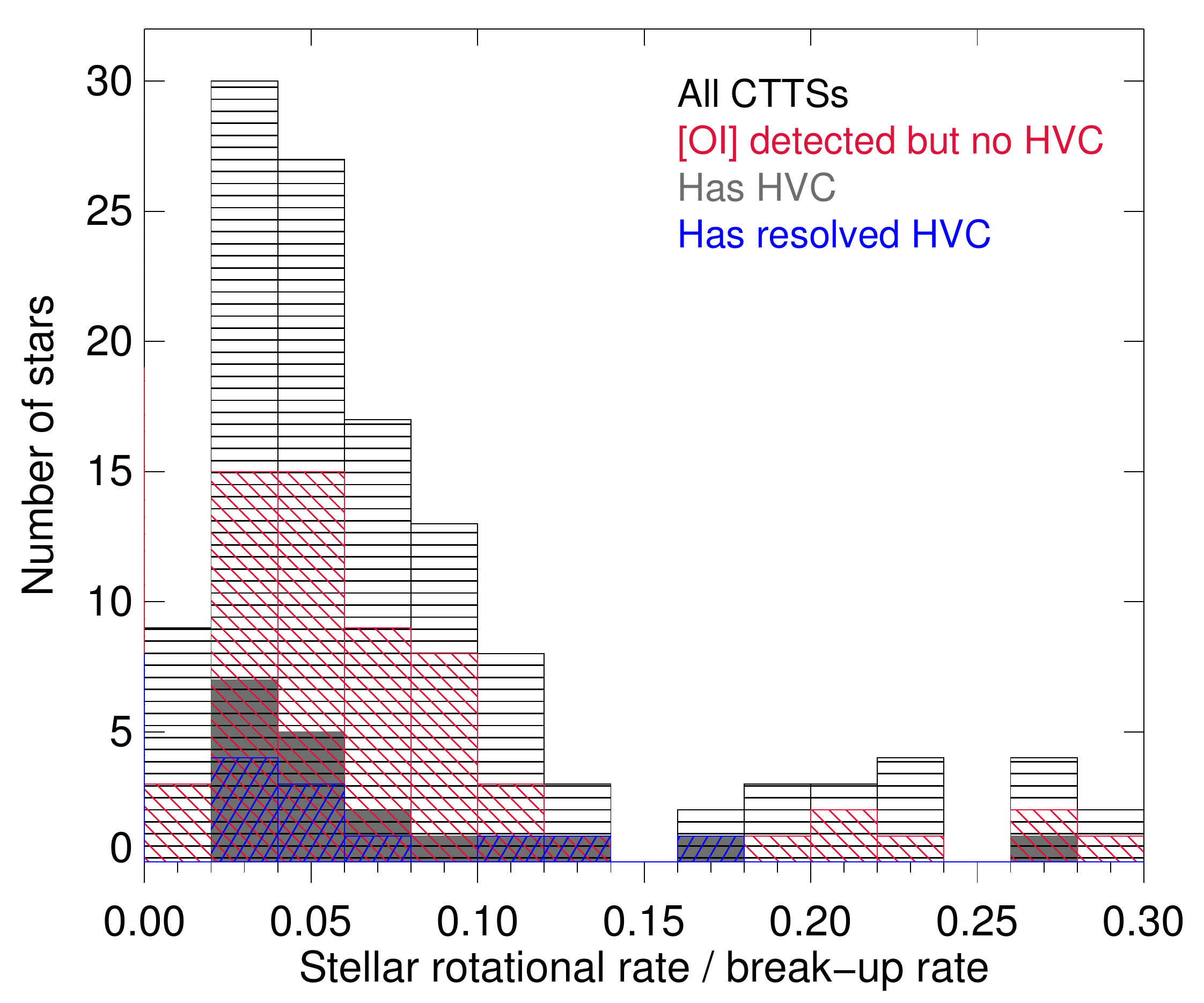}} 
    \caption{Distribution of rotation rates as a fraction of break-up rate for all CTTS (black), 
stars with [OI]$\lambda$6300 emission but no HVC (red), stars with a HVC (gray) and stars with 
a spectrally resolved HVC (blue). 
}\label{fig:rotdist}
\end{figure}

The occurrence of a HVC seems to be independent of the stellar rotation rate. Figure 
\ref{fig:rotdist} shows a histogram of the rotation rates (normalized by break-up rates) of 
all of the CTTSs in our sample, of those that have detected [OI]$\lambda$6300 emission but show 
no HVC, and of those for which an [OI]$\lambda$6300 HVC was detected. We can see in this figure that 
the HVCs are mostly associated with slow rotators ($0.02 \lesssim v_{rot}/v_{brk} \lesssim 0.06$) 
but a KS test shows that their distribution is not statistically different than the distribution 
for sources without a HVC.  
This suggests that stellar rotation alone may not play a key role in launching jets and that an 
extra energy source, namely the tapping of accretion power, is required. 

One last aspect we wish to investigate is the detection of jets among sources of different internal 
structures, and therefore likely possessing different stellar magnetic field topologies, as a star's 
magnetic field configuration is believed to change considerably as its internal structure changes. 
It seems that the stellar magnetic field increases in complexity and the dipole component decreases 
in strength as a radiative core develops, as argued by \citet{gregory12} for accreting T Tauri stars 
and Fig. 12 of \citet{hill17} for young stars. While fully convective stars have substantial dipolar 
components of 0.3 to a few KG, no strong dipolar component above a few tenths of a kG has yet been 
detected for M$_{core}/$M$_* \gtrsim 0.4$ or $T_{eff} \gtrsim 4\,500$K, however it is unclear if this 
is due to a true evolution or to the current low number statistics. The limit of 
M$_{core}/$M$_* \approx 0.4$ is empirical and more observations are required to confirm, and 
properly determine, the exact boundary. 

\begin{figure*}[t!]
    \sidecaption 
    \includegraphics[width=12cm]{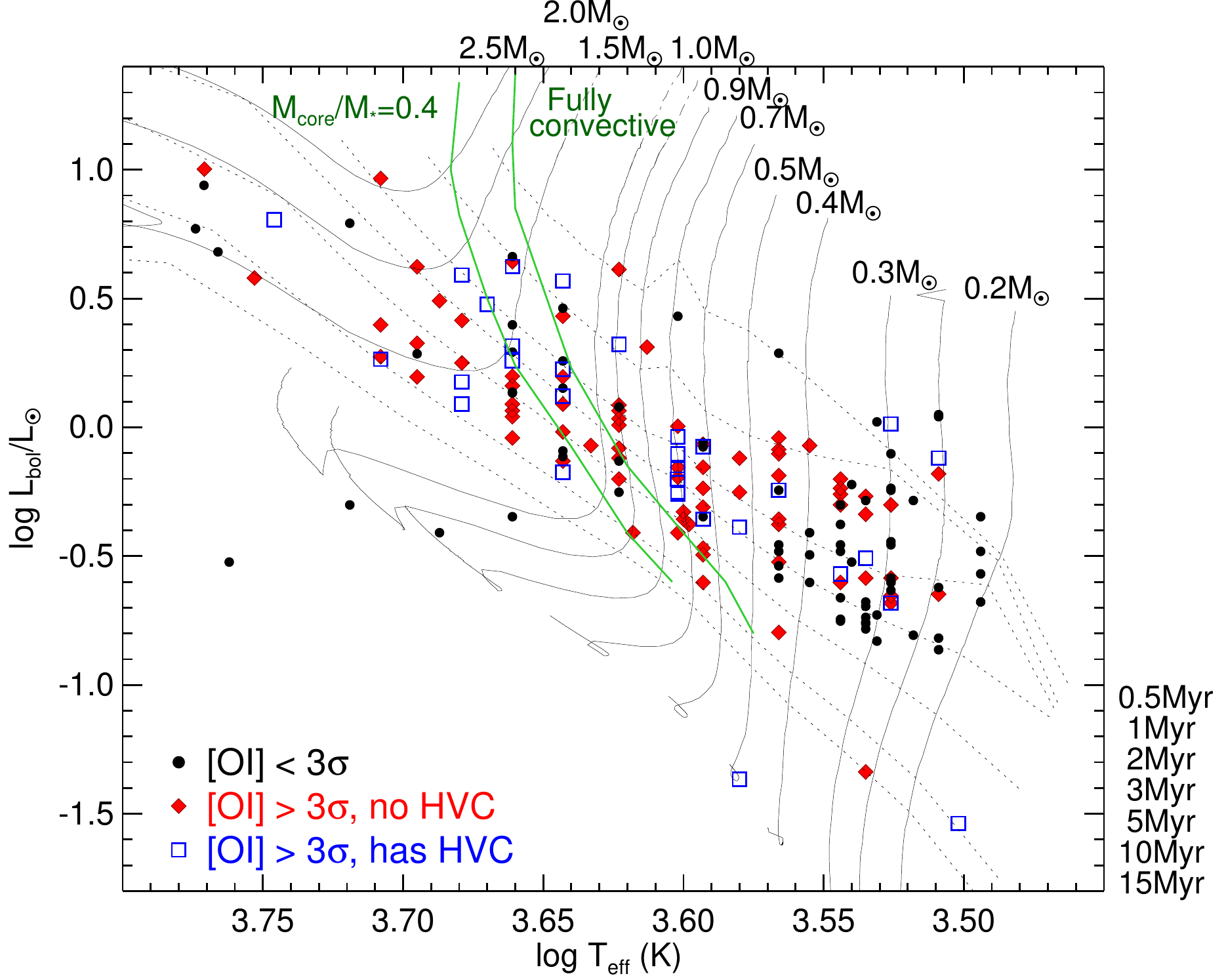} 
    \caption{HR diagram of the CTTSs in our sample where no emission was detected (black dots), 
those with detected [OI]$\lambda$6300 emission but no HVC (red diamonds), and those with a HVC 
(blue squares). Solid and dotted black lines represent mass tracks and isochrones, respectively, 
of \citet{siess00}. Solid green lines separate the regions in which a star is expected to be 
fully convective (right) and possess a radiative core of \mbox{M$_{core} > 0.4$ M$_*$} (left), 
according to \citet{gregory12}. }
    \label{fig:hrd-hvc}
\end{figure*}

Figure \ref{fig:hrd-hvc} shows that our sample contains many sources whose position on the HR 
diagram indicates that they should be fully convective (right of the green lines in Fig. 
\ref{fig:hrd-hvc}), and also a number of sources in the region where a significant radiative core 
is expected to have developed \citep[the left of the green lines in Fig. \ref{fig:hrd-hvc} 
corresponds to the region where M$_{core}/$M$_* \gtrsim 0.4$, according to][]{gregory12}.  
We do not see a difference in the fraction of stars with detected [OI]$\lambda$6300 HVC to stars 
with detected [OI]$\lambda$6300 emission (N(HVC)/N([OI])) between the fully convective range 
($N(HVC)/N([OI]) = 29\substack{+6\\-5}\%$) and the range where a significant ($\gtrsim 0.4\mathrm{M}_*$) 
radiative core is expected to have developed ($N(HVC)/N([OI]) = 30\substack{+10\\-7}\%$). 
If there truly is an evolution of the stellar magnetic field topology in this region of the HR 
diagram, then this result shows that the topology of the stellar magnetic field does not strongly 
affect the occurrence of protostellar jets and that kG-order dipoles are not necessary to launch them. 

\citet{villebrun16} show that stellar magnetic fields of strength $\gtrsim 200$G are commonly 
detected in T Tauri stars and Herbig Ae/Be stars of effective temperatures up to around 6\,000K, 
beyond which very few magnetic stars are found. 
Our sample does not probe well the region beyond this threshold, 
so we cannot fully exclude the influence of stellar magnetic fields in launching jets. 
We can however conclude from Figure \ref{fig:hrd-hvc} that, if stellar magnetic fields are 
necessary to power protostellar jets, then fields of a few 100G are sufficient to do so. 

\section{Summary and conclusions}\label{sec:conc}

We have searched for [OI]$\lambda$6300 emission in the spectra of 182 CTTSs and two HAeBe 
stars in the young, open cluster and well-known star forming region NGC~2264. We calculated the 
luminosity of these emission line profiles in order to compare them with previously determined 
stellar and accretion properties. We also compared the [OI]$\lambda$6300 line profiles of 
different groups separated according to their CoRoT light curve morphologies, in order to 
analyze the relation between photometric variability and [OI]$\lambda$6300 emission. In this 
section, we emphasize our main results. 

We detected [OI]$\lambda$6300 emission in 108 CTTSs and two HAeBe stars. Our detection rate 
among accreting T Tauri stars is low (59\%) compared to previous studies of this forbidden 
emission line and decreases toward lower masses, reaching only 40\% for 
$\mathrm{M}_* \leq 0.5 \mathrm{M}_{\odot}$ (for $\mathrm{M}_* > 0.5 \mathrm{M}_{\odot}$ this 
rate is 75\%). We find no statistical difference between the ages or radial velocities of the 
systems in which we detect [OI]$\lambda$6300 and those in which we do not. 
Since other studies have detected [OI]$\lambda$6300 emission in stars of similar L$_*$ and 
$\dot{\mathrm{M}}_{acc}$ as ours and find [OI]$\lambda$6300 line luminosities in the same 
range as the upper limits of our nondetections, we are led to believe that our low detection 
rate is likely due to a sensitivity issue. 

A Gaussian decomposition of our [OI]$\lambda$6300 line profiles led us 
to identify up to four different components: 
a narrow low-velocity component (NLVC), a broad low-velocity component (BLVC), a blueshifted 
high-velocity component (HVC), and a redshifted HVC. The NLVC was detected in 64 stars (among 
184 CTTSs and HAeBe stars analyzed), with an average FWHM of 41.6($\pm$10.6)~km~s$^{-1}$ and centroid 
velocity ($v_c$) of -3.5($\pm$9.1)~km~s$^{-1}$. The BLVC was detected in 75 stars, with an average 
FWHM of 112.5($\pm$41.7)~km~s$^{-1}$ and average $v_c$ of -12.4($\pm$16.5)~km~s$^{-1}$. The HVC was detected 
in 31 stars, 20 of which showed only a blueshifted HVC, 4 only a redshifted HVC, and 7 showed 
both a red- and blueshifted HVC. The average HVC has a FWHM of 61.7($\pm$31.2)~km~s$^{-1}$ and $|v_c|$ 
of 94.3($\pm$38.8)~km~s$^{-1}$. Both LVCs appear to be more blueshifted when a HVC is present. 

The luminosities of both LVCs and of the HVC all correlate with the stellar luminosity 
L$_*$ and with the accretion luminosity L$_{acc}$. The luminosity of the HVC does not 
correlate with the stellar X-ray luminosity L$_X$, but the luminosities of the NLVC and the 
BLVC weakly do. The luminosities of both the NLVC and BLVC also correlate positively with the 
UV flux. However, these correlations with X-ray luminosity and UV flux disappear after 
normalizing by the stellar luminosity L$_*$, and thus could be driven by an underlying 
correlation with L$_*$. 

We find that systems that present aperiodic photometric variability, especially those dominated 
by flux bursts, tend to present stronger and more frequent HVCs than systems that show periodic 
(spot-like or AA Tau-like) behavior, whose [OI]$\lambda$6300 line profiles are usually dominated 
by the LVC. These aperiodic light curves have been associated with stars accreting in an unstable 
regime, which generally present higher mass accretion rates and more complex stellar magnetic 
fields than those accreting in a stable regime.  
We also find that systems whose light curves are dominated by stable spots or flux bursts show 
much more pronounced HVCs in their [OI]$\lambda$6300 line profile, often with a resolved peak, 
than the ones with light curves dominated by AA Tau-like or aperiodic flux dips. This is 
consistent with a scenario in which the latter are viewed at much higher inclinations than the 
former, as was proposed when interpreting these light curve classifications 
\citep[e.g.,][]{mcginnis15}. 

Assuming that the LVC broadening is dominated by Keplerian rotation, we infer launching radii 
between 0.05~au and 0.5~au for the BLVC and between 0.5~au and 5~au for the NLVC. 
This is the same as what S16 find for their sources in Taurus.  
The centroid velocities of the NLVCs in our sample are in reasonable agreement with current 
photoevaporative disk winds models \citep{ercolano16}, however the FWHM of at least 42\% (27/64) 
of the NLVCs in our sample cannot be reproduced by these models. 
Meanwhile, the larger blueshifts and FWHM of our BLVCs are much more consistent with inner 
disk winds. These winds may represent the base of the same winds that accelerate and 
collimate on larger scales, to form the HVC jets, but no concluding evidence was found to 
support this hypothesis. It is necessary to study a large sample of systems that present both 
the BLVC and HVC and have accurate system inclination measurements, in order to confirm this 
scenario.  

We detected a HVC in the [OI]$\lambda$6300 line only among objects with significant 
infrared excess ($\mathrm{K} - 3.6\mu m > 0.3$), as was noted also by HEG95. This shows that it 
is necessary for the system to have an optically thick inner disk in order to be able to power 
strong bipolar jets, from which this component is emitted. The NLVC, on the other hand, was 
detected in many sources with optically thin inner disks ($\mathrm{K} - 3.6\mu m$ down to 0 and 
$\alpha_{IRAC}$ indices that would classify the systems as naked photospheres), while the BLVC 
is much more common among thick disk systems than among those with anemic disks, 
indicating that as the disk evolves, the BLVC disappears and the NLVC tends to dominate the 
[OI]$\lambda$6300 emission. The evolution of the [OI]$\lambda$6300 line profile with disk 
evolution is also demonstrated by the higher detection of NLVCs than BLVCs among systems with 
transition disks (also observed by S16 in Taurus). 
Since the BLVC has been shown to originate closer to the star than the NLVC, these 
findings support the scenario of inside-out gas dissipation in the inner disk. 

The maximum velocities of the jets in our sample do not correlate with any stellar or 
accretion properties. We are, however, subject to large uncertainties because of the lack of 
directly measured system inclinations and to low number statistics, so we cannot exclude 
the possibility that a correlation may exist, masked within the uncertainties.  
The average deprojected jet velocity in our sample is $v_{jet} = 140(\pm76)$~km~s$^{-1}$, 
which agrees with the average jet velocities measured in the younger star forming region 
Taurus \citep[$196 \pm 16$~km~s$^{-1}$, according to][]{appenzeller13}. 

The HVC was detected across the full T Tauri range of the HR diagram, with no apparent 
difference in the detection rates among stars that should be fully convective and those that 
have probably already developed a significant radiative core \mbox{(M$_{core} > 0.4$ M$_*$)}, 
suggesting that the topology of the stellar magnetic field may not play a significant role in 
the jet launching process. It is important to confirm this with proper mapping of stellar 
magnetic fields and to extend this study to larger stellar masses, where a significant drop 
of stellar magnetic field detection has been shown to occur. 

\begin{acknowledgements}
The authors acknowledge the anonymous referee for their contribution to the quality of 
this paper. 
We thank E. Flaccomio for providing us with X-ray luminosities and A.P. Sousa for 
information on transition disks. 
P.M. would like to thank M. Guimar\~aes for helpful discussions and the physics 
department at the Universidade Federal de Sergipe, Brazil, for support during visits. 
This work is based on data products from observations made with ESO Telescopes 
under program ID 094.C-0467(A).
This work was partly supported by CAPES, within the CAPES/Cofecub program, and by 
FAPEMIG under grant number 23893. 
P.M. and S.H.P.A also acknowledge financial support from CNPq. 
\end{acknowledgements}

\bibliographystyle{aa}   % abnt-alf or abnt-num unsrt
\bibliography{references}

\begin{appendix}

\section{Completion}\label{sec:compl}

\begin{figure*}[t]
    \sidecaption 
    \includegraphics[width=12cm]{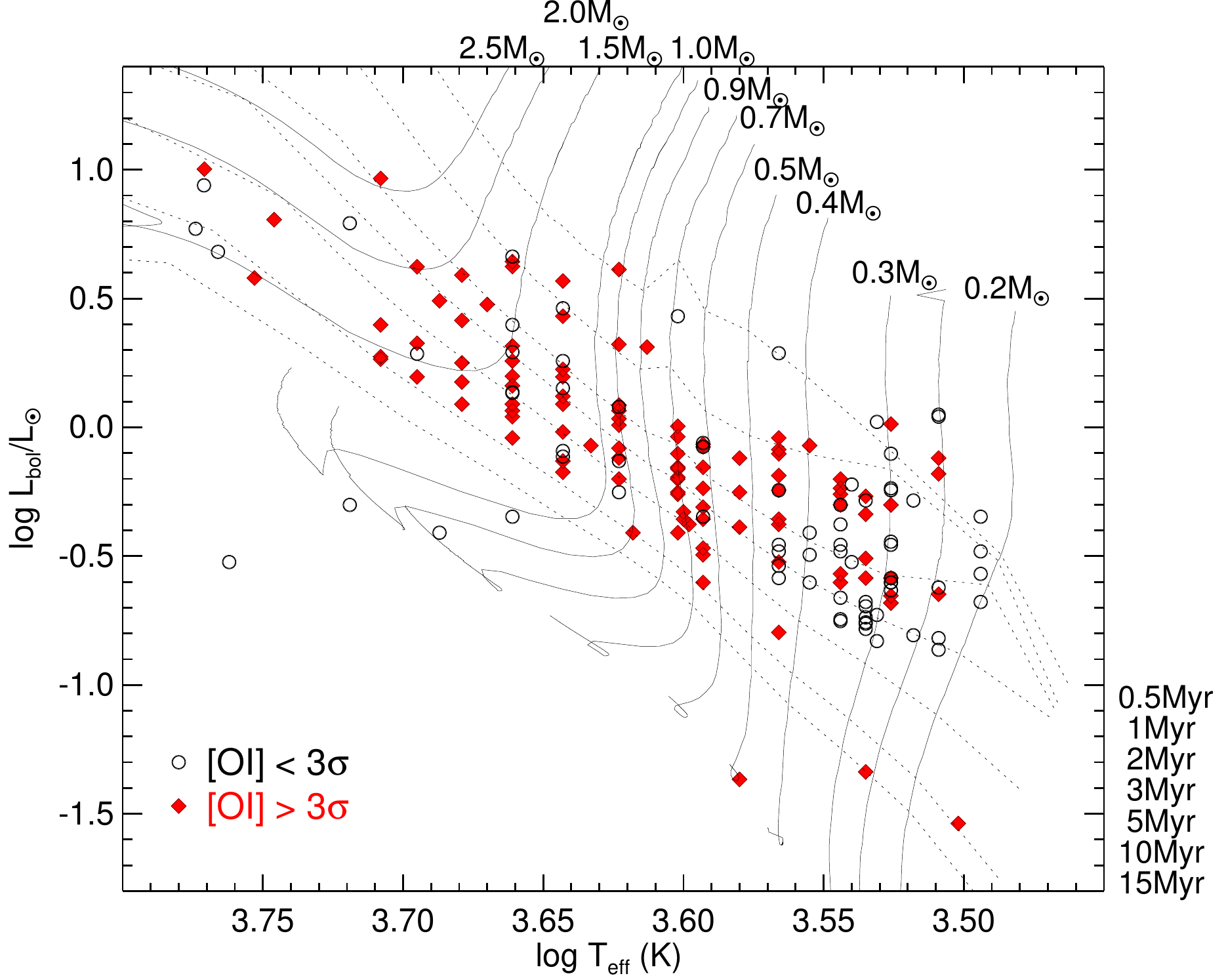}
    \caption{HR Diagram of the CTTSs in our sample that have detected [OI]$\lambda$6300 emission 
(red filled diamonds) and that do not (black open circles). Mass tracks and isochrones of 
\citet{siess00} are shown as solid and dashed lines, respectively. }
    \label{fig:hrd}
\end{figure*}

As has been mentioned throughout this work, our detection of the [OI]$\lambda$6300 emission 
line is much lower among CTTSs than in previous studies (we find a detection rate of 59\%, 
compared to 100\% in HEG95, 84\% in N14, 91\% in S16 and 77\% in N18).  
On the one hand, the sample studied by HEG95 (and R13, who obtained modern estimates of 
accretion luminosities for this sample) was composed of stars with on average higher mass 
accretion rates than ours, which could explain their much higher detection rate\footnote{See 
Sect. \ref{sec:oilumcor} for a discussion on the connection between mass accretion and 
[OI]$\lambda$6300 emission.}. 
Among stars of L$_{acc}>10^{-2.6} \mathrm{L}_{\odot}$ (where 90\% of their sample lies), our 
detection rate is 69\%. This is much higher than the global detection rate we find, but still 
lower than HEG95, who detected the [OI]$\lambda$6300 line in all CTTSs analyzed. On the other 
hand, N14 analyzed a sample of CTTSs with lower mass accretion rates than ours and still found 
a 91\% detection rate in the [OI]$\lambda$6300 line among CTTSs, meaning that mass accretion 
rates alone cannot account for our low detection rate. 

Our global detection rate of the [OI]$\lambda$6300 emission line 
is not constant throughout the HR diagram, but is much higher in the upper left portion  
and decreases toward lower effective temperatures (see Fig. \ref{fig:hrd}). Among 
stars of M$_* \leq 0.5$M$_{\odot}$ (or T$_{eff} \lesssim 3800$K) the detection rate is only 
40\% (32/80), while for stars of M$_* > 0.5$M$_{\odot}$ (T$_{eff} \gtrsim 3800$K) 
it is 75\% (78/104). This could be the result of an observational bias if the cooler 
stars tend to present emission of lower intensity compared to the level of noise in their 
spectra, which fall below our detection threshold. 
One or more of the following factors could cause this:
lower mass stars tend to have lower mass accretion rates \citep{venuti14}, which 
would result in lower intensity [OI]$\lambda$6300 emission lines;
cooler stars have more numerous and deeper photospheric absorption lines near the 
[OI]$\lambda$6300 line that, even after subtracting a photospheric template, lead to higher 
levels of noise in the spectra; and 
stars of lower luminosity tend to have spectra with lower signal to noise ratios 
(S/N), since all of the spectra were observed with the same integration time\footnote{The stars 
that were observed in configuration A1 are an exception (see Sect. \ref{sec:obs}), which was 
observed a total of four times, resulting in spectra of better S/N than those observed in the other 
configurations. However the detection rate in this configuration was no different than the 
overall detection rate.}. 

The third factor should be countered by a contrast effect: the lower a photosphere's luminosity, 
the higher the contrast between a jet's emission and the stellar continuum. This may be partly why 
we still detect emission in the lower right corner of the HR diagram, despite the sources being 
fainter, though this contrast effect does not seem to be enough to counter the other factors. 
Another effect that can be interfering with the overall [OI]$\lambda$6300 detection is the telluric 
[OI]$\lambda$6300 emission line. If a star has only a weak component of low radial velocity which 
coincides closely with the radial velocity of the telluric emission line on the night of the 
observation, it will be impossible to recover the intrinsic profile.

In an attempt to assess the level of observational bias, we plot in Fig. \ref{fig:hrdflux} 
the equivalent width of the measured [OI]$\lambda$6300 emission line versus the flux of the 
continuum at 6300\AA. 
The flux is given in instrumental counts, since our data are not flux calibrated. We separate the 
stars observed in configuration A1 (right panel) from those observed in all other configurations 
(left panel), because we expect there may be a difference in the detectability of the 
[OI]$\lambda$6300 line in configuration A1, since it was observed four times and therefore these 
spectra show better S/N than the spectra that were observed only once. 
Detections are shown as red filled diamonds, while the upper limits of nondetections are shown 
as black open diamonds. Upper limits are defined as 3$\sigma$ times the average FWHM of all 
detected components of the [OI]$\lambda$6300 line profiles (79~km~s$^{-1}$), where $\sigma$ was 
measured from two regions of the spectrum 5\AA \space before and after the [OI]$\lambda$6300 line. 
Though the detections generally lie well above the 3$\sigma$ threshold, 
there are several that fall in the same range as the upper limits. When analyzing only 
configuration A1 the outcome is similar. Though in this case the detections separate more from 
the nondetections, there are still some very close to the region where upper limits fall. 
The global [OI]$\lambda$6300 detection rate in this configuration is very similar to the overall 
detection rate. From this diagram, we judge that it is not possible to state for certain that the 
nondetections in our sample truly have no [OI]$\lambda$6300 emission. Our heterogeneous detection 
rate across the HR diagram, as well as our overall low detection rate compared with other works, 
are likely caused by an observational bias. 

\begin{figure*}[t!]
    \centering
    \includegraphics[width=17cm]{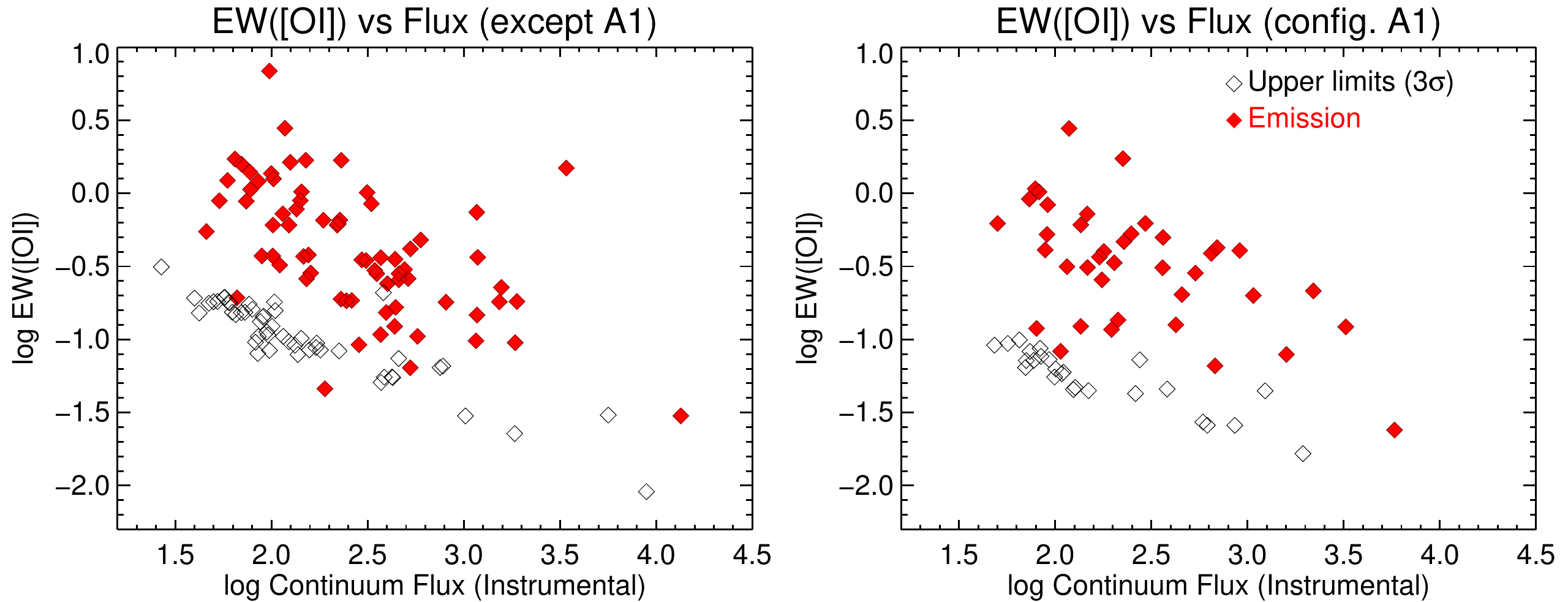}
    \caption{Equivalent width of the [OI]$\lambda$6300 line versus the instrumental flux of the 
continuum at 6300\AA. Detections are shown as red filled diamonds, while upper limits of 
nondetections (taken as 3$\sigma$ times the average FWHM of detections in our sample) are shown 
as black open diamonds. The right panel shows the A1 configuration, which was observed four times 
and has better S/N, while the left panel shows all other configurations. 
}\label{fig:hrdflux}
\end{figure*}

\section{Calculation of uncertainties}\label{sec:errors}

Uncertainties in stellar radial velocities were estimated by assuming that photospheric 
absorption lines are approximately Gaussian and using the formula for the uncertainty in the 
centroid of a Gaussian fit, according to \citet{porter04}: 

\begin{equation}\label{eq:errvr} 
\delta v_{r0} = \frac{FWHM_{phot}}{2\sqrt{2 \ln{2}} \ S/N_{phot}} ,
\end{equation}

where $FWHM_{phot}$ corresponds to a typical width of the photospheric lines of a given spectrum 
and $S/N_{phot}$ is the signal-to-noise ratio of that spectrum, both of which were measured 
directly in the spectra. For the CTTSs, we added to the uncertainty the $\delta v_{r0}$ of 
the WTTS template that was used to determine its stellar rest velocity, therefore   
\mbox{$\delta v_{r0,CTTS} = \sqrt{\delta v_{r0}^2 + \delta v_{r0,WTTS}^2 }$}. 

For the absolute values of $v_{rad}$, given in Table \ref{table:ctts}, we considered an additional 
uncertainty corresponding to a systematic shift of \mbox{$\Delta\lambda=$ 2~km~s$^{-1}$}, which was 
observed between these values of $v_{rad}$ and values we derived from FLAMES spectra of previous 
campaigns, which may be due to wavelength calibration issues. The final error bars on $v_{rad}$ 
are thus $\delta v_{rad} = \sqrt{ \delta v_{r0}^2 + \Delta \lambda ^2}$. 
For most ($\sim$80\%) of our sample, \mbox{$\delta v_{rad} \leq$ 2.6~km~s$^{-1}$}, but it reaches 
13~km~s$^{-1}$ for the spectrum that presented the worst S/N in the photospheric lines (that of star 
\mbox{\object{CSIMon-000423}}). 

Errors in the centroid velocities of the Gaussian components ($\delta v_c$) were estimated using 
the formula 

\begin{equation}\label{eq:errvc}  
\delta v_c = \sqrt{ \left( \frac{FWHM}{2\sqrt{2 \ln{2}} \  S/N} \right) ^2 + \delta v_{r0}^2 }, 
\end{equation} 

where $S/N$ is the signal to noise ratio of the component (its peak intensity divided by the noise 
of the residual spectrum) and $\delta v_{r0}$ is the uncertainty of the stellar rest velocity, 
disregarding the shift due to wavelength calibration, which would not affect the centroid velocity 
of the [OI]$\lambda$6300 line relative to the stellar rest frame measured in the same spectrum. 

For the FWHM, errors were estimated using the formula adapted from \citet{porter04}, 

\begin{equation}\label{eq:errfwhm}  
\delta FWHM = \frac{FWHM}{\sqrt{2} \  S/N} . 
\end{equation} 

\section{Tables of stellar and line parameters}\label{sec:tables}

Tables C.1 - C.4 are available at the CDS. 

\onecolumn 

\begin{small}

\begin{table*}[ht]
\begin{center}
\caption{Stellar parameters of the WTTSs used as photospheric templates
}
\label{table:wtts}
%\small{
\begin{tabular}{l c c c c c}
\hline
\hline
 CSIMon ID & 2MASS ID & Sp type      &   L$_{bol}$   &   M$_*$       &   R$_*$       \\
           &          & (Literature) & (L$_{\odot}$) & (M$_{\odot}$) & (R$_{\odot}$) \\
\hline
% MonID  2MASSID  SpT  Lbol  Mass  Rad 
\object{CSIMon-000020} & J06420924+0944034 &     K7 &    0.44 & 0.71 & 1.39 \\ 
\object{CSIMon-000029} & J06410328+0957549 &     K7 &    0.64 & 0.70 & 1.68 \\ 
\object{CSIMon-000057} & J06410393+0958094 &     M3 &    0.53 & 0.30 & 2.16 \\ 
\object{CSIMon-000066} & J06410357+1000353 &     M1 &    0.39 & 0.45 & 1.53 \\ 
\object{CSIMon-000071} & J06410025+0958496 &   K1.5 &    2.60 & 1.69 & 2.14 \\ 
\object{CSIMon-000104} & J06410417+0952020 &     K6 &    0.98 & 0.91 & 1.87 \\
\hline
\end{tabular}
\tablefoot{
Spectral types were taken from \citet{venuti14}, who adopted the values of \citet{dahm05}, 
\citet{rebull02} or \citet{walker56}, in that order of preference, 
or derived them from CFHT colors when no classification was available in the literature. 
Stellar luminosities, masses and radii were also adopted from \citet{venuti14}. 
} 
\end{center}
\end{table*}
\setlength\tabcolsep{0.175cm}

\begin{sidewaystable*}[p]
\begin{center}
\begin{small}
\caption{Stellar parameters of the CTTSs
}
\ContinuedFloat
\label{table:ctts}
\begin{tabular}{l c c c c c c c c c c c c c c c}
\hline
\hline
 CSIMon ID & 2MASS ID & Sp type & Sp type &   L$_{bol}$   &   M$_*$       &   R$_*$       & $\dot{\mathrm{M}}_{acc}$ &   L$_{acc}$   & Veiling & $\alpha_{IRAC}$ & E(K-3.6$\mu$m) & Period & $v_{rad}$ & $v\sin i$ & $i$        \\
           &          & (Lit)   & (WTTS)  & (L$_{\odot}$) & (M$_{\odot}$) & (R$_{\odot}$) &  (M$_{\odot}$/yr)        & (L$_{\odot}$) &  ($r$)  &                 &                & (days) &  (km/s)   &  (km/s)   & $^{\circ}$ \\
\hline
% MonID  2MASSID  SpT(lit)  SpT(us)  Lbol  Mass  Rad  Macc  Lacc  Veil  aIRAC  IRexcess  Prot  vsini  inc
\object{CSIMon-000011} & J06411725$+$0954323 &      K7 &    K7 &  0.79 & 0.70 & 1.86 & 1.86e-07 & 9.74e-02 & 0.30$\pm$0.02 &   ...  &    1.15 &  ...  &  21.2$\pm$  2.1 & 14.5$\pm$ 2.0 &     ...    \\
\object{CSIMon-000021} & J06405944$+$0959454 &      K5 &    K5 &  0.96 & 1.20 & 1.69 & 9.50e-10 & 1.70e-02 & 0.06$\pm$0.02 &   ...  &    0.48 &  3.15 &  18.5$\pm$  2.1 & 21.7$\pm$ 4.2 & 53$\substack{+37\\-25}$ \\
\object{CSIMon-000056} & J06415315$+$0950474 &      K5 &    K5 &  0.77 & 1.16 & 1.52 & 5.75e-09 & 2.18e-03 & 0.02$\pm$0.02 &  -1.36 &    0.25 &  5.83 &  20.2$\pm$  2.0 & 15.8$\pm$ 2.7 & 61$\substack{+29\\-13}$ \\
\object{CSIMon-000062} & J06422190$+$0948329 &      K0 &   ... &  0.50 & 0.90 & 0.86 &    ...   &    ...   &      ...      &   ...  &    0.12 &  ...  &      ...        &     ...       &     ...    \\
\object{CSIMon-000063} & J06411193$+$0959412 &    M2.5 &  M2.5 &  0.17 & 0.32 & 1.19 & 6.17e-09 & 2.89e-03 & 0.35$\pm$0.03 &  -1.37 &    ...  &  ...  &  17.9$\pm$  2.0 & 16.0$\pm$ 6.2 &     ...    \\
\object{CSIMon-000064} & J06411070$+$0957424 &    M1.5 &    M1 &  0.39 & 0.40 & 1.62 & 4.68e-09 & 1.27e-03 & 0.03$\pm$0.02 &  -1.15 &    0.43 &  9.44 &  23.1$\pm$  2.0 &  9.5$\pm$ 2.0 & 55$\substack{+35\\-15}$ \\
\hline
\end{tabular}
%\tablefoot{
%Caption 
%}
\end{small}
\end{center}
\end{sidewaystable*}

\begin{sidewaystable*}[p]
\begin{center}
\begin{small}
\caption{Stellar parameters of the CTTSs - continued
}
\ContinuedFloat
\begin{tabular}{l c c c c c c c c c c c c c c c}
\hline
\hline
 CSIMon ID & 2MASS ID & Sp type & Sp type &   L$_{bol}$   &   M$_*$       &   R$_*$       & $\dot{\mathrm{M}}_{acc}$ &   L$_{acc}$   & Veiling & $\alpha_{IRAC}$ & E(K-3.6$\mu$m) & Period & $v_{rad}$ & $v\sin i$ & $i$        \\
           &          & (Lit)   & (WTTS)  & (L$_{\odot}$) & (M$_{\odot}$) & (R$_{\odot}$) &  (M$_{\odot}$/yr)        & (L$_{\odot}$) &  ($r$)  &                 &                & (days) &  (km/s)   &  (km/s)   & $^{\circ}$ \\
\hline
\object{CSIMon-000318} & J06404458$+$0948126 &      K6 &    K6 &  1.20 & 0.91 & 2.07 & 4.70e-10 & 5.20e-03 & 0.00$\pm$0.06 &  -0.97 &    ...  &  4.95 &  20.8$\pm$  2.0 & 14.3$\pm$ 2.9 & 43$\substack{+30\\-19}$ \\
\object{CSIMon-000326} & J06405882$+$0939187 &      M0 &    K7 &  0.34 & 0.66 & 1.27 & 1.70e-09 & 1.16e-02 & 0.26$\pm$0.05 &  -1.39 &    0.06 &  6.64 &  20.3$\pm$  2.1 & 13.1$\pm$ 6.5 & 44$\substack{+46\\-21}$ \\
\object{CSIMon-000328} & J06412700$+$0930131 &      M1 &    M2 &  0.30 & 0.45 & 1.34 & 2.40e-09 & 1.02e-02 & 0.51$\pm$0.06 &  -1.88 &    0.02 & 10.00 &  20.9$\pm$  2.4 & 11.5$\pm$ 2.4 & 86$\substack{+ 4\\- 2}$ \\
\object{CSIMon-000330} & J06413160$+$0948328 &      K2 &    K2 &  1.93 & 1.55 & 1.89 & 1.34e-10 & 2.76e-03 & 0.00$\pm$0.04 &  -2.84 &   -0.10 &  5.05 &  17.7$\pm$  2.0 & 23.2$\pm$ 1.6 & 70$\substack{+20\\- 8}$ \\
\object{CSIMon-000335} & J06413728$+$0945066 &      K4 &    K4 &  1.45 & 1.40 & 1.92 & 2.31e-09 & 4.45e-02 & 0.00$\pm$0.02 &  -0.93 &    0.36 &  4.58 &  16.8$\pm$  2.1 & 15.2$\pm$ 2.1 & 46$\substack{+29\\-18}$ \\
\object{CSIMon-000340} & J06415593$+$0940468 &    M2.5 &  M1.5 &  0.52 & 0.33 & 2.05 & 3.55e-09 & 1.24e-03 & 0.00$\pm$0.02 &  -2.82 &    0.11 &  1.23 &  17.3$\pm$  2.0 & 12.4$\pm$ 4.2 &  8$\substack{+ 5\\- 5}$ \\
\hline
\end{tabular}
%\tablefoot{
%Caption 
%}
\end{small}
\end{center}
\end{sidewaystable*}

\begin{sidewaystable*}[p]
\begin{center}
\begin{small}
\caption{Stellar parameters of the CTTSs - continued
}
\ContinuedFloat
\begin{tabular}{l c c c c c c c c c c c c c c c}
\hline
\hline
 CSIMon ID & 2MASS ID & Sp type & Sp type &   L$_{bol}$   &   M$_*$       &   R$_*$       & $\dot{\mathrm{M}}_{acc}$ &   L$_{acc}$   & Veiling & $\alpha_{IRAC}$ & E(K-3.6$\mu$m) & Period & $v_{rad}$ & $v\sin i$ & $i$        \\
           &          & (Lit)   & (WTTS)  & (L$_{\odot}$) & (M$_{\odot}$) & (R$_{\odot}$) &  (M$_{\odot}$/yr)        & (L$_{\odot}$) &  ($r$)  &                 &                & (days) &  (km/s)   &  (km/s)   & $^{\circ}$ \\
\hline
\object{CSIMon-000510} & J06410429$+$0924521 &      M0 &    K7 &  0.84 & 0.62 & 1.99 & 9.12e-08 & 2.29e-01 & 0.78$\pm$0.04 &  -0.97 &    0.85 & 14.99 &  18.1$\pm$  2.2 & 14.3$\pm$ 2.1 &     90     \\
\object{CSIMon-000525} & J06405118$+$0944461 &      G0 &    G0 &  8.70 & 1.80 & 2.82 & 1.23e-08 & 1.97e-01 & 0.27$\pm$0.20 &  -1.69 &    0.94 &  1.99 &  18.1$\pm$  2.0 & 51.8$\pm$ 4.6 & 46$\substack{+26\\-17}$ \\
\object{CSIMon-000530} & J06411938$+$0921469 &    K2.5 &    K2 &  0.39 & 0.82 & 0.89 & 1.35e-08 & 3.02e-01 & 0.14$\pm$0.02 &  -1.59 &    ...  &  ...  &  39.3$\pm$  2.0 & 13.1$\pm$ 2.1 &     ...    \\
\object{CSIMon-000553} & J06410303$+$0937357 &      M1 &    M1 &  1.94 & 0.45 & 3.44 & 9.75e-10 & 3.21e-03 & 1.02$\pm$0.05 &  -1.22 &    0.95 &  ...  &  19.8$\pm$  2.0 & 39.8$\pm$ 5.8 &     ...    \\
\object{CSIMon-000558} & J06413974$+$0940279 &      K4 &    K4 &  4.40 & 1.60 & 3.35 & 6.06e-08 & 9.00e-01 & 0.25$\pm$0.02 &  -1.46 &    0.45 & 11.71 &  18.4$\pm$  2.0 & 15.4$\pm$ 2.1 & 58$\substack{+32\\-14}$ \\
\object{CSIMon-000566} & J06405275$+$0943004 &    M3.5 &    M5 &  0.16 & 0.26 & 1.21 & 2.29e-09 & 1.92e-03 & 0.09$\pm$0.03 &   ...  &    0.37 &  ...  &  18.8$\pm$  2.0 & 18.0$\pm$ 3.9 &     ...    \\
\hline  
\end{tabular}
%\tablefoot{
%Caption 
%}
\end{small}
\end{center}
\end{sidewaystable*}

\begin{sidewaystable*}[p]
\begin{center}
\begin{small}
\caption{Stellar parameters of the CTTSs - continued
}
\ContinuedFloat
\begin{tabular}{l c c c c c c c c c c c c c c c}
\hline
\hline
 CSIMon ID & 2MASS ID & Sp type & Sp type &   L$_{bol}$   &   M$_*$       &   R$_*$       & $\dot{\mathrm{M}}_{acc}$ &   L$_{acc}$   & Veiling & $\alpha_{IRAC}$ & E(K-3.6$\mu$m) & Period & $v_{rad}$ & $v\sin i$ & $i$        \\
           &          & (Lit)   & (WTTS)  & (L$_{\odot}$) & (M$_{\odot}$) & (R$_{\odot}$) &  (M$_{\odot}$/yr)        & (L$_{\odot}$) &  ($r$)  &                 &                & (days) &  (km/s)   &  (km/s)   & $^{\circ}$ \\
\hline
\object{CSIMon-000798} & J06405694$+$0948407 &      F9 &    G0 &  5.90 & 1.57 & 2.30 & 9.33e-09 & 1.60e-01 & 0.34$\pm$0.04 &  -3.73 &   -0.03 &  3.81 &  25.5$\pm$  2.0 & 20.2$\pm$ 5.3 & 41$\substack{+36\\-21}$ \\
\object{CSIMon-000804} & J06405571$+$0951138 &    K5.5 &    K6 &  0.85 & 1.06 & 1.67 & 6.76e-09 & 1.74e-02 & 0.08$\pm$0.06 &  -0.29 &    0.23 &  3.27 &  26.3$\pm$  2.1 & 18.3$\pm$ 2.0 & 45$\substack{+25\\-17}$ \\
\object{CSIMon-000808} & J06405159$+$0928445 &      K4 &    K5 &  0.91 & 1.24 & 1.52 & 1.70e-08 & 1.17e-01 & 0.14$\pm$0.02 &  -1.85 &    0.53 &  ...  &  20.3$\pm$  2.1 & 13.0$\pm$ 4.0 &     ...    \\
\object{CSIMon-000810} & J06410982$+$0927122 &      K5 &    K5 &  2.90 & 1.20 & 2.95 & 1.24e-09 & 1.41e-02 & 0.03$\pm$0.03 &  -2.84 &    0.04 &  2.92 &  22.0$\pm$  2.0 & 56.7$\pm$ 3.1 & 65$\substack{+25\\-11}$ \\
\object{CSIMon-000811} & J06404321$+$0947072 &      K6 &    K6 &  1.08 & 0.91 & 1.97 & 6.75e-09 & 6.29e-02 & 0.10$\pm$0.02 &  -1.64 &    0.77 &  7.84 &  19.4$\pm$  2.1 & 15.1$\pm$ 1.7 & 65$\substack{+25\\-11}$ \\
\object{CSIMon-000823} & J06405014$+$0957039 &      M3 &  M2.5 &  0.25 & 0.30 & 1.49 & 7.76e-10 & 9.01e-04 & 2.94$\pm$0.26 &   ...  &   -0.00 &  ...  &  22.2$\pm$  2.0 &     ...       &     ...    \\
\hline
\end{tabular}
%\tablefoot{
%Caption 
%}
\end{small}
\end{center}
\end{sidewaystable*}

\begin{sidewaystable*}[p]
\begin{center}
\begin{small}
\caption{Stellar parameters of the CTTSs - continued}
\begin{tabular}{l c c c c c c c c c c c c c c c}
\hline
\hline
 CSIMon ID & 2MASS ID & Sp type & Sp type &   L$_{bol}$   &   M$_*$       &   R$_*$       & $\dot{\mathrm{M}}_{acc}$ &   L$_{acc}$   & Veiling & $\alpha_{IRAC}$ & E(K-3.6$\mu$m) & Period & $v_{rad}$ & $v\sin i$ & $i$        \\
           &          & (Lit)   & (WTTS)  & (L$_{\odot}$) & (M$_{\odot}$) & (R$_{\odot}$) &  (M$_{\odot}$/yr)        & (L$_{\odot}$) &  ($r$)  &                 &                & (days) &  (km/s)   &  (km/s)   & $^{\circ}$ \\
\hline
\object{CSIMon-001064} & J06403518$+$0951567 &      M1 &    M1 &  0.91 & 0.45 & 2.35 & 1.45e-08 & 2.06e-02 & 0.17$\pm$0.07 &  -0.68 &    0.32 &  2.70 &  22.9$\pm$ 3.5 & 25.1$\pm$ 6.9 & 35$\substack{+24\\-18}$ \\
\object{CSIMon-001076} & J06403819$+$0929524 &      M1 &    M2 &  0.26 & 0.45 & 1.25 & 4.57e-10 & 4.15e-04 & 0.39$\pm$0.02 &  -1.39 &    0.25 &  5.88 &  17.2$\pm$ 2.0 & 15.3$\pm$11.0 & 33$\substack{+57\\-28}$ \\
\object{CSIMon-001094} & J06403164$+$0948233 &      K5 &    K5 &  0.74 & 1.14 & 1.49 & 1.58e-08 & 1.62e-01 & 0.22$\pm$0.02 &  -1.96 &    0.11 &  4.25 &  17.5$\pm$ 2.4 & 21.3$\pm$ 2.1 & 66$\substack{+24\\-10}$ \\
\object{CSIMon-001099} & J06404136$+$0954138 &      G1 &    G0 &  4.80 & 1.52 & 2.16 & 3.23e-09 & 1.14e-01 & 0.00$\pm$0.04 &   ...  &    0.78 &  3.42 &  26.9$\pm$ 2.0 & 26.8$\pm$ 3.1 & 57$\substack{+33\\-23}$ \\
\object{CSIMon-001100} & J06403934$+$0934455 &      K6 &    K6 &  2.10 & 0.90 & 2.75 & 3.19e-09 & 2.62e-02 & 0.72$\pm$0.04 &  -0.42 &    1.63 &  ...  &  20.3$\pm$ 5.1 & 23.1$\pm$ 3.2 &     ...    \\
\object{CSIMon-001114} & J06393339$+$0952017 &    M1.5 &    M2 &  0.85 & 0.40 & 2.40 & 4.27e-08 & 1.77e-01 & 0.33$\pm$0.02 &   ...  &    0.02 &  2.58 &  13.5$\pm$ 2.7 & 32.6$\pm$ 4.5 & 44$\substack{+26\\-17}$ \\
\hline
\end{tabular}
\tablefoot{
The CSIMon ID is a naming scheme devised for the CSI2264 campaign \citep{cody14}. 2MASS IDs are given
for comparison.
Spectral types were taken from \citet{venuti14}, who adopted the values of \citet{dahm05},
\citet{rebull02} or \citet{walker56}, or derived them from CFHT colors.
Stellar luminosity (L$_{bol}$), mass (M$_*$) and radius (R$_*$) were determined by \citet{venuti14}.
Mass accretion rates ($\dot{\mathrm{M}}_{acc}$) and accretion luminosity (L$_{acc}$) were derived
from UV excess by \citet{venuti14} or from H$\alpha$ line profiles.
Veiling, radial velocities ($v_{rad}$) and $v\sin i$ were determined using FLAMES spectra.
The periods given are those from CoRoT observations \citep{alencar10,venuti17}. $\alpha_{IRAC}$ and IR excess
were determined using photometry from the literature or the CSI2264 campaign. Inclinations were
estimated by us from the rotational properties of the stars.
}
\end{small}
\end{center}
\end{sidewaystable*}

\setlength\tabcolsep{0.2cm}
\begin{sidewaystable*}[p]
\begin{center}
%\begin{small}
\caption{Equivalent width, centroid velocity and FWHM of the components of the [OI] 6300\AA \space line}\label{table:oi_comp_par} %\\
\ContinuedFloat
%\vspace{-0.2cm}
\begin{tabular}{l|c c c|c c c|c c c|c c c}
\hline
\hline
CSIMon ID & \multicolumn{3}{|c|}{Narrow LVC} & \multicolumn{3}{|c|}{Broad LVC} & \multicolumn{3}{|c|}{Blueshifted HVC} & \multicolumn{3}{|c}{Redshifted HVC} \\ 
                       &  EW   & $v_c$          & FWHM          &  EW   & $v_c$          &  FWHM          &  EW   &  $v_c$          & FWHM           &  EW   & $v_c$          & FWHM           \\ 
\hline
\object{CSIMon-000011} & 0.195 &  -4.7$\pm$ 1.3 & 38.1$\pm$ 1.8 & 0.127 &  -1.1$\pm$ 6.2 &  72.7$\pm$10.3 &  ...  &        ...      &       ...      & 0.032 & 114.0$\pm$ 3.1 &  31.6$\pm$ 5.0 \\ 
\object{CSIMon-000021} & 0.039 &  17.3$\pm$ 2.6 & 28.3$\pm$ 4.3 &  ...  &       ...      &       ...      &  ...  &        ...      &       ...      &  ...  &       ...      &       ...      \\ 
\object{CSIMon-000103} & 0.099 &  -6.5$\pm$ 3.9 & 50.1$\pm$ 5.8 & 0.186 & -14.5$\pm$ 9.3 & 110.6$\pm$15.1 &  ...  &        ...      &       ...      &  ...  &       ...      &       ...      \\ 
\object{CSIMon-000119} & 0.053 &   3.4$\pm$ 2.6 & 38.4$\pm$ 4.1 & 0.100 & -36.1$\pm$56.0 & 250.4$\pm$93.2 &  ...  &        ...      &       ...      &  ...  &       ...      &       ...      \\ 
\object{CSIMon-000120} & 0.653 &   2.5$\pm$ 3.5 & 43.2$\pm$ 1.5 &  ...  &       ...      &       ...      &  ...  &        ...      &       ...      &  ...  &       ...      &       ...      \\ 
\object{CSIMon-000126} & 1.008 & -29.2$\pm$23.1 & 66.4$\pm$ 2.4 & 0.665 & -20.8$\pm$24.5 & 130.3$\pm$14.2 &  ...  &        ...      &       ...      &  ...  &       ...      &       ...      \\
\hline
\end{tabular}
%\tablefoot{
%Caption 
%}
%\end{small}
\end{center}
\end{sidewaystable*}

\begin{sidewaystable*}[p]
\begin{center} 
%\begin{small}
\caption{Equivalent width, centroid velocity and FWHM of the components of the [OI] 6300\AA \space line - continued}
\ContinuedFloat
%\vspace{-0.2cm}
\begin{tabular}{l|c c c|c c c|c c c|c c c}
\hline
\hline
CSIMon ID & \multicolumn{3}{|c|}{Narrow LVC} & \multicolumn{3}{|c|}{Broad LVC} & \multicolumn{3}{|c|}{Blueshifted HVC} & \multicolumn{3}{|c}{Redshifted HVC} \\ 
%\hline
                       &  EW   & $v_c$          & FWHM          &  EW   & $v_c$          &  FWHM          &  EW   &  $v_c$          & FWHM           &  EW   & $v_c$          & FWHM           \\ 
\hline
\object{CSIMon-000434} &  ...  &       ...      &      ...      & 0.125 & -20.8$\pm$16.4 & 110.5$\pm$27.2 &  ...  &        ...      &       ...      &  ...  &       ...      &       ...      \\ 
\object{CSIMon-000441} & 0.451 &   0.9$\pm$ 8.2 & 52.1$\pm$ 6.8 &  ...  &       ...      &       ...      &  ...  &        ...      &       ...      &  ...  &       ...      &       ...      \\ 
\object{CSIMon-000448} &  ...  &       ...      &      ...      & 0.928 & -17.4$\pm$ 6.2 & 136.6$\pm$ 8.1 &  ...  &        ...      &       ...      &  ...  &       ...      &       ...      \\ 
\object{CSIMon-000456} &  ...  &       ...      &      ...      & 0.080 &   6.4$\pm$ 9.6 &  90.4$\pm$16.0 &  ...  &        ...      &       ...      &  ...  &       ...      &       ...      \\ 
\object{CSIMon-000457} &  ...  &       ...      &      ...      & 0.940 & -40.3$\pm$ 3.0 & 116.5$\pm$ 3.4 & 0.624 & -115.8$\pm$ 2.4 &  57.4$\pm$ 1.4 & 0.164 & 101.8$\pm$ 3.5 &  72.6$\pm$ 4.5 \\ 
\object{CSIMon-000462} & 0.031 &  -2.7$\pm$ 1.3 & 33.8$\pm$ 1.8 &  ...  &       ...      &       ...      &  ...  &        ...      &       ...      &  ...  &       ...      &       ...      \\
\hline
\end{tabular}
%\tablefoot{
%Caption 
%}
%\end{small}
\end{center}
\end{sidewaystable*}

\begin{sidewaystable*}[p]
\begin{center} 
%\begin{small}
\caption{Equivalent width, centroid velocity and FWHM of the components of the [OI] 6300\AA \space line - continued}
%\vspace{-0.2cm}
\begin{tabular}{l|c c c|c c c|c c c|c c c}
\hline
\hline
CSIMon ID & \multicolumn{3}{|c|}{Narrow LVC} & \multicolumn{3}{|c|}{Broad LVC} & \multicolumn{3}{|c|}{Blueshifted HVC} & \multicolumn{3}{|c}{Redshifted HVC} \\ 
%\hline
                       &  EW   & $v_c$          & FWHM          &  EW   & $v_c$          &  FWHM          &  EW   &  $v_c$          & FWHM           &  EW   & $v_c$          & FWHM           \\ 
\hline
\object{CSIMon-000877} & 0.251 &  -2.5$\pm$ 1.4 & 59.2$\pm$ 2.1 & 0.240 &  -1.3$\pm$ 5.6 & 121.4$\pm$ 9.3 &  ...  &        ...      &       ...      &  ...  &       ...      &       ...      \\ 
\object{CSIMon-000879} & 0.092 &  10.7$\pm$ 5.3 & 51.8$\pm$ 8.8 &  ...  &       ...      &       ...      &  ...  &        ...      &       ...      &  ...  &       ...      &       ...      \\ 
\object{CSIMon-000904} &  ...  &       ...      &      ...      & 1.292 &  -4.4$\pm$ 2.6 &  64.1$\pm$ 2.2 &  ...  &        ...      &       ...      &  ...  &       ...      &       ...      \\ 
\object{CSIMon-000919} &  ...  &       ...      &      ...      & 0.818 & -13.5$\pm$ 6.9 &  91.7$\pm$ 4.2 &  ...  &        ...      &       ...      & 0.205 &  49.9$\pm$ 6.9 &  32.7$\pm$ 4.2 \\ 
\object{CSIMon-000928} & 0.120 & -10.1$\pm$ 4.3 & 51.1$\pm$ 7.0 &       &       ...      &       ...      &  ...  &        ...      &       ...      &  ...  &       ...      &       ...      \\ 
\object{CSIMon-000936} & 0.146 & -15.4$\pm$ 2.4 & 41.1$\pm$ 3.6 & 0.478 & -37.1$\pm$11.3 & 168.7$\pm$18.8 & 0.225 & -124.9$\pm$ 2.1 &  46.2$\pm$ 3.1 &  ...  &       ...      &       ...      \\
\hline
\end{tabular}
\tablefoot{Equivalent widths are given in \AA, $v_c$ and FWHM are given in km/s. 
}
%\end{small}
\end{center}
\end{sidewaystable*}

\begin{table*}[p]
\centering
\begin{small}
\caption{[OI] 6300\AA \space line parameters}\label{table:ctts_oi} %\\
\ContinuedFloat
%\vspace{-0.2cm}
\begin{tabular}{l c c c c c c c}
\hline
\hline
 CSIMon ID & 2MASS ID &  L$_{[OI]}$   &  L$_{NLVC}$   &  L$_{BLVC}$   & L$_{HVC,blue}$ & L$_{HVC,red}$ & v$_{jet}$  \\
           &          & (L$_{\odot}$) & (L$_{\odot}$) & (L$_{\odot}$) & (L$_{\odot}$)  & (L$_{\odot}$) &  (km/s)    \\
\hline
% MonID  2MASSID  LOI, LNLVC  LBLVC  LbHVC  LrHVC  vjet 
\object{CSIMon-000011} & J06411725+0954323 &    1.54e-05 & 8.48e-06 & 5.52e-06 &   ...    & 1.39e-06 &   ...    \\ 
\object{CSIMon-000021} & J06405944+0959454 &    3.55e-06 & 3.55e-06 &   ...    &   ...    &   ...    &   ...    \\ 
\object{CSIMon-000056} & J06415315+0950474 & $<$1.91e-05 &   ...    &   ...    &   ...    &   ...    &   ...    \\ 
\object{CSIMon-000062} & J06422190+0948329 & $<$2.19e-05 &   ...    &   ...    &   ...    &   ...    &   ...    \\ 
\object{CSIMon-000063} & J06411193+0959412 & $<$3.93e-06 &   ...    &   ...    &   ...    &   ...    &   ...    \\ 
\object{CSIMon-000064} & J06411070+0957424 & $<$8.53e-06 &   ...    &   ...    &   ...    &   ...    &   ...    \\
\hline
\end{tabular}
\end{small}
\end{table*}

\begin{table*}[p]
\centering
\begin{small}
\caption{[OI] 6300\AA \space line parameters - continued 
}
\ContinuedFloat
%\vspace{-0.2cm}
\begin{tabular}{l c c c c c c c}
\hline
\hline
 CSIMon ID & 2MASS ID &  L$_{[OI]}$   &  L$_{NLVC}$   &  L$_{BLVC}$   & L$_{HVC,blue}$ & L$_{HVC,red}$ & v$_{jet}$  \\
           &          & (L$_{\odot}$) & (L$_{\odot}$) & (L$_{\odot}$) & (L$_{\odot}$)  & (L$_{\odot}$) &  (km/s)    \\
\hline
\object{CSIMon-000425} & J06411668+0929522 &    9.96e-05 & 5.10e-05 &   ...    & 3.59e-05 & 1.27e-05 & -89/+96  \\ 
\object{CSIMon-000426} & J06411945+0930286 & $<$4.13e-05 &    ...   &   ...    &   ...    &   ...    &   ...    \\ 
\object{CSIMon-000433} & J06410111+0934522 &    1.67e-05 & 8.78e-07 & 1.58e-05 &   ...    &   ...    &   ...    \\ 
\object{CSIMon-000434} & J06410406+0935211 &    3.00e-06 &    ...   & 3.00e-06 &   ...    &   ...    &   ...    \\ 
\object{CSIMon-000441} & J06405809+0936533 &    2.99e-06 & 2.99e-06 &   ...    &   ...    &   ...    &   ...    \\ 
\object{CSIMon-000448} & J06410360+0930290 &    1.19e-05 &   ...    & 1.19e-05 &   ...    &   ...    &   ...    \\
\hline
\end{tabular}
\end{small}
\end{table*}

\begin{table*}[p]
\centering
\begin{small}
\caption{[OI] 6300\AA \space line parameters - continued 
}
%\vspace{-0.2cm}
\begin{tabular}{l c c c c c c c}
\hline
\hline
 CSIMon ID & 2MASS ID &  L$_{[OI]}$   &  L$_{NLVC}$   &  L$_{BLVC}$   & L$_{HVC,blue}$ & L$_{HVC,red}$ & v$_{jet}$  \\
           &          & (L$_{\odot}$) & (L$_{\odot}$) & (L$_{\odot}$) & (L$_{\odot}$)  & (L$_{\odot}$) &  (km/s)    \\
\hline
\object{CSIMon-000861} & J06410080+0952075 &    4.45e-05 &   ...    & 4.45e-05 &   ...    &   ...    &   ...    \\ 
\object{CSIMon-000873} & J06411320+0926103 &    1.48e-04 & 3.21e-05 & 1.16e-04 &   ...    &   ...    &   ...    \\ 
\object{CSIMon-000877} & J06411678+0927301 &    5.88e-05 & 3.01e-05 & 2.88e-05 &   ...    &   ...    &   ...    \\ 
\object{CSIMon-000879} & J06410338+0940448 &    4.34e-06 & 4.34e-06 &   ...    &   ...    &   ...    &   ...    \\ 
\object{CSIMon-000893} & J06410457+0938308 & $<$3.47e-06 &   ...    &   ...    &   ...    &   ...    &   ...    \\ 
\object{CSIMon-000904} & J06404278+0933350 &    3.98e-05 &   ...    & 3.98e-05 &   ...    &   ...    &   ...    \\
\hline
\end{tabular} 
\end{small}
\end{table*}
\end{small} 

\section{Additional figures}\label{sec:figures}

\begin{figure*}[h]
 \centering
 \includegraphics[width=17cm]{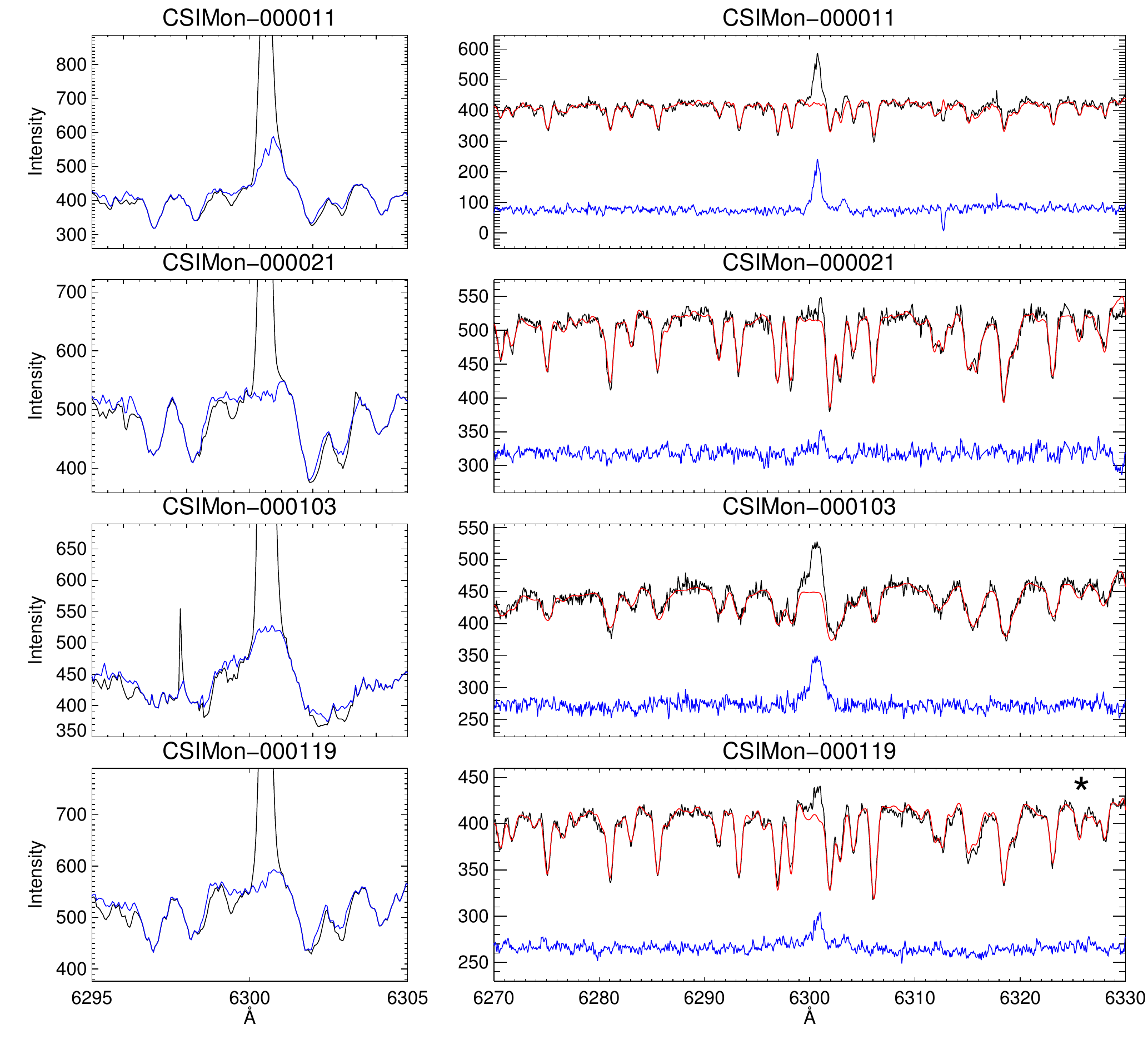}
 \caption{Correction of the [OI]$\lambda$6300 line profiles for contamination by telluric emission 
and absorption lines (left panels) and photospheric absorption lines (right panels) for all stars 
where emission was detected in [OI]$\lambda$6300. \\ Left panels: the original spectra are plotted 
in black, while the blue shows the spectrum after corrections for telluric emission and absorption, 
as well as the removal of cosmic rays. 
Right panels: the spectrum after telluric corrections is shown in black, while red represents the 
photospheric template that was used (including veiling and rotational broadening). 
The final, residual profile is shown in blue, shifted in the vertical axis for better viewing. \\  
\textbf{Notes.} 
Profiles marked with an asterisk on the top right corner of the right panel are 
those that were observed more than once. In these cases, telluric corrections were applied for 
each night individually, then the mean spectrum was taken in order to increase the S/N. This mean 
spectrum (shown in black on the right panels) was then corrected for the photospheric contribution 
and the residual profile was recovered. For these stars, the left panel shows the telluric 
corrections of one night. \\ 
Some stars did not present photospheric features in the region close to the [OI]$\lambda$6300 line, 
either due to their effective temperature (the B-type star \object{CSIMon-000392}) or to insufficient S/N 
in the photospheric lines (\object{CSIMon-000423}, \object{CSIMon-000632}, and \object{CSIMon-001011}). 
For these stars, we fitted the continuum and subtracted it to recover the residual profile. \\ 
The star \object{CSIMon-000631} is an A-type star for which a good template was not found since few hot 
stars were included in our observations. We used as its photospheric template an F-type star with 
a similar absorption feature close to the [OI]$\lambda$6300 line in order to remove this 
feature, ignoring the rest of the spectrum. This is why the absorption line close to 6320\AA 
\space is not well fitted. \\ 
The star \object{CSIMon-001234} is a spectroscopic binary, therefore it was fitted with a combination 
of two photospheric templates.  
}\ContinuedFloat
\label{fig:speccor}
\end{figure*}

\begin{figure*}[p]
 \centering
 \includegraphics[width=17cm]{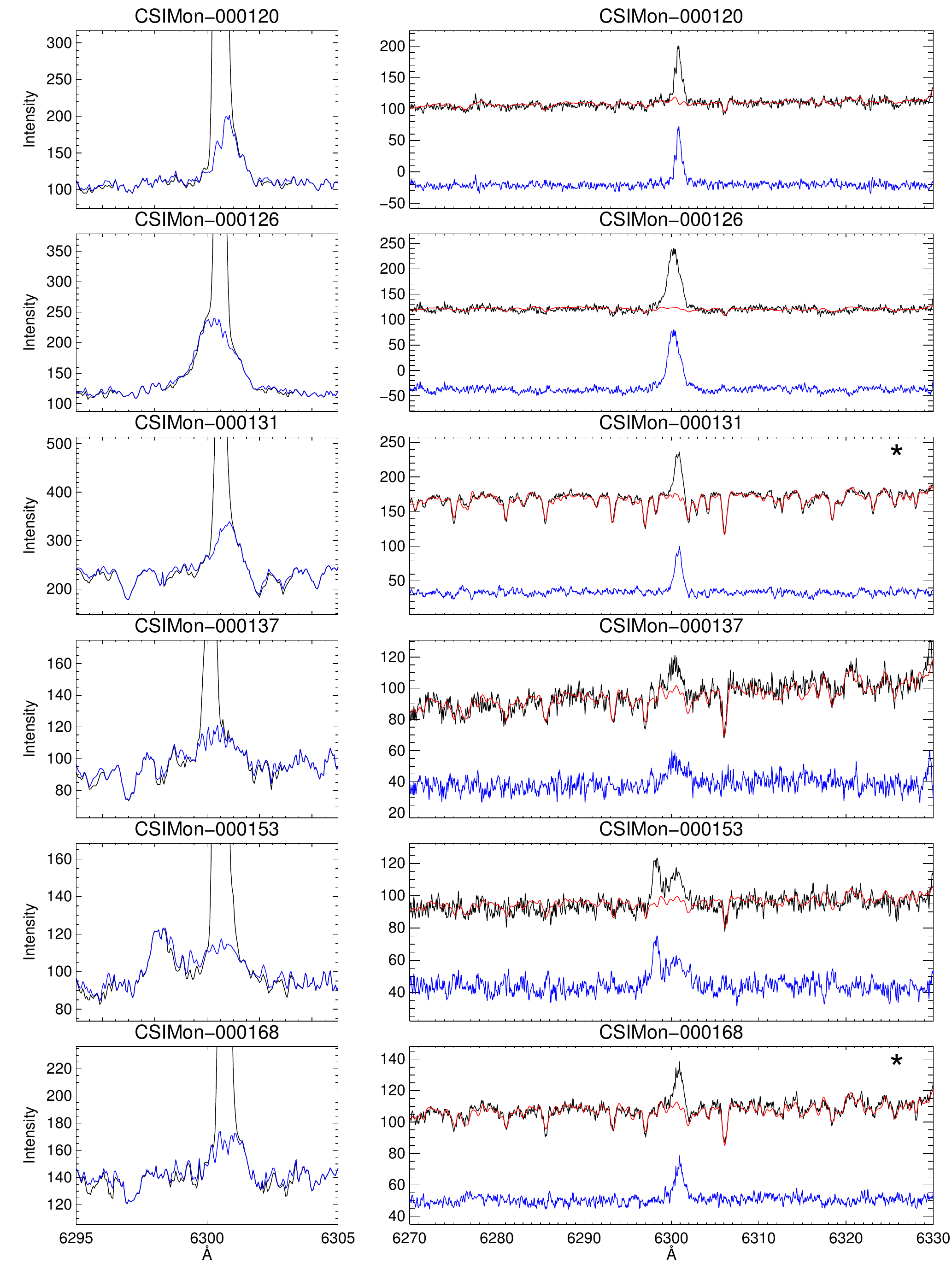}
 \caption{Continued. 
}\ContinuedFloat
\end{figure*}

\begin{figure*}[p]
 \centering
 \includegraphics[width=17cm]{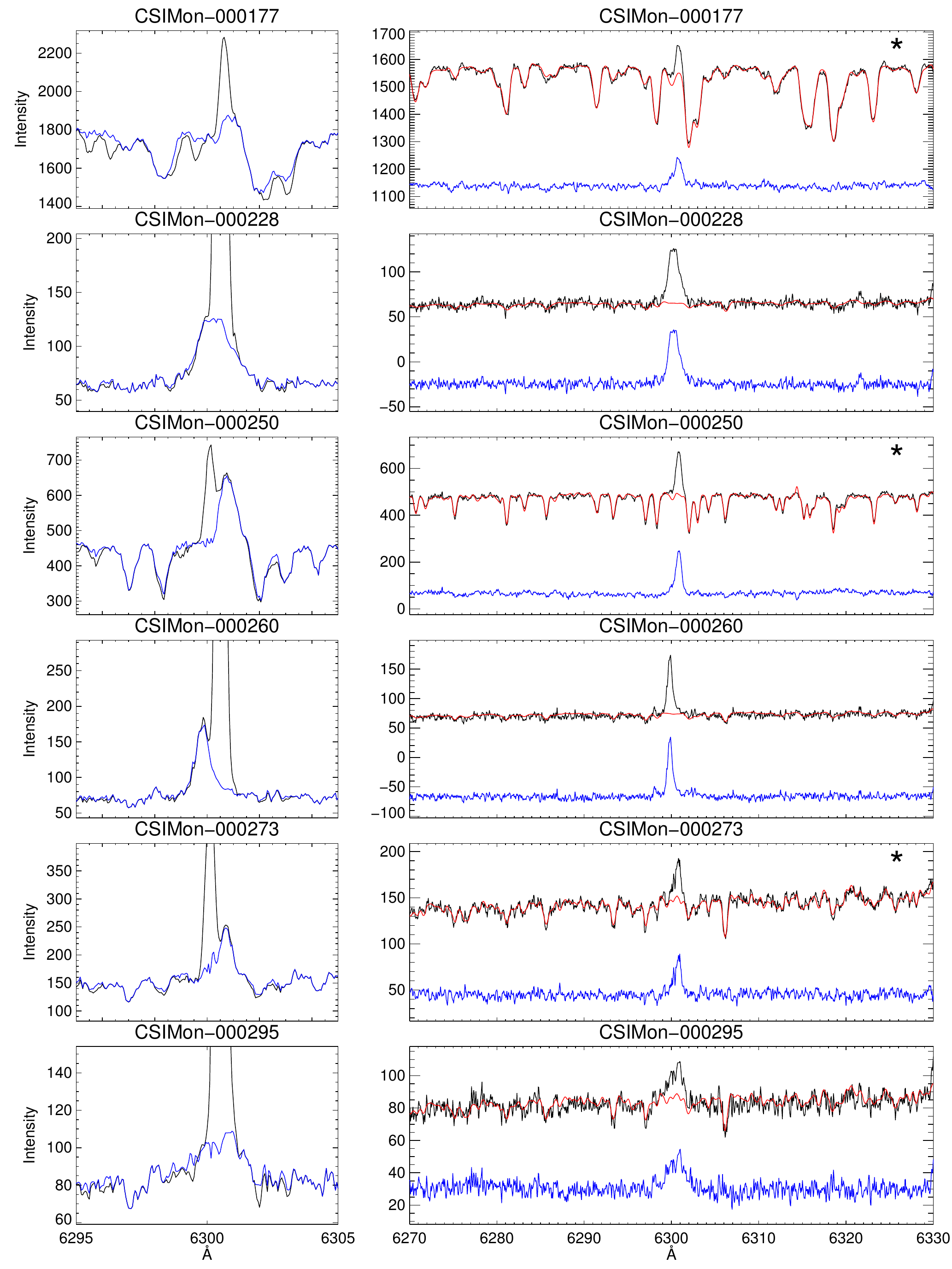}
 \caption{Continued. 
}\ContinuedFloat
\end{figure*}

\begin{figure*}[p]
 \centering
 \includegraphics[width=17cm]{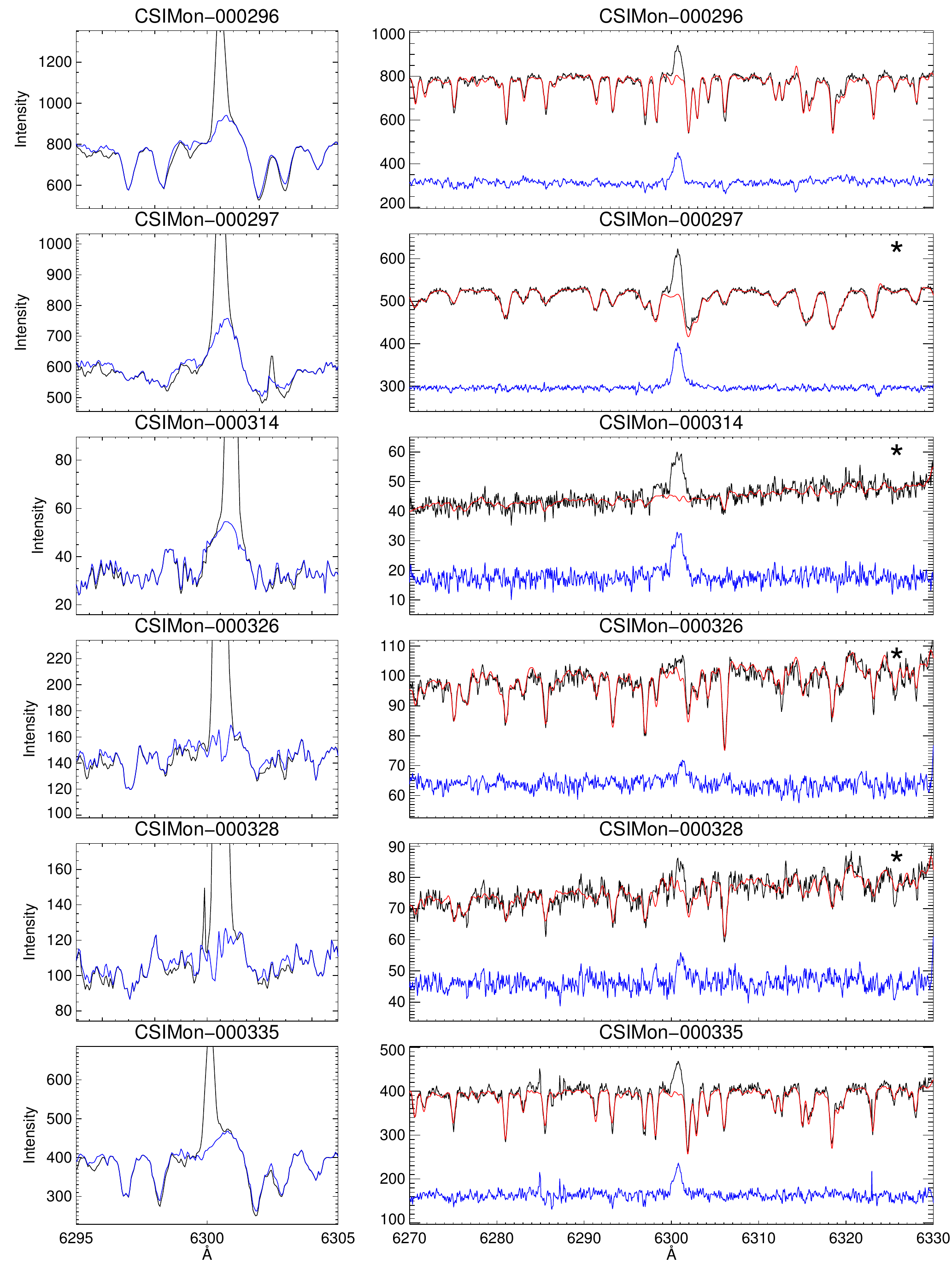}
 \caption{Continued. 
}\ContinuedFloat
\end{figure*}

\begin{figure*}[p]
 \centering
 \includegraphics[width=17cm]{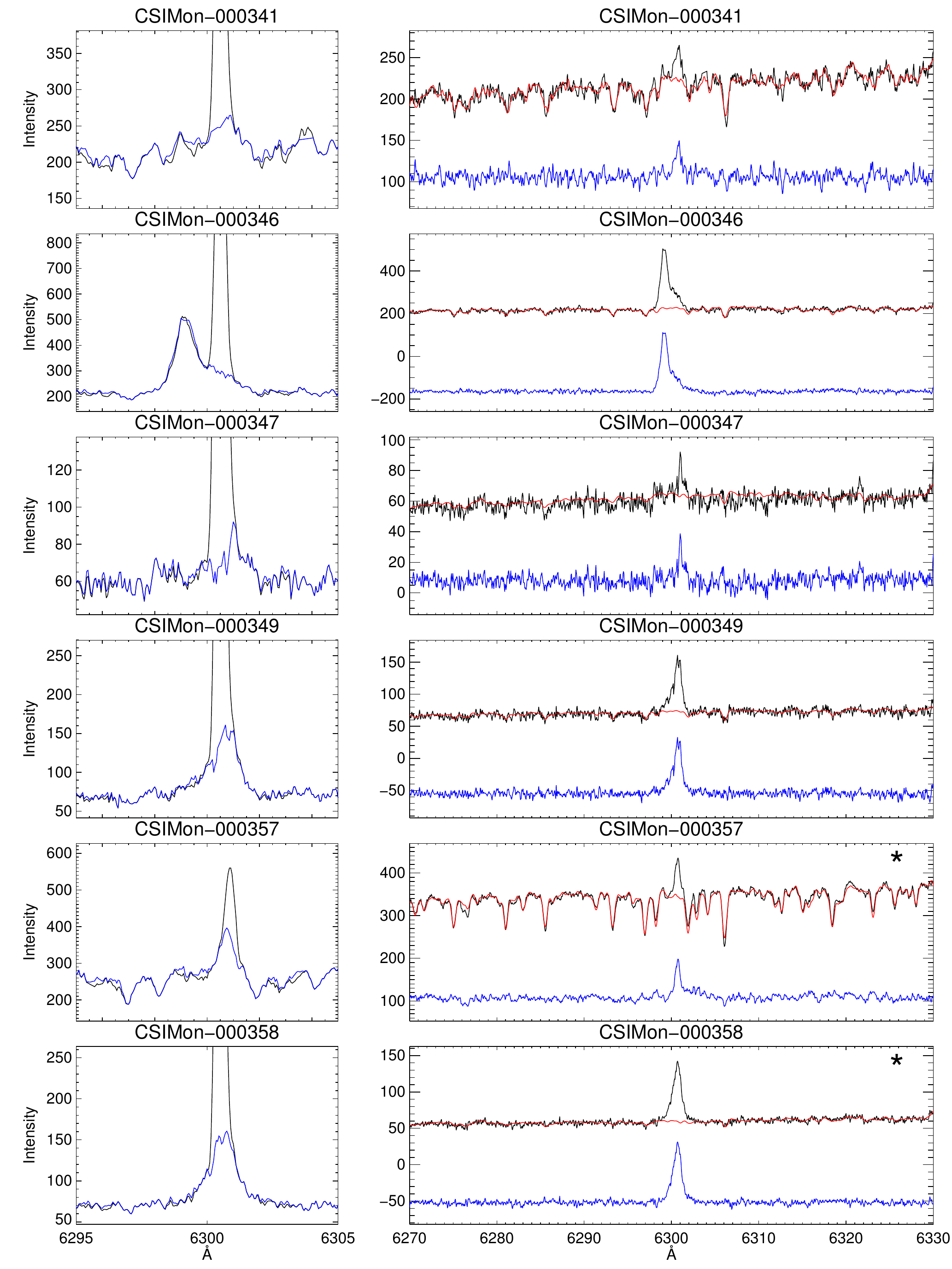}
 \caption{Continued. 
}\ContinuedFloat
\end{figure*}

\begin{figure*}[p]
 \centering
 \includegraphics[width=17cm]{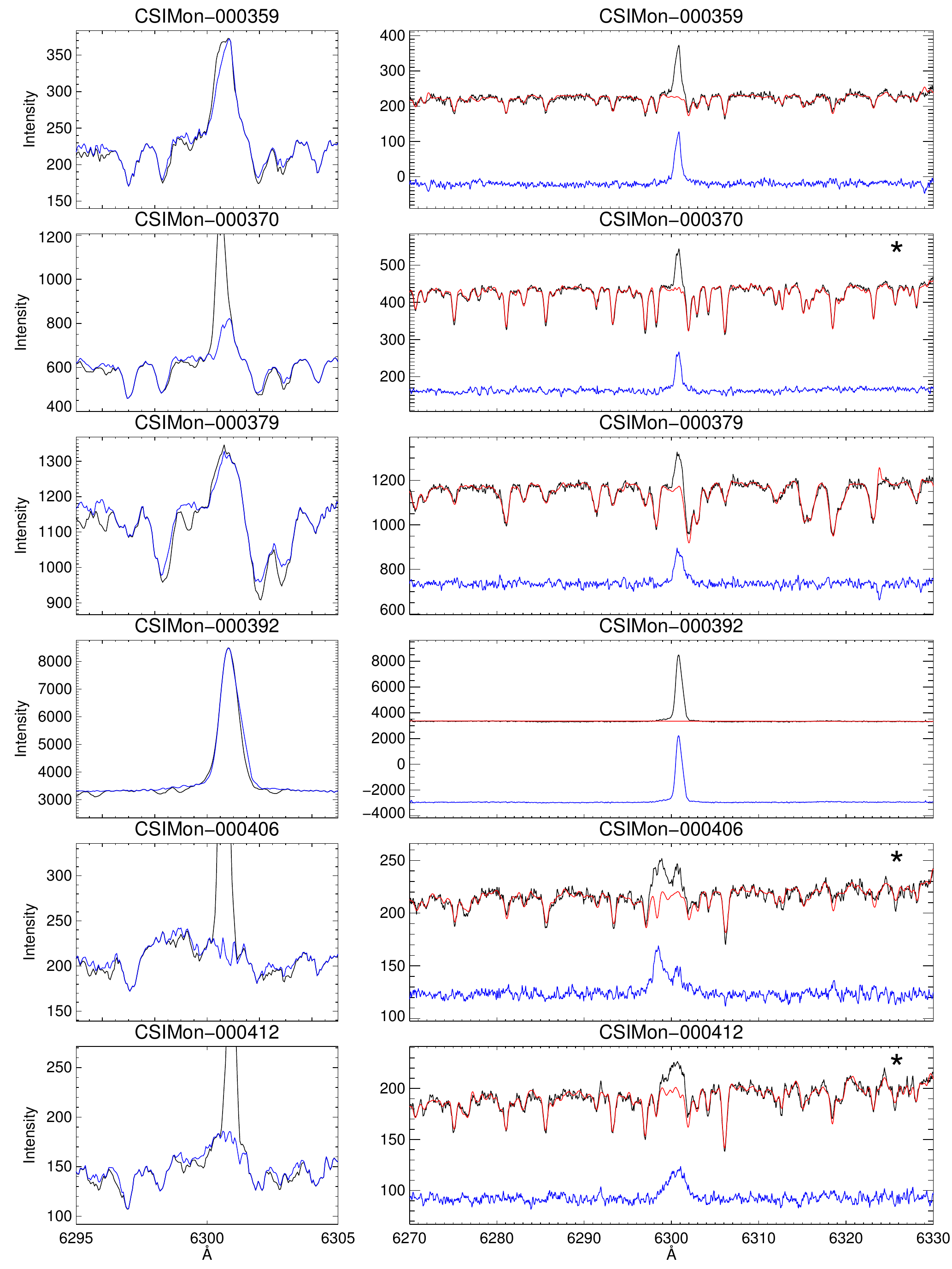}
 \caption{Continued. 
}\ContinuedFloat
\end{figure*}

\begin{figure*}[p]
 \centering
 \includegraphics[width=17cm]{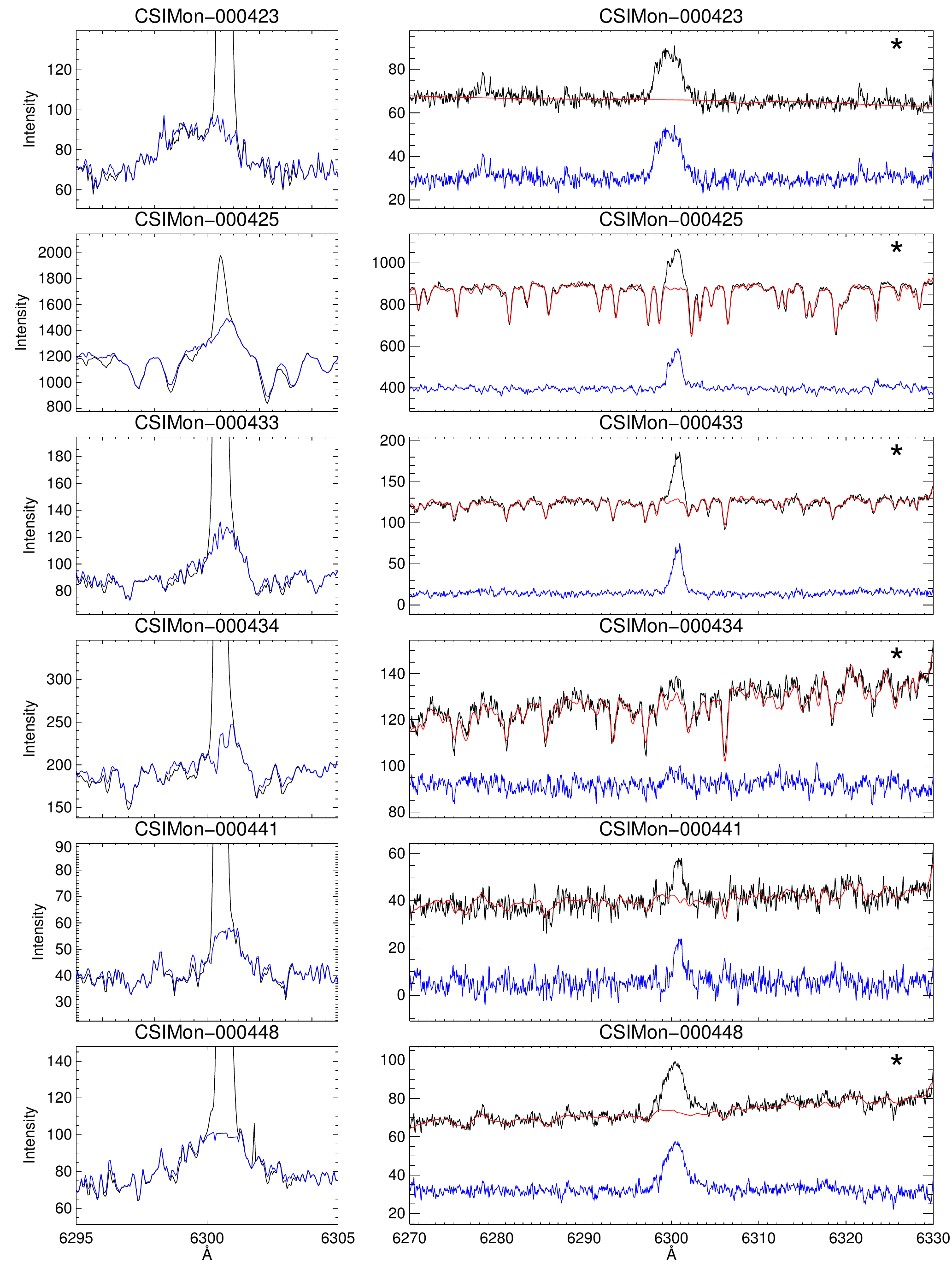}
 \caption{Continued. 
}\ContinuedFloat
\end{figure*}

\begin{figure*}[p]
 \centering
 \includegraphics[width=17cm]{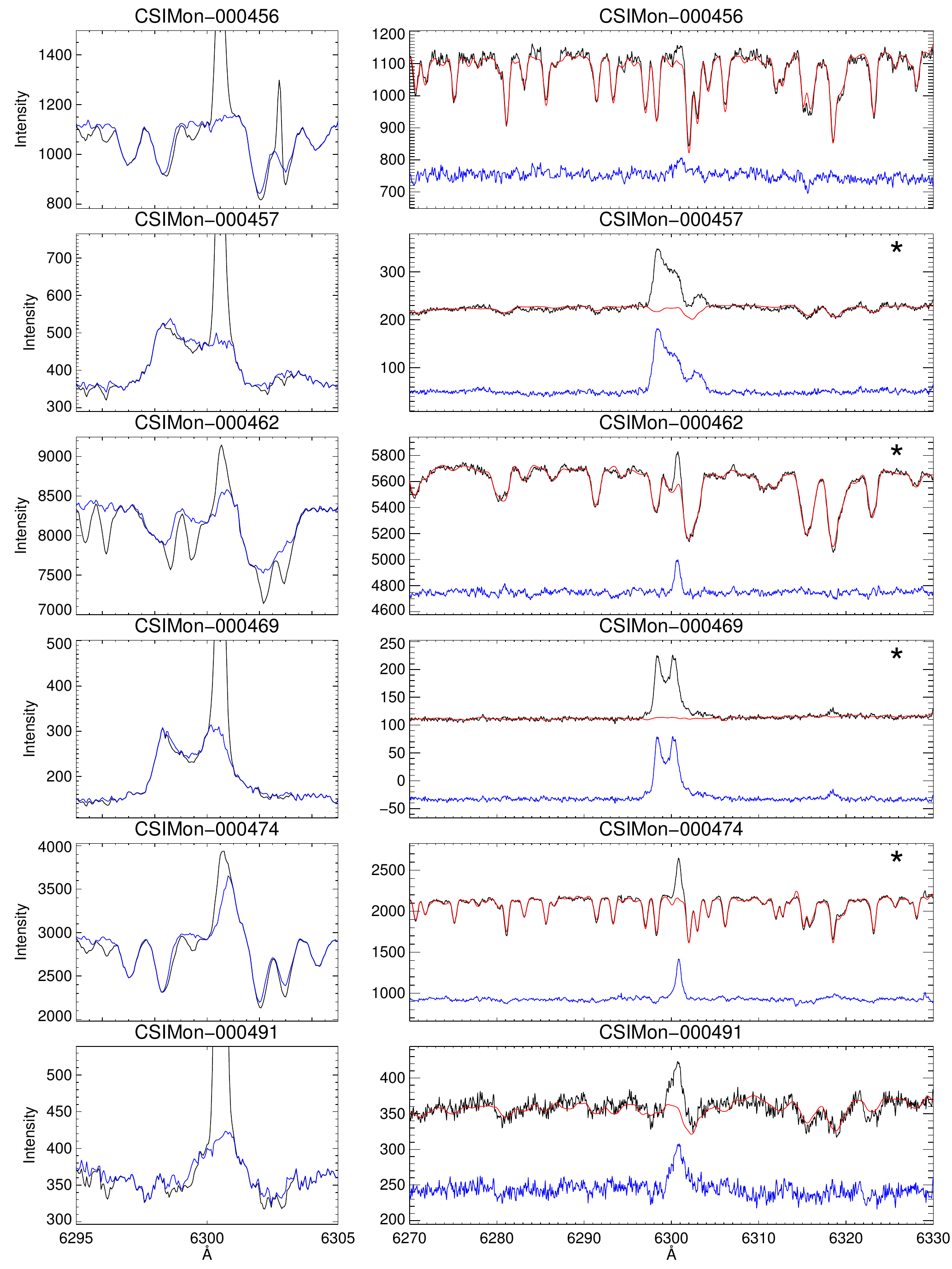}
 \caption{Continued. 
}\ContinuedFloat
\end{figure*}

\begin{figure*}[p]
 \centering
 \includegraphics[width=17cm]{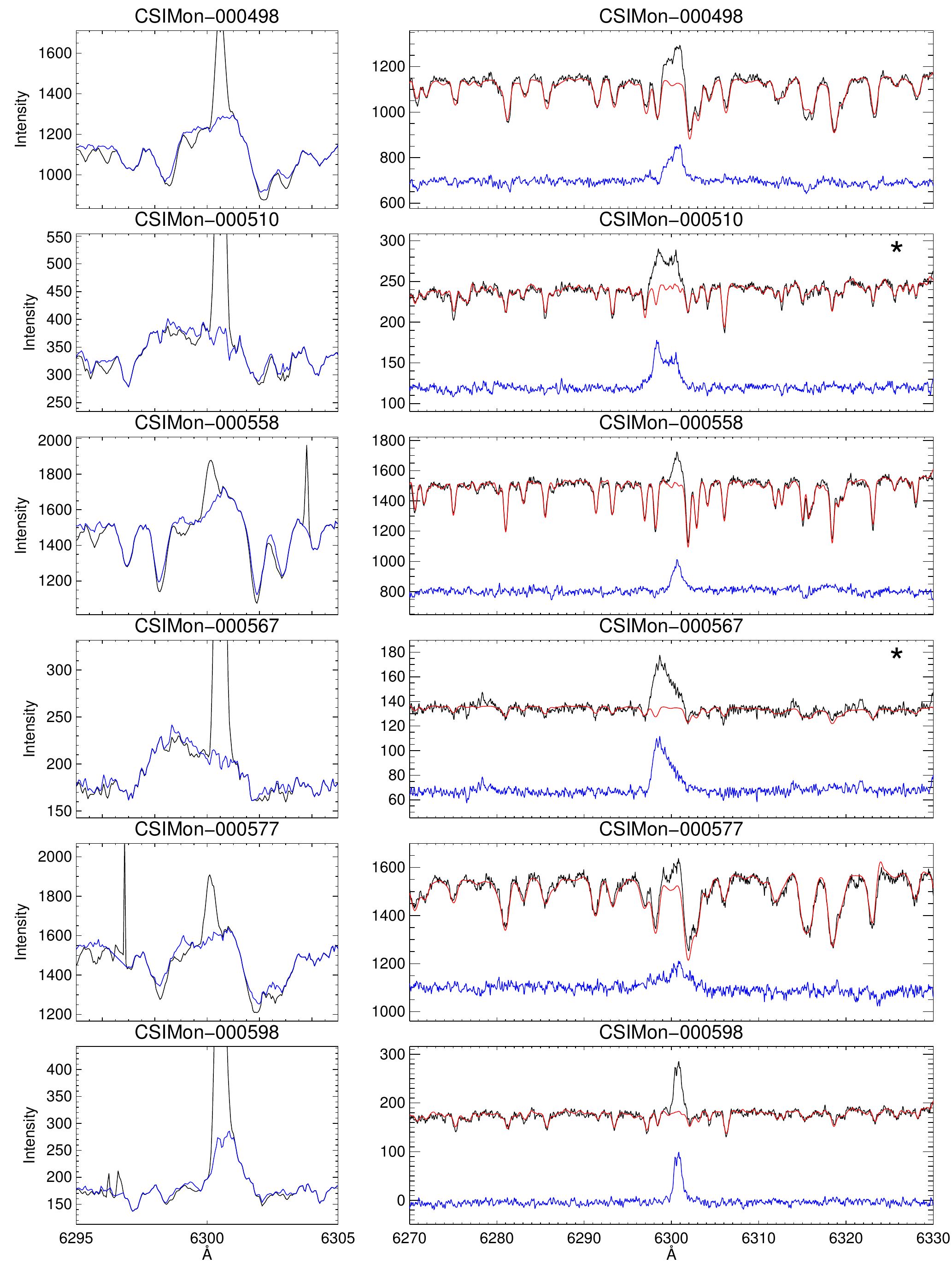}
 \caption{Continued. 
}\ContinuedFloat
\end{figure*}

\begin{figure*}[p]
 \centering
 \includegraphics[width=17cm]{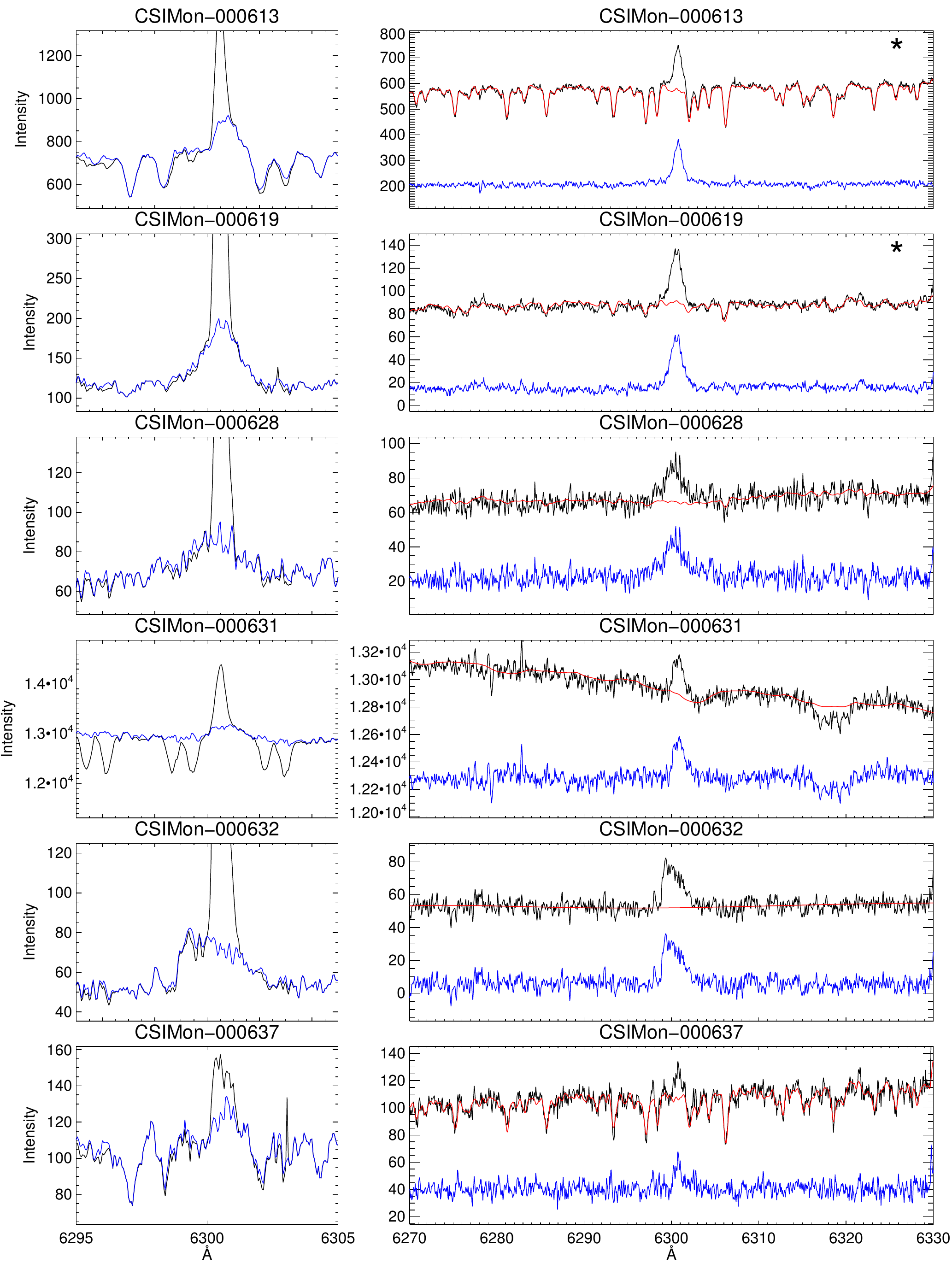}
 \caption{Continued. 
}\ContinuedFloat
\end{figure*}

\begin{figure*}[p]
 \centering
 \includegraphics[width=17cm]{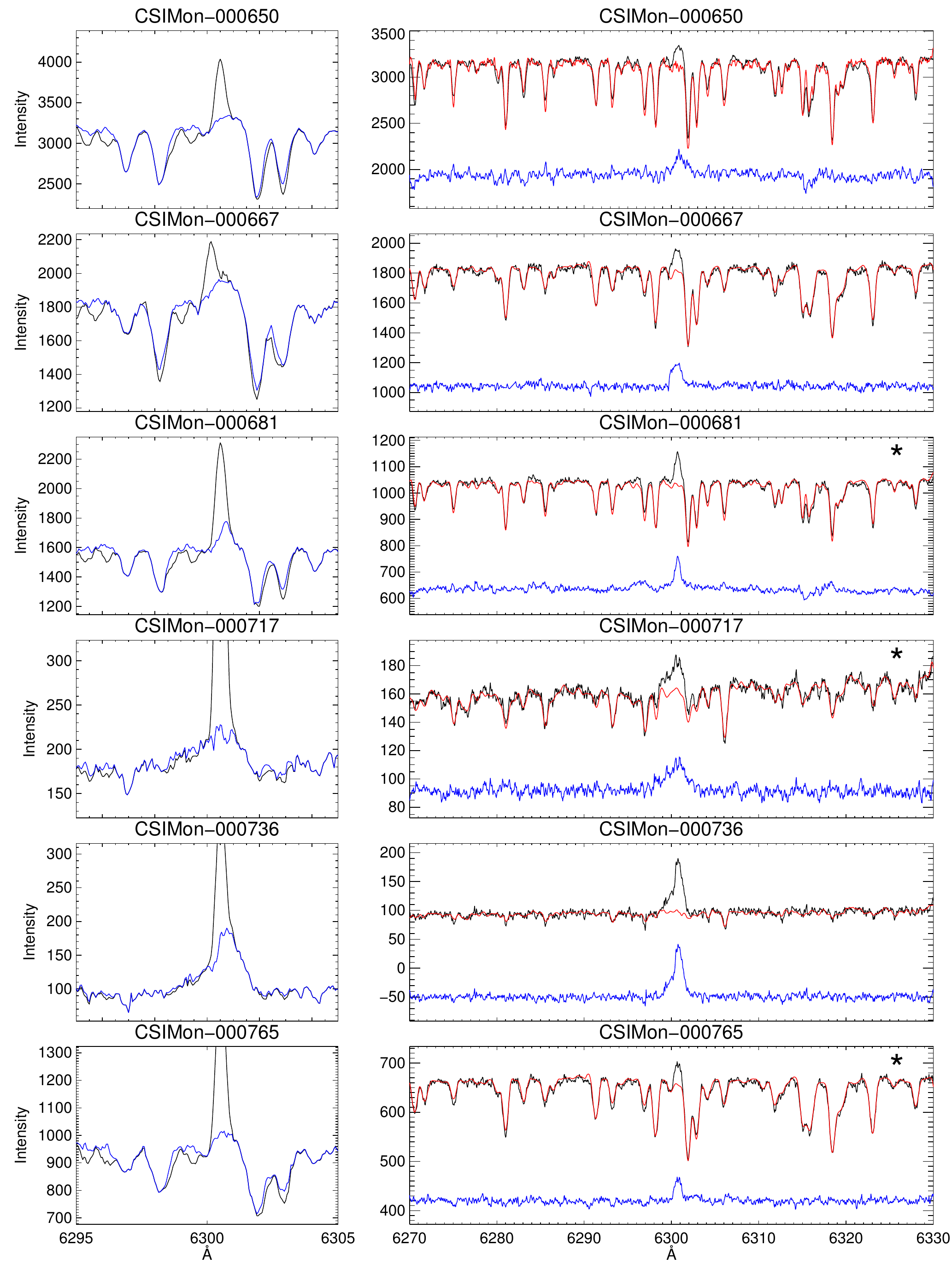}
 \caption{Continued. 
}\ContinuedFloat
\end{figure*}

\begin{figure*}[p]
 \centering
 \includegraphics[width=17cm]{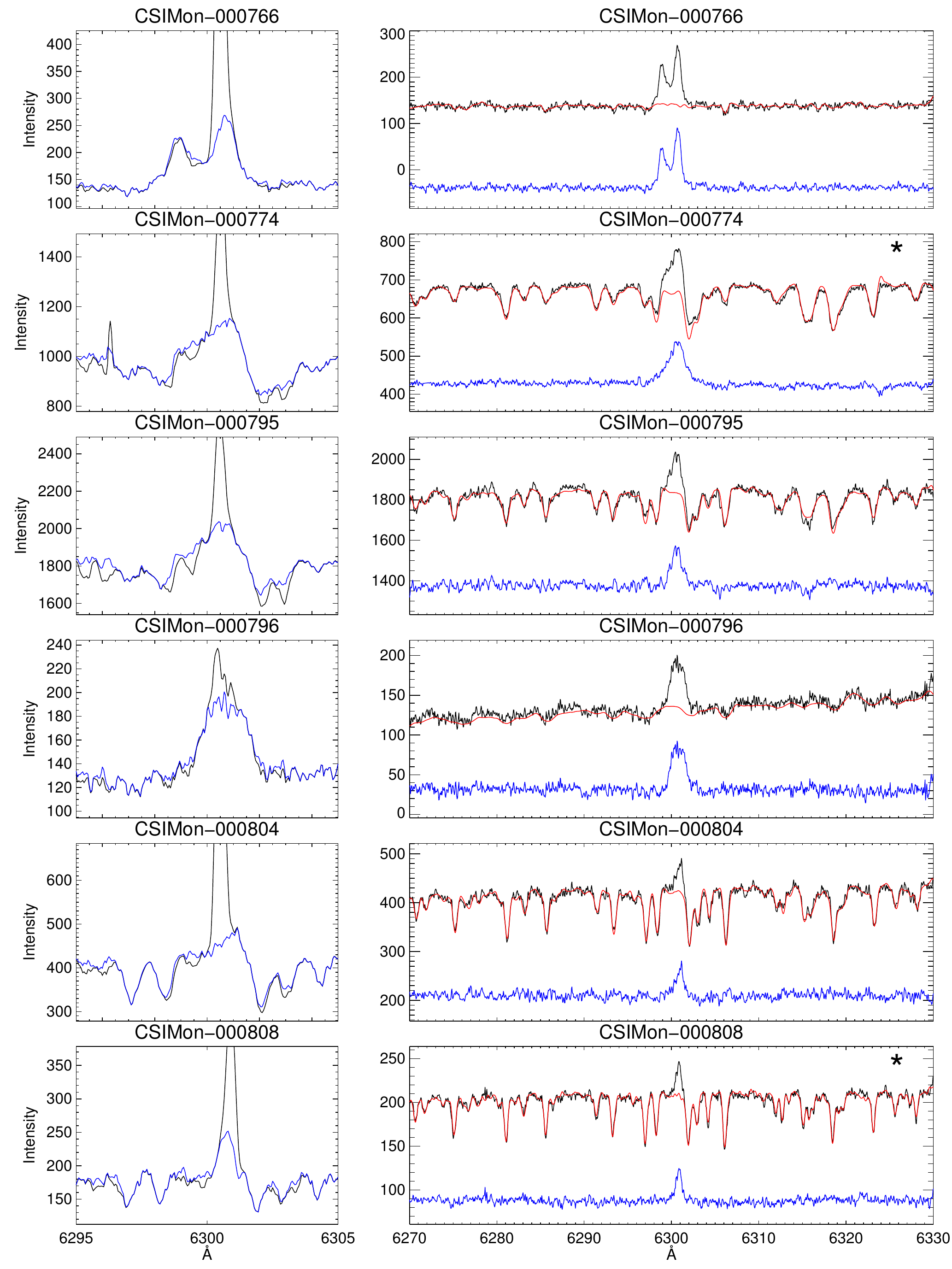}
 \caption{Continued. 
}\ContinuedFloat
\end{figure*}

\begin{figure*}[p]
 \centering
 \includegraphics[width=17cm]{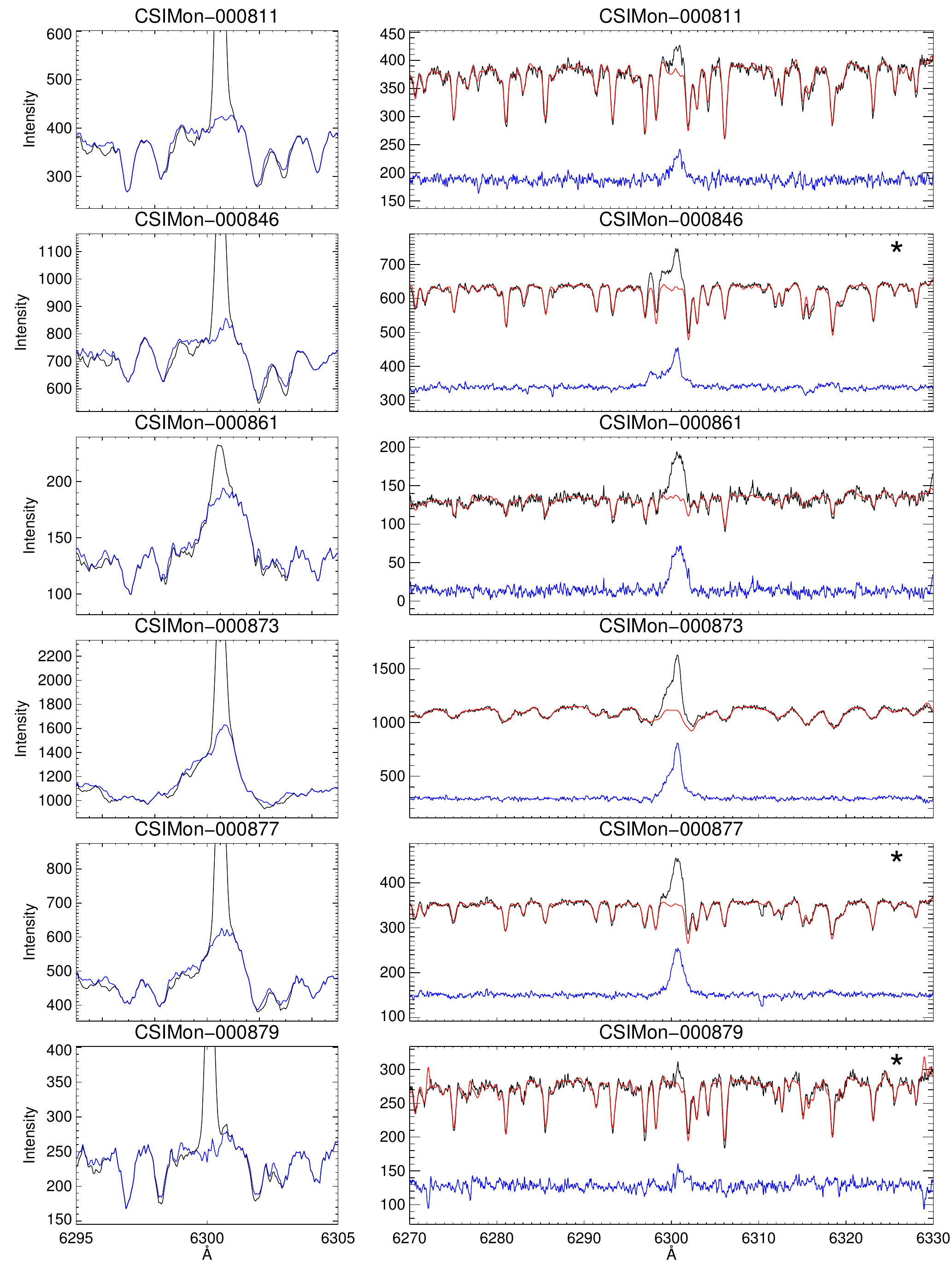}
 \caption{Continued. 
}\ContinuedFloat
\end{figure*}

\begin{figure*}[p]
 \centering
 \includegraphics[width=17cm]{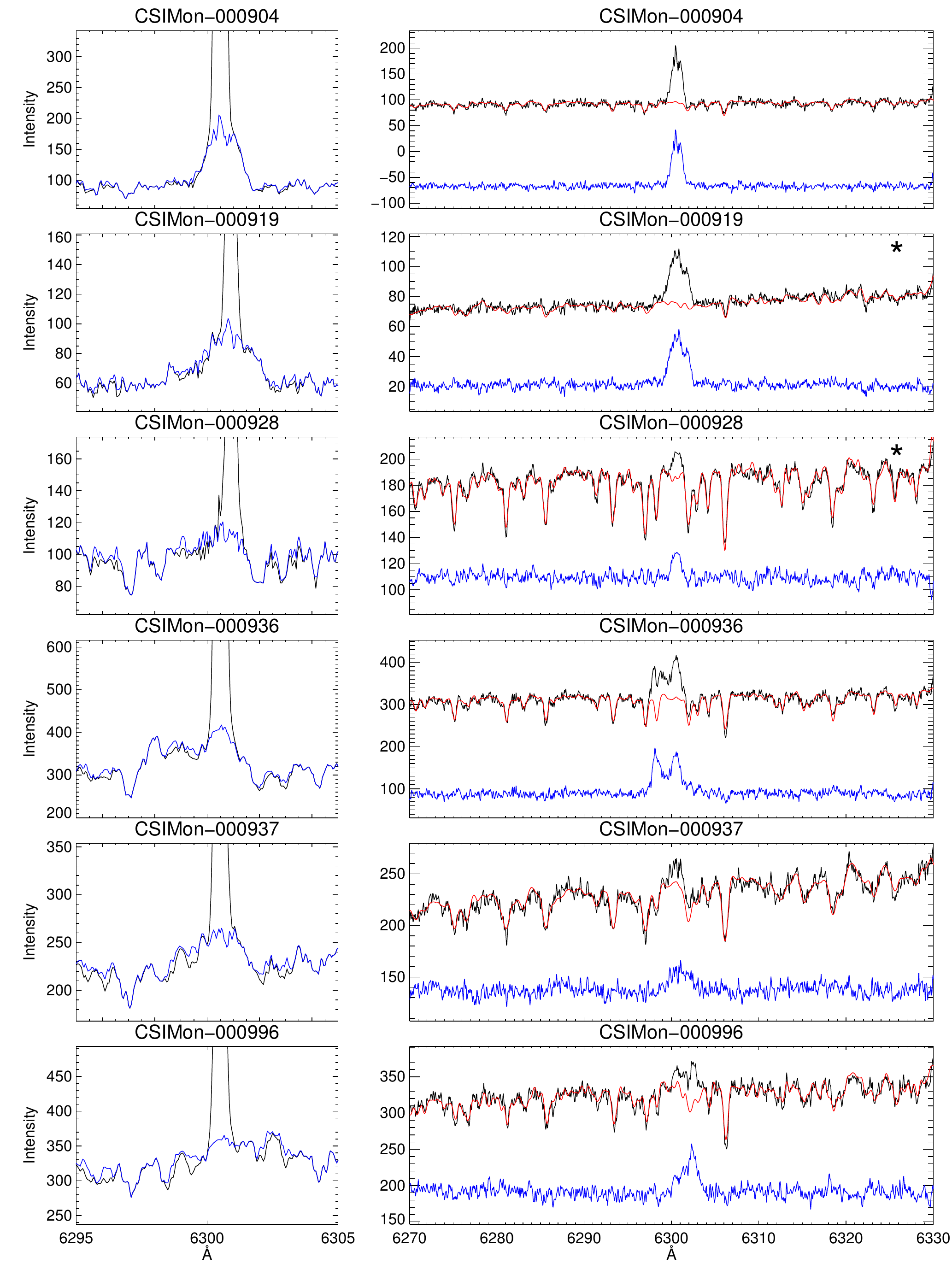}
 \caption{Continued. 
}\ContinuedFloat
\end{figure*}

\begin{figure*}[p]
 \centering
 \includegraphics[width=17cm]{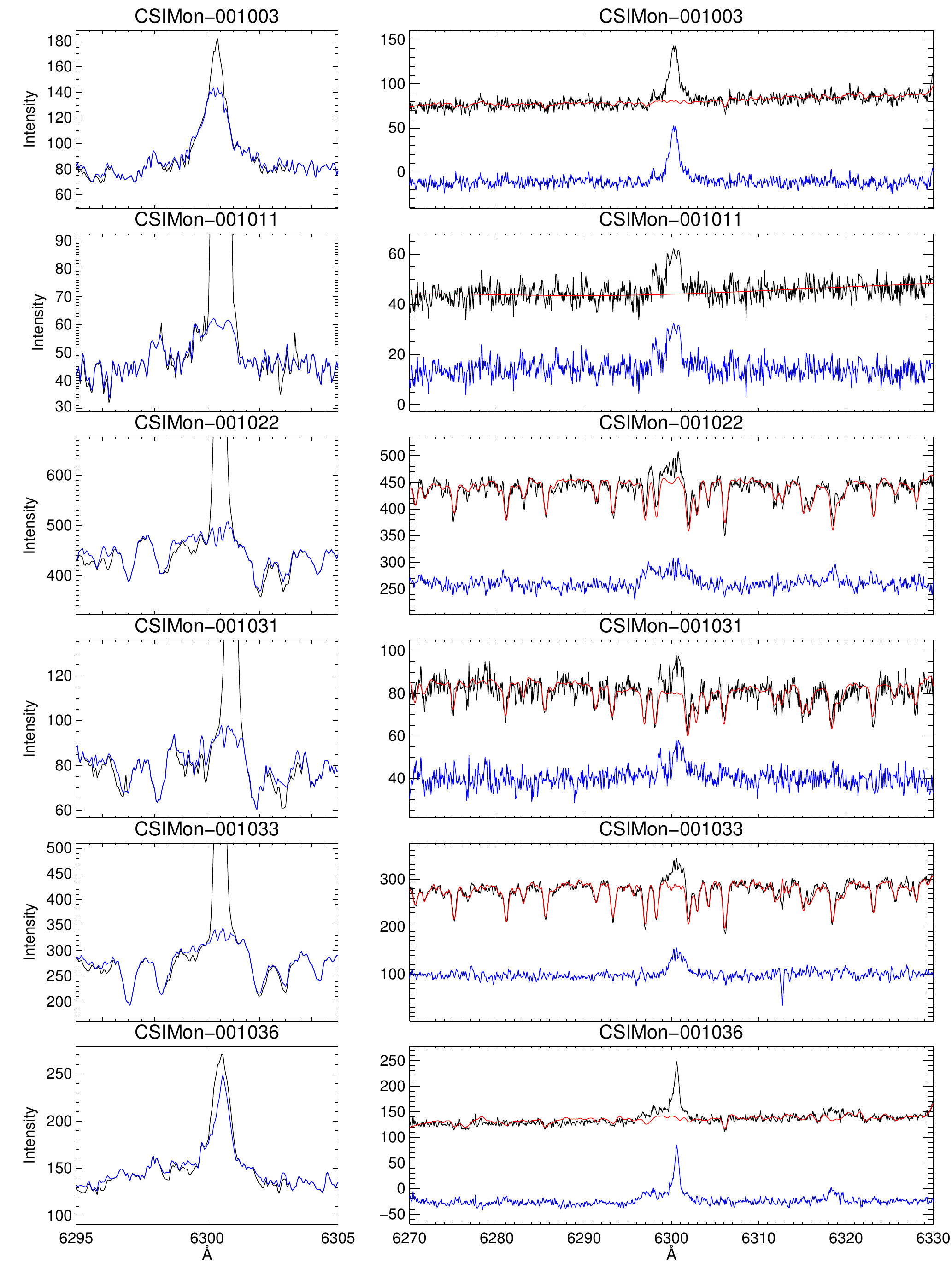}
 \caption{Continued. 
}\ContinuedFloat
\end{figure*}

\begin{figure*}[p]
 \centering
 \includegraphics[width=17cm]{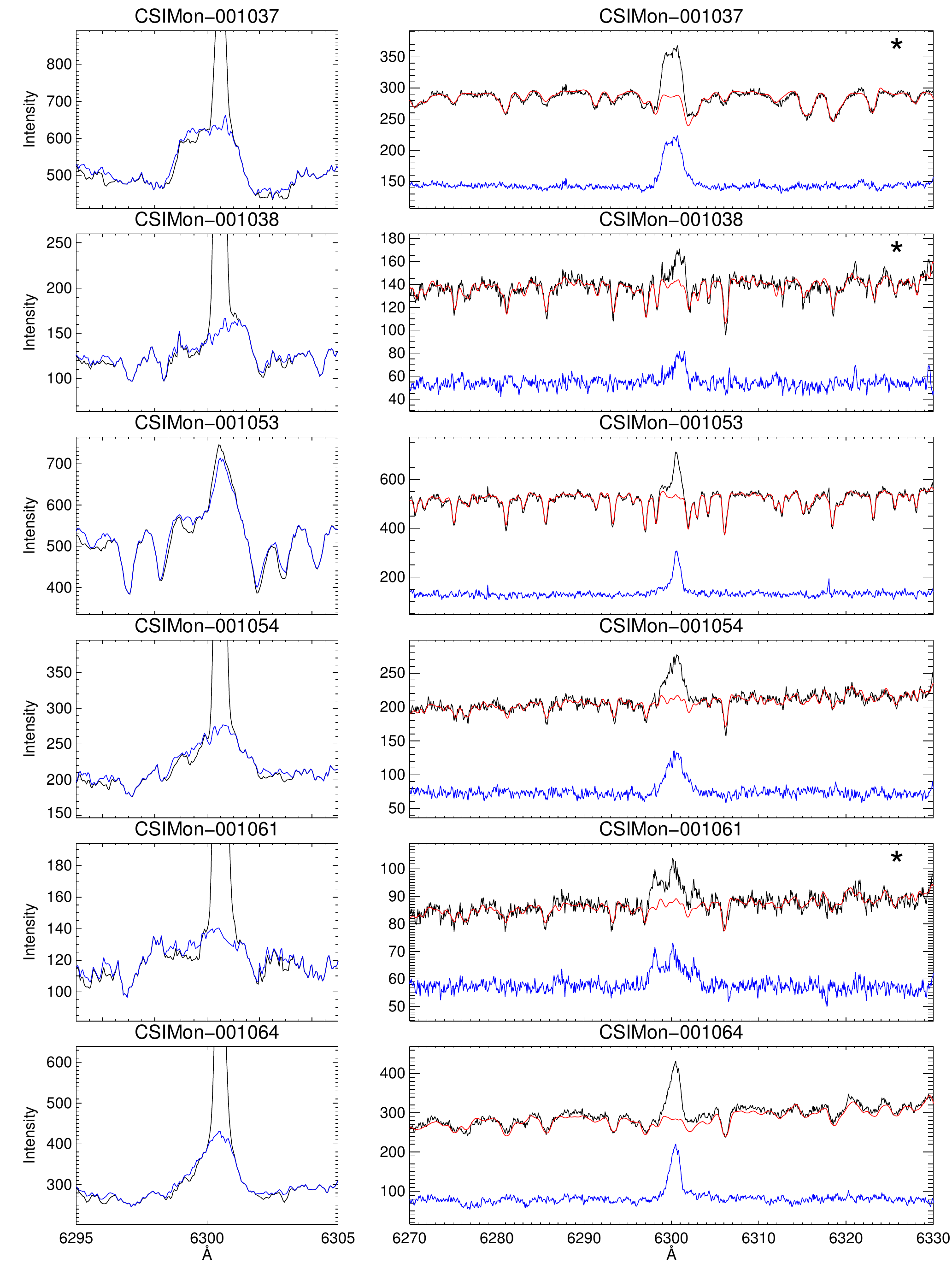}
 \caption{Continued. 
}\ContinuedFloat
\end{figure*}

\begin{figure*}[p]
 \centering
 \includegraphics[width=17cm]{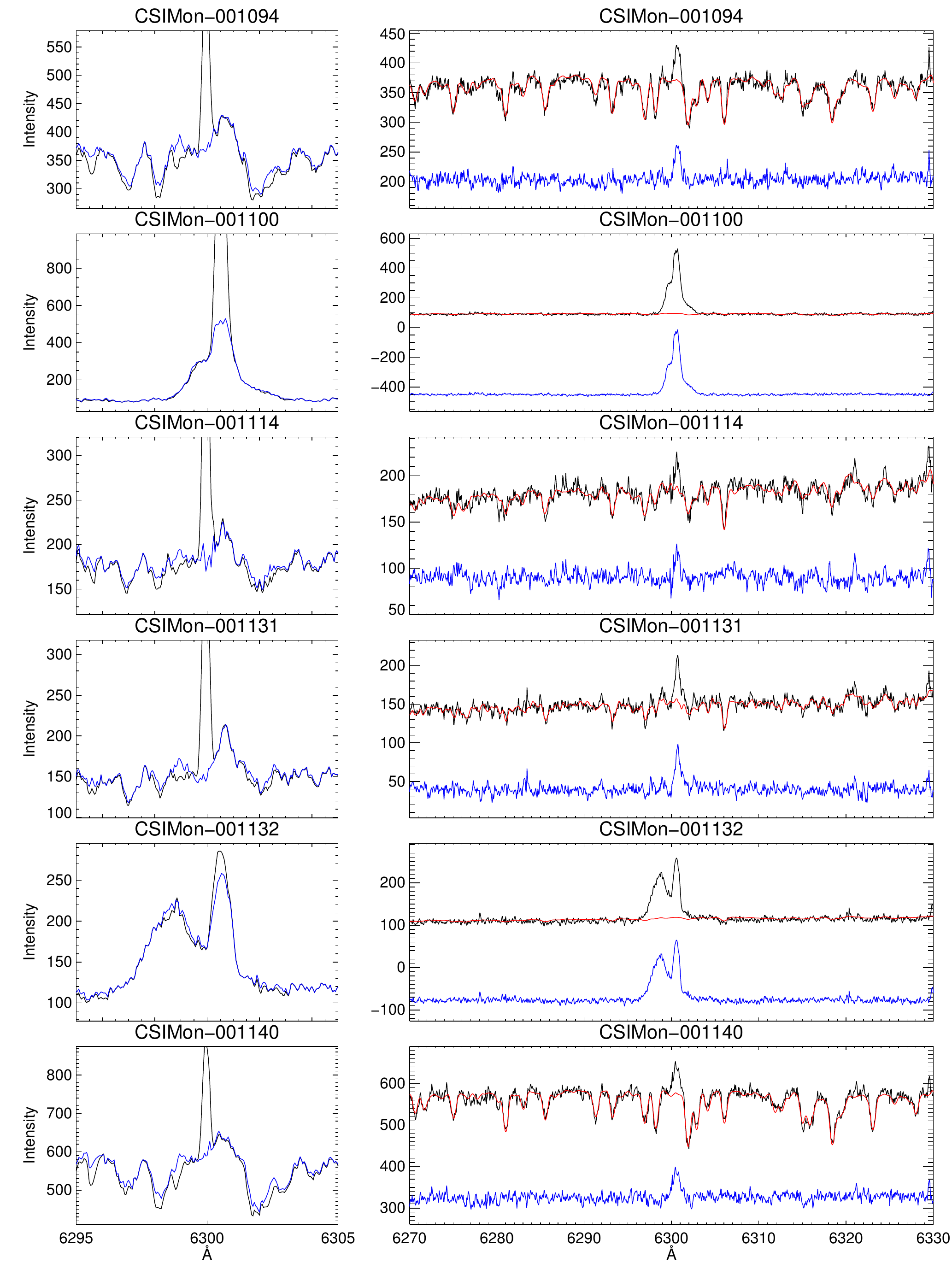}
 \caption{Continued. 
}\ContinuedFloat
\end{figure*}

\begin{figure*}[p]
 \centering
 \includegraphics[width=17cm]{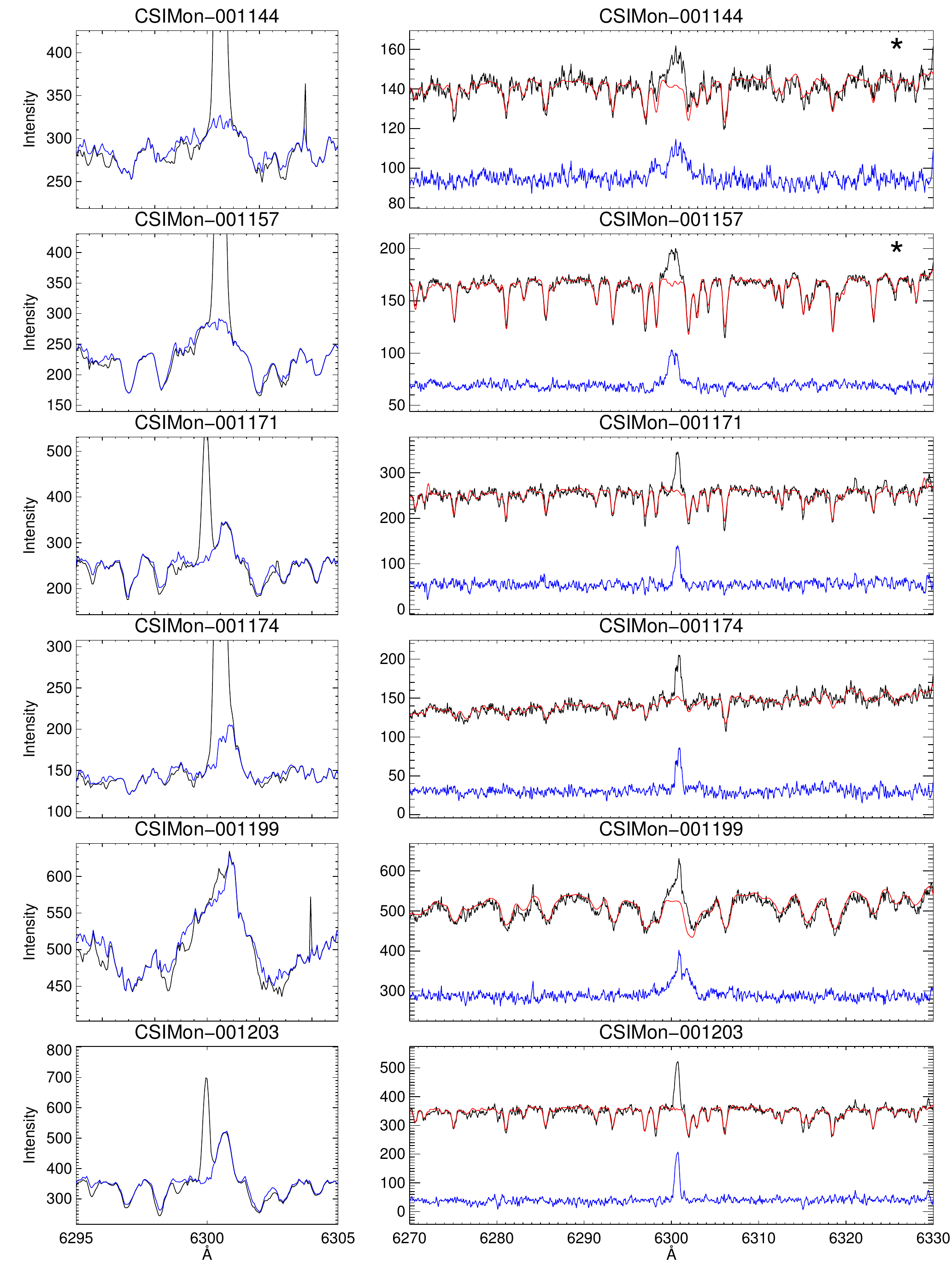}
 \caption{Continued. 
}\ContinuedFloat
\end{figure*}

\begin{figure*}[p]
 \centering
 \includegraphics[width=17cm]{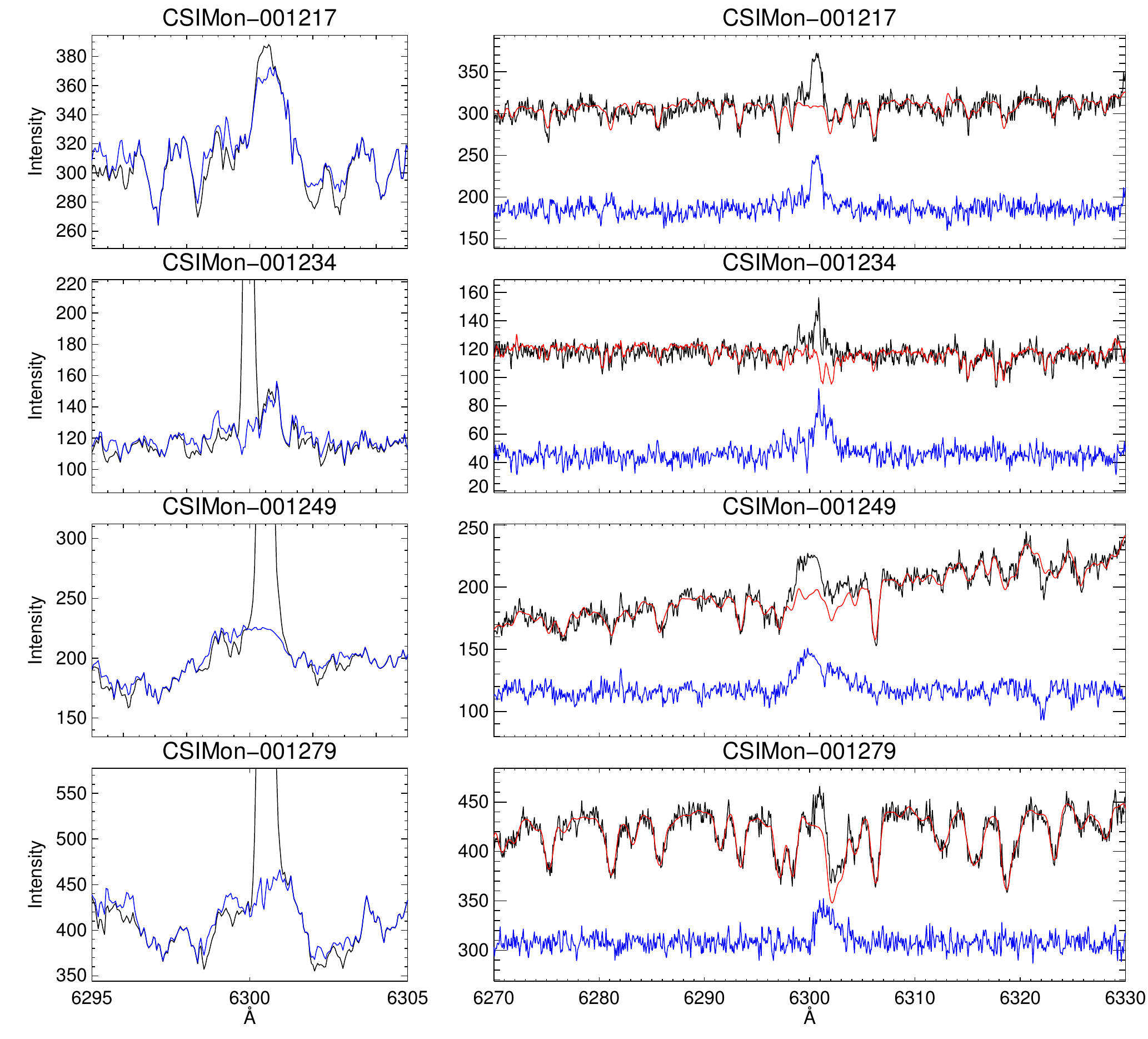}
 \caption{Continued. 
}
\end{figure*}

\begin{figure*}[p]
 \centering
 \includegraphics[width=17cm]{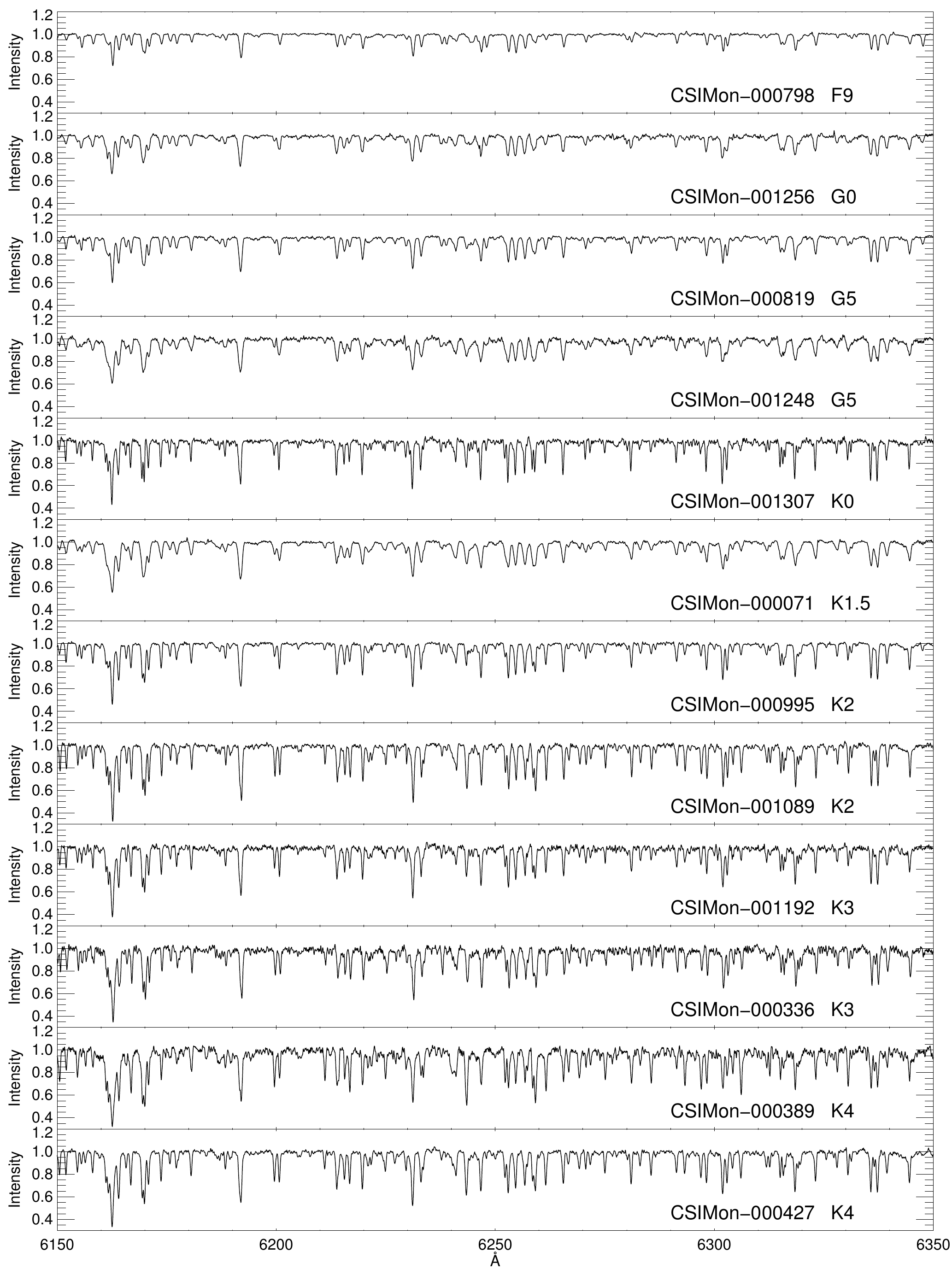}
 \caption{WTTSs used as photospheric templates, ordered by spectral type (as given in the literature). 
}\ContinuedFloat
\label{fig:specwtts}
\end{figure*}

\begin{figure*}[p]
 \centering
 \includegraphics[width=17cm]{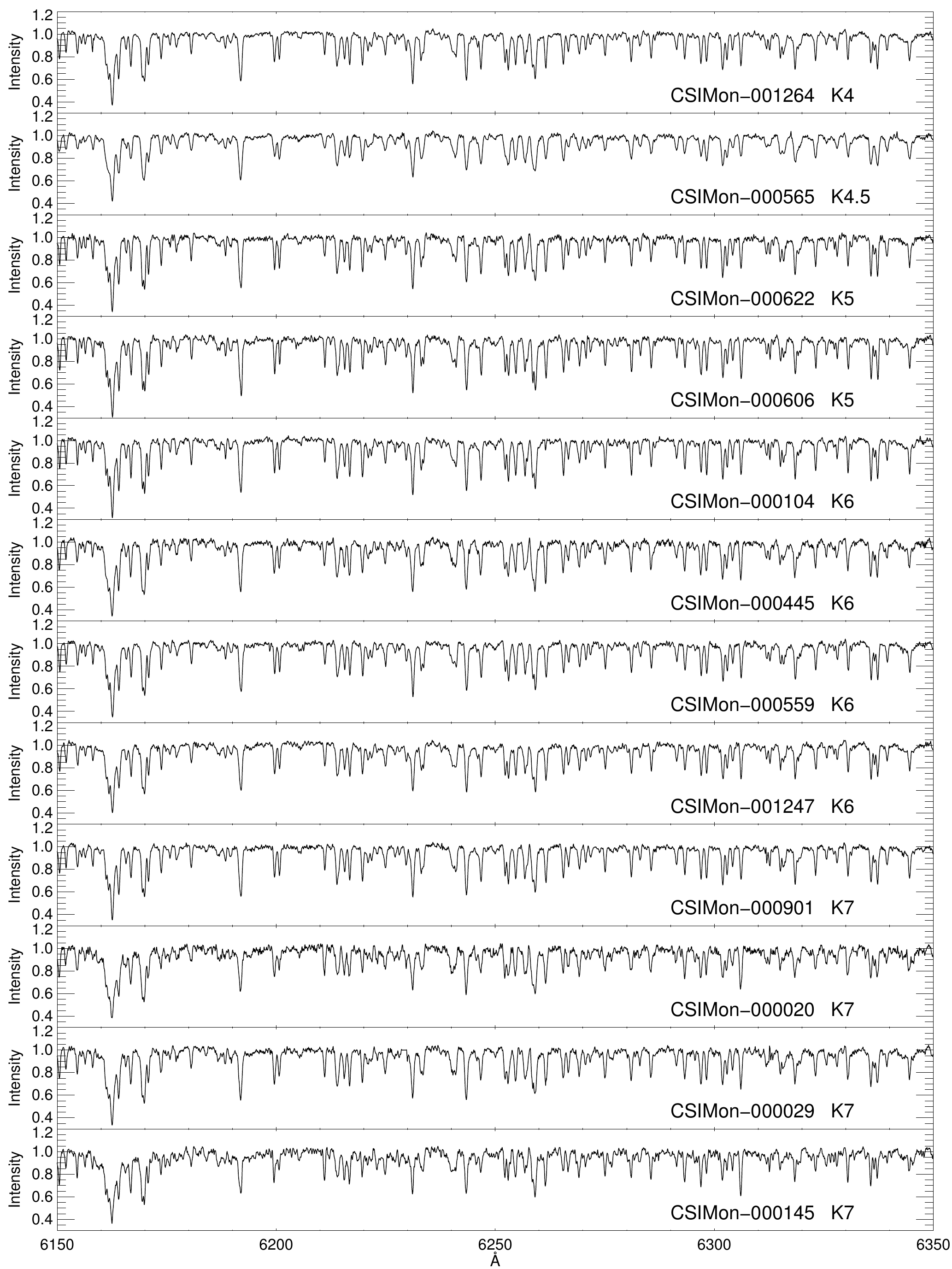}
 \caption{Continued.  
}\ContinuedFloat
\end{figure*}

\begin{figure*}[p]
 \centering
 \includegraphics[width=17cm]{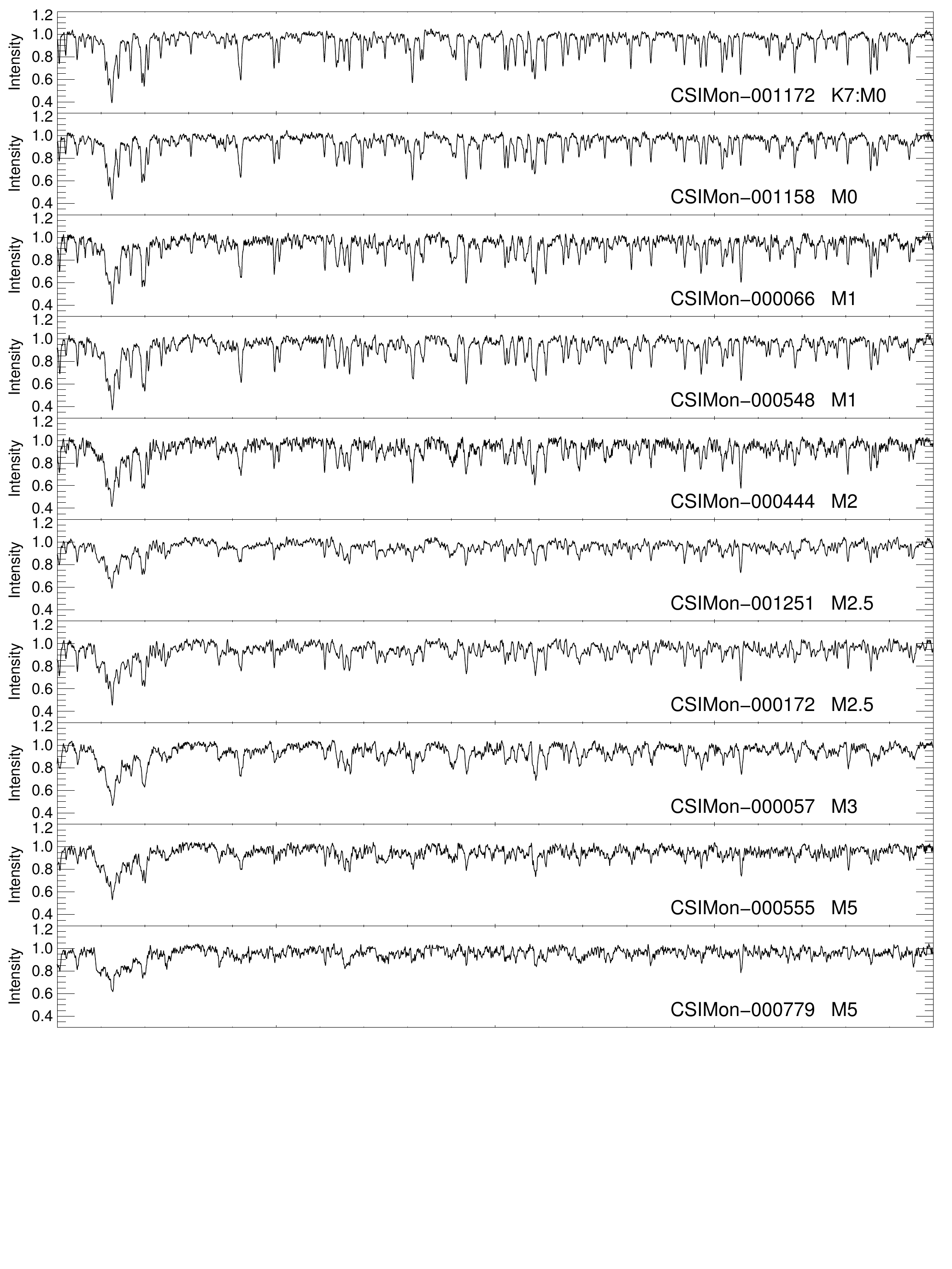}
 \caption{Continued.  
}
\end{figure*}

\begin{figure*}[p]
 \centering
 \includegraphics[width=17cm]{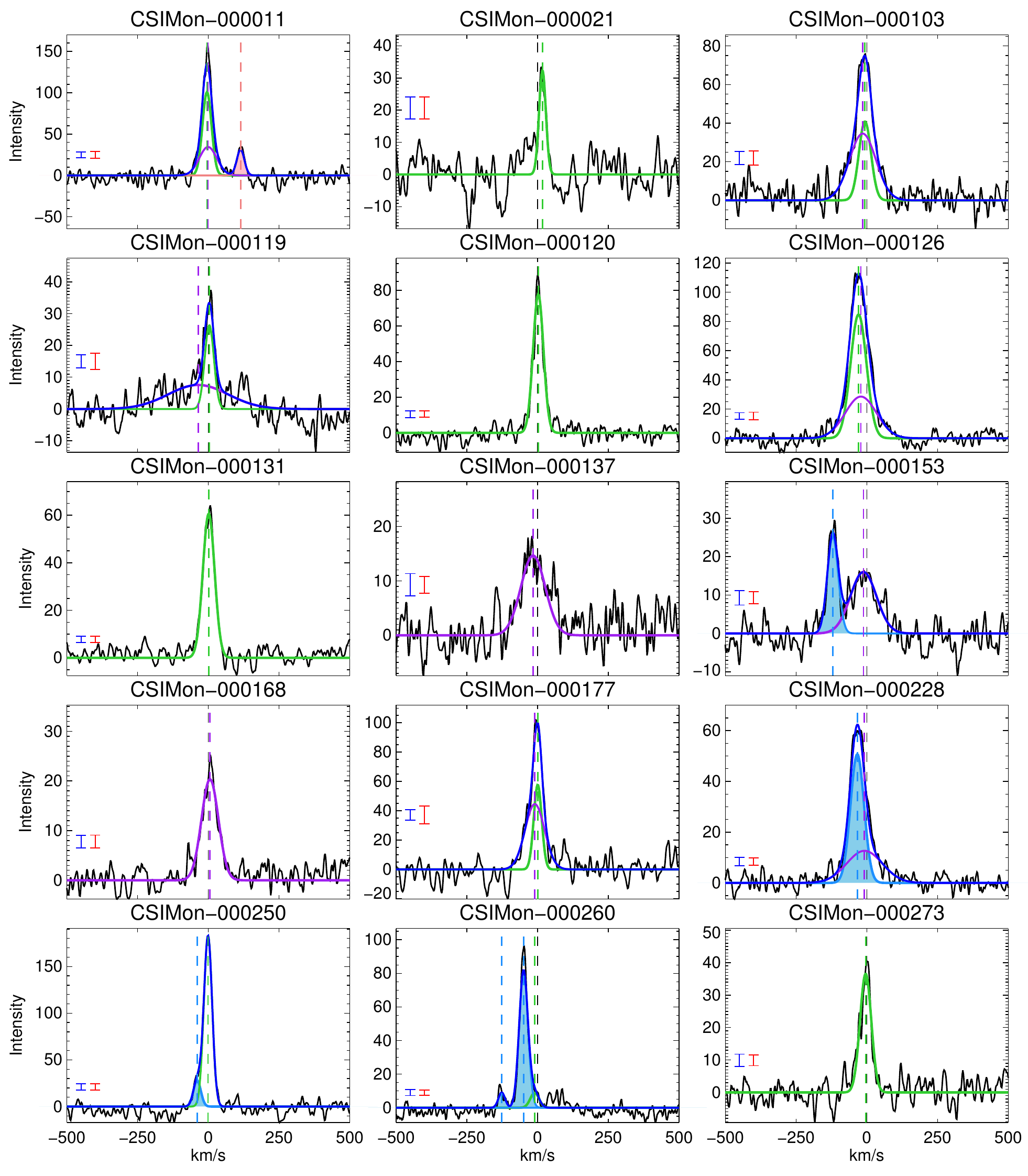}
 \caption{Gaussian decomposition of the [OI]$\lambda$6300 line profiles. Individual Gaussians are 
shown in different colors: in green are the narrow LVCs; in purple the broad LVCs; light blue shaded 
components represent blueshifted HVCs; and pink shaded components represent redshifted HVCs. The sum 
of all Gaussians is shown in blue. These are all plotted over the observed profile, which is in black. 
A red error bar in the bottom left corner of each panel shows the 1$\sigma$ level of residue 
of the final spectrum (corrected for telluric and photospheric contribution), while the blue error 
bar shows the 1$\sigma$ level of the noise in the continuum of the original, uncorrected spectrum. 
}\ContinuedFloat
\label{fig:gausdecomp}
\end{figure*}

\begin{figure*}[p]
 \centering
 \includegraphics[width=17cm]{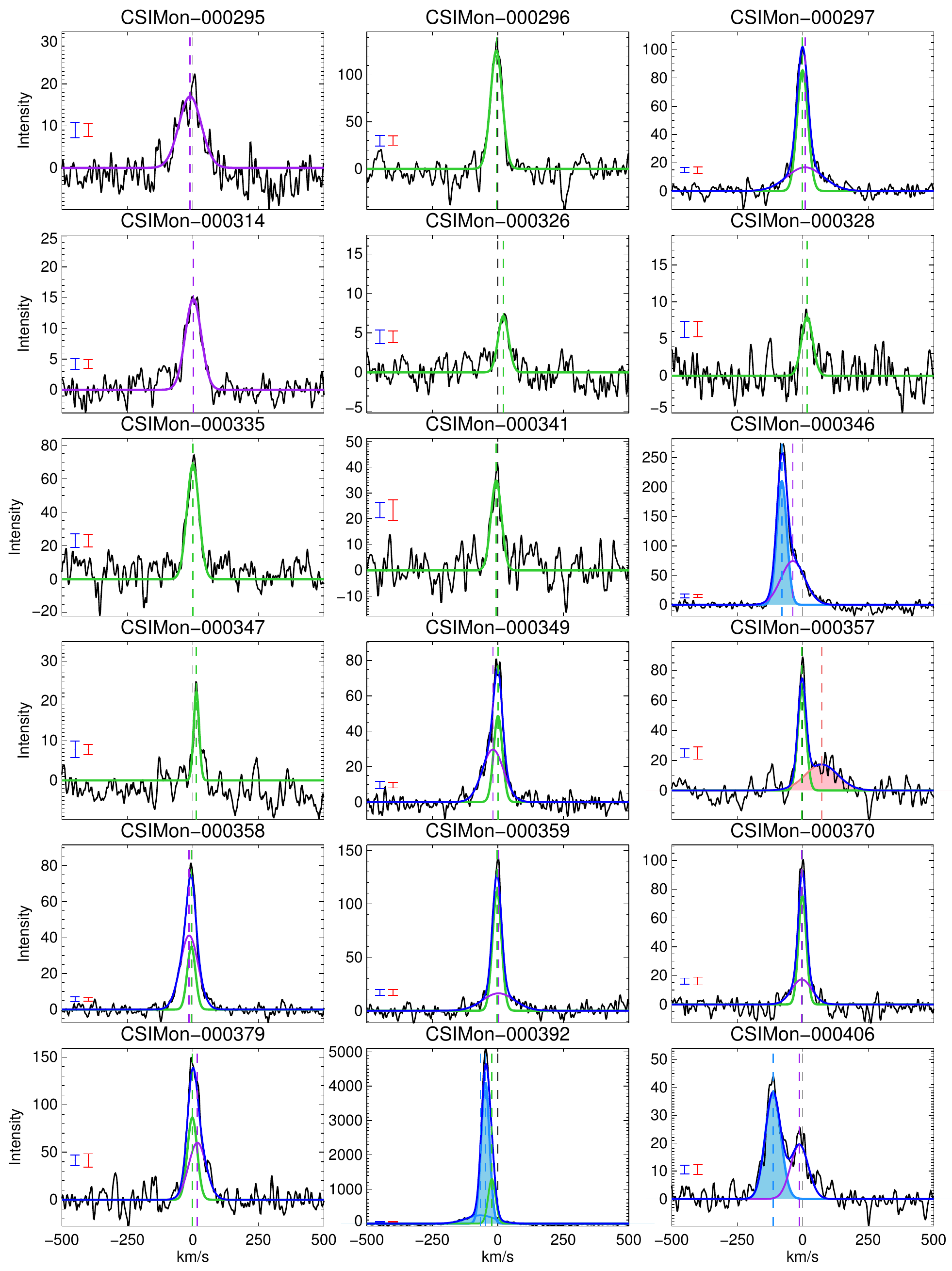}
 \caption{Continued. 
}\ContinuedFloat
\end{figure*}

\begin{figure*}[p]
 \centering
 \includegraphics[width=17cm]{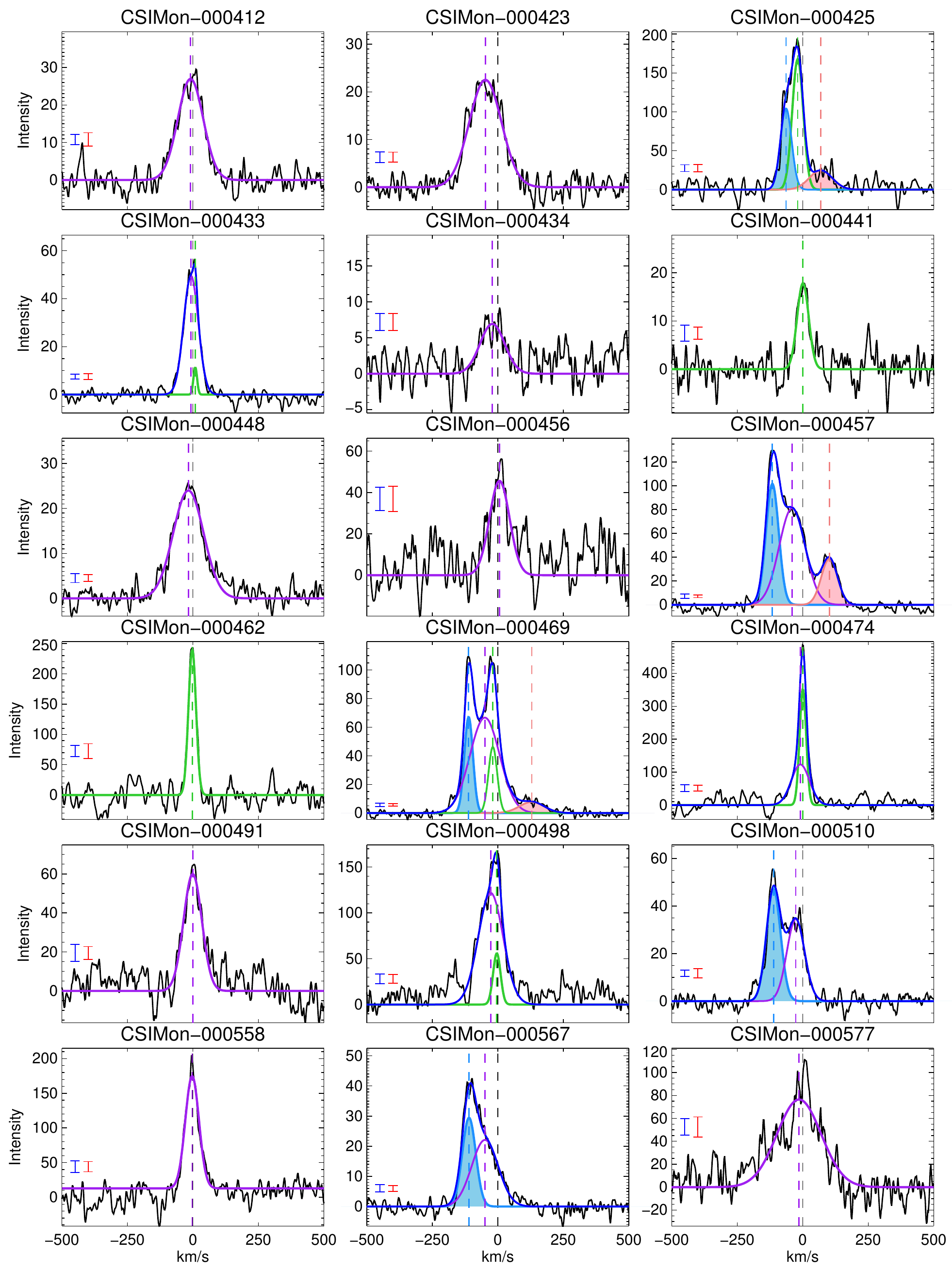}
 \caption{Continued. 
}\ContinuedFloat
\end{figure*}

\begin{figure*}[p]
 \centering
 \includegraphics[width=17cm]{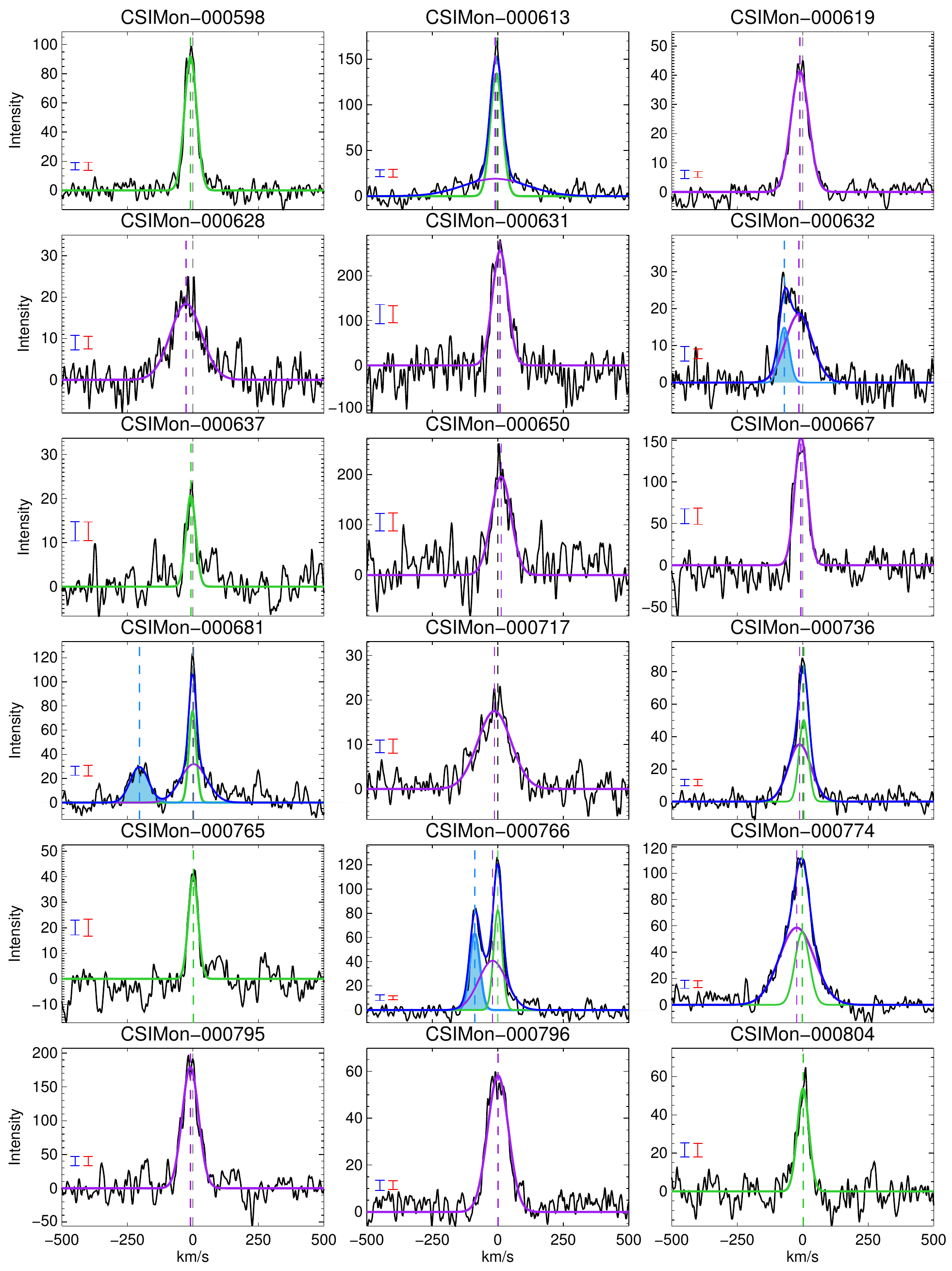}
 \caption{Continued. 
}\ContinuedFloat
\end{figure*}

\begin{figure*}[p]
 \centering
 \includegraphics[width=17cm]{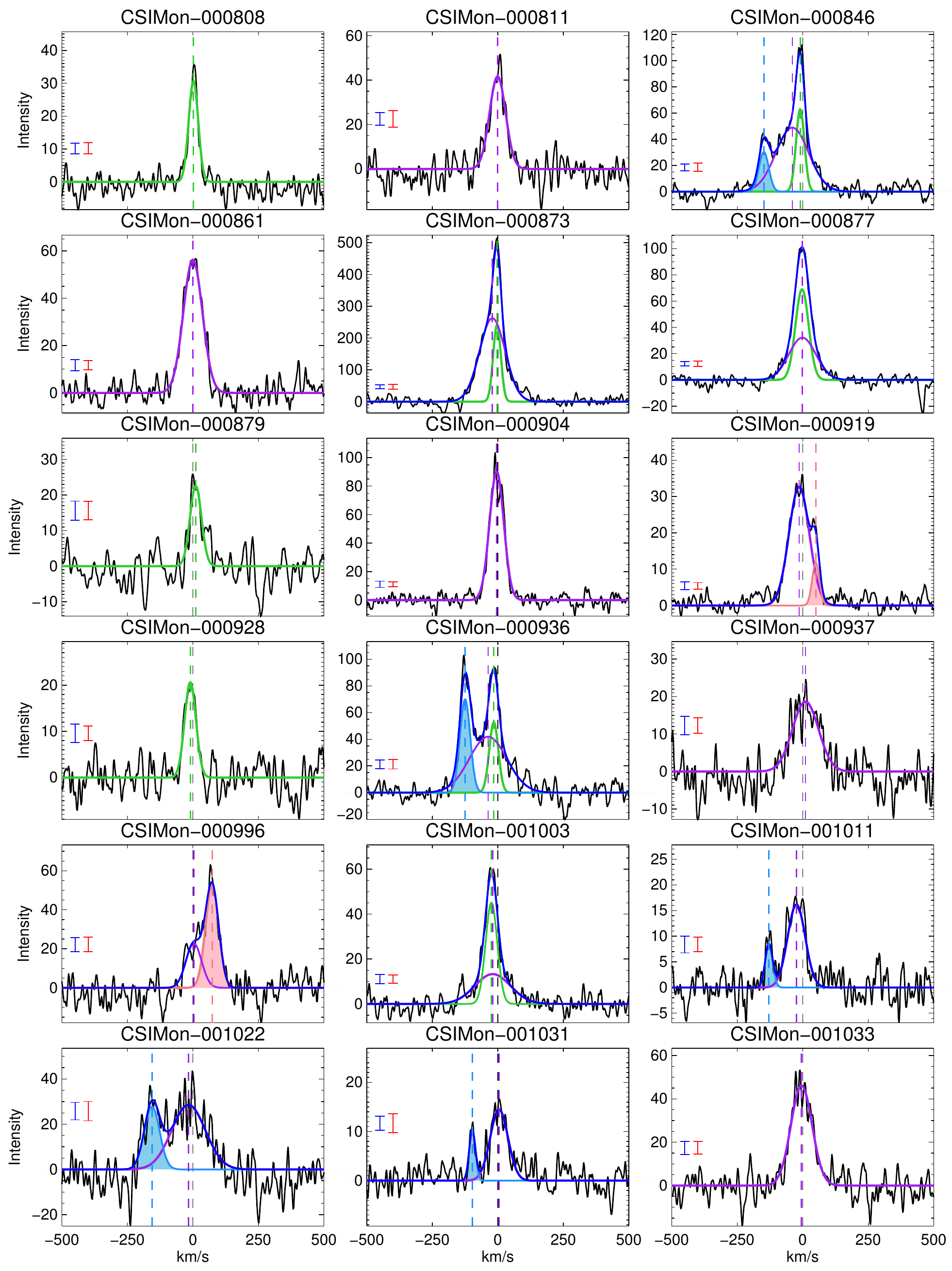}
 \caption{Continued. 
}\ContinuedFloat
\end{figure*}

\begin{figure*}[p]
 \centering
 \includegraphics[width=17cm]{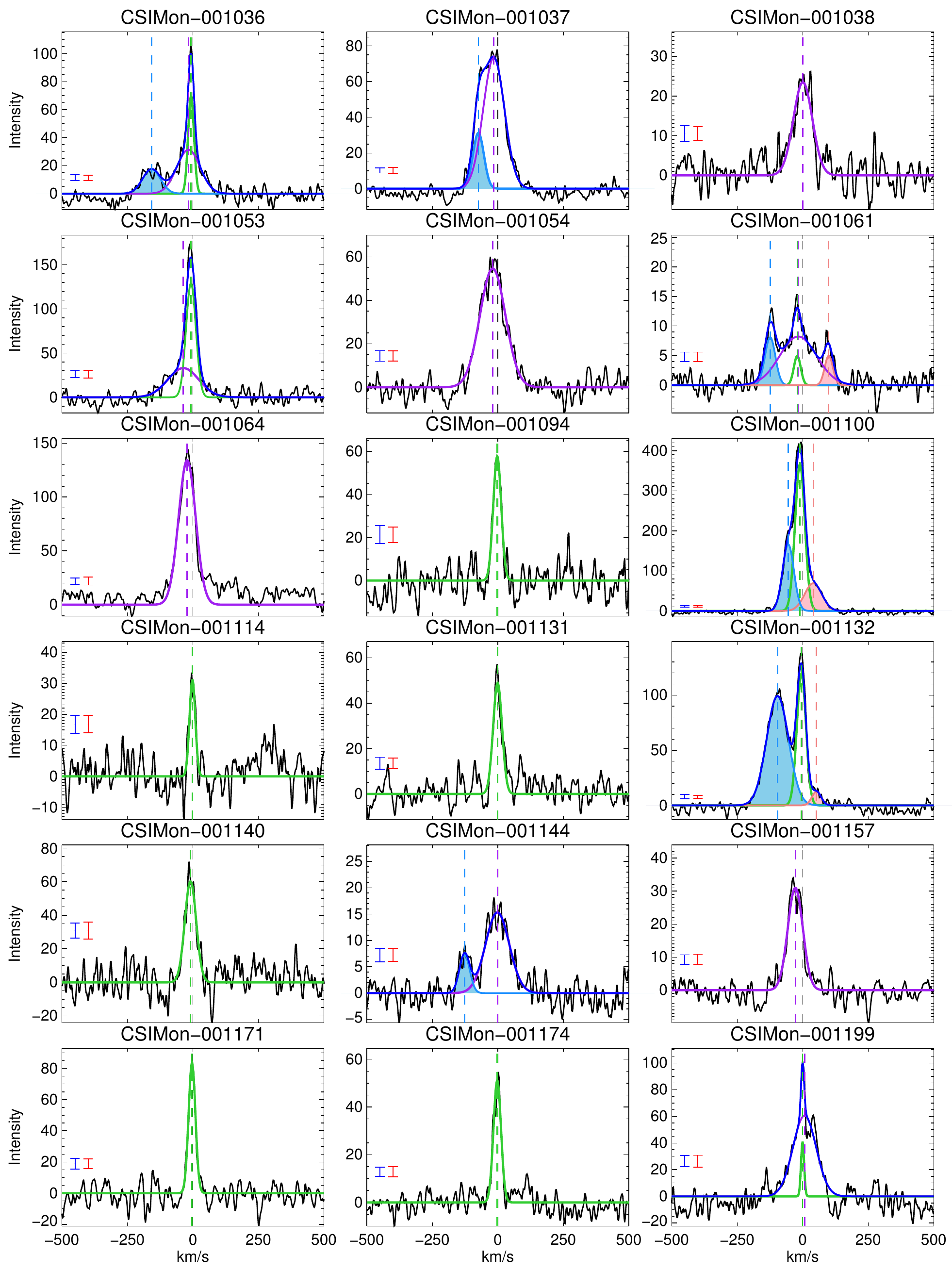}
 \caption{Continued. 
}\ContinuedFloat
\end{figure*}

\begin{figure*}[p]
 \centering
 \includegraphics[width=17cm]{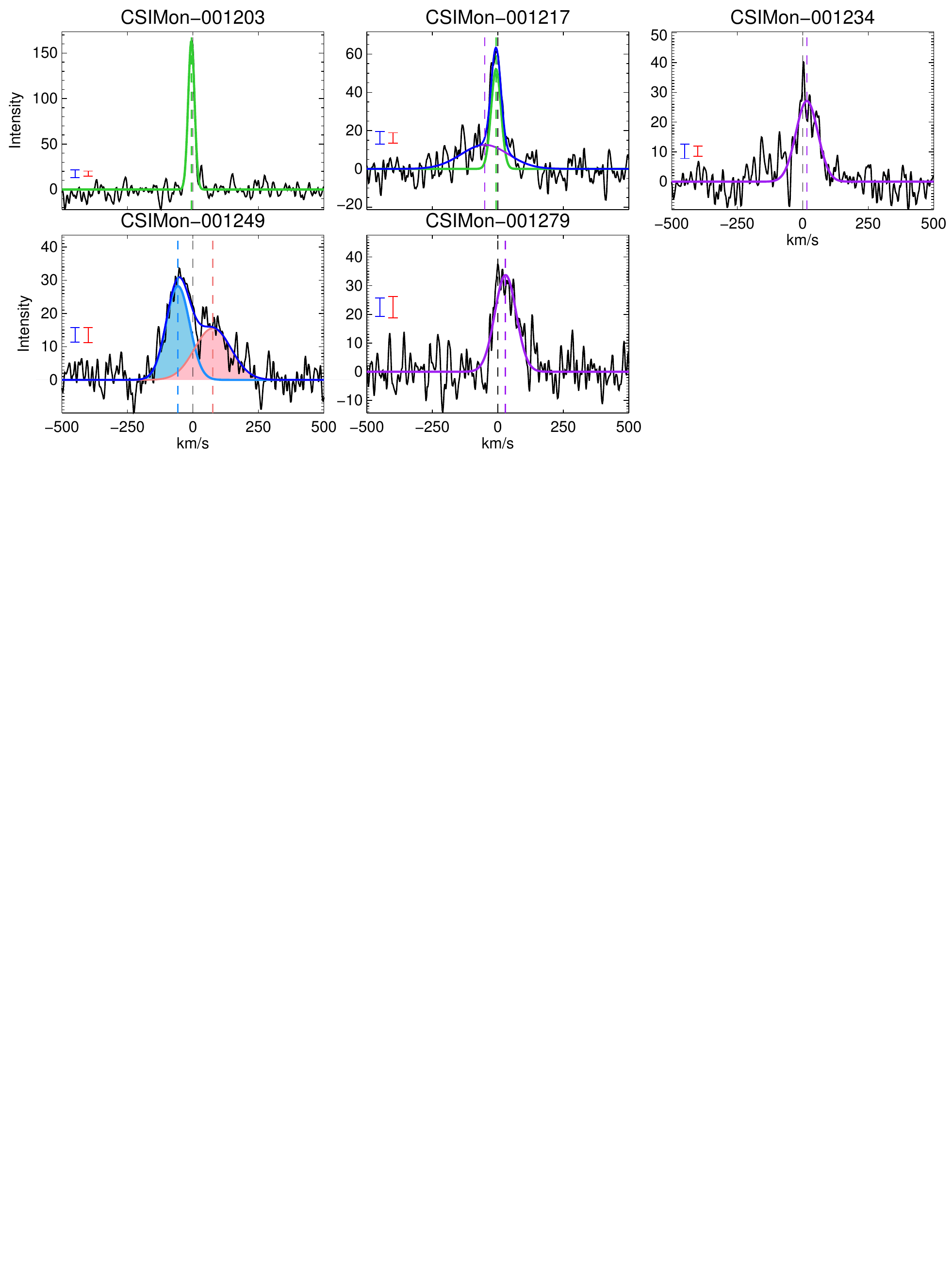}
 \caption{Continued. 
}
\end{figure*}

\begin{figure*}[p]
    \centering
    \includegraphics[width=17cm]{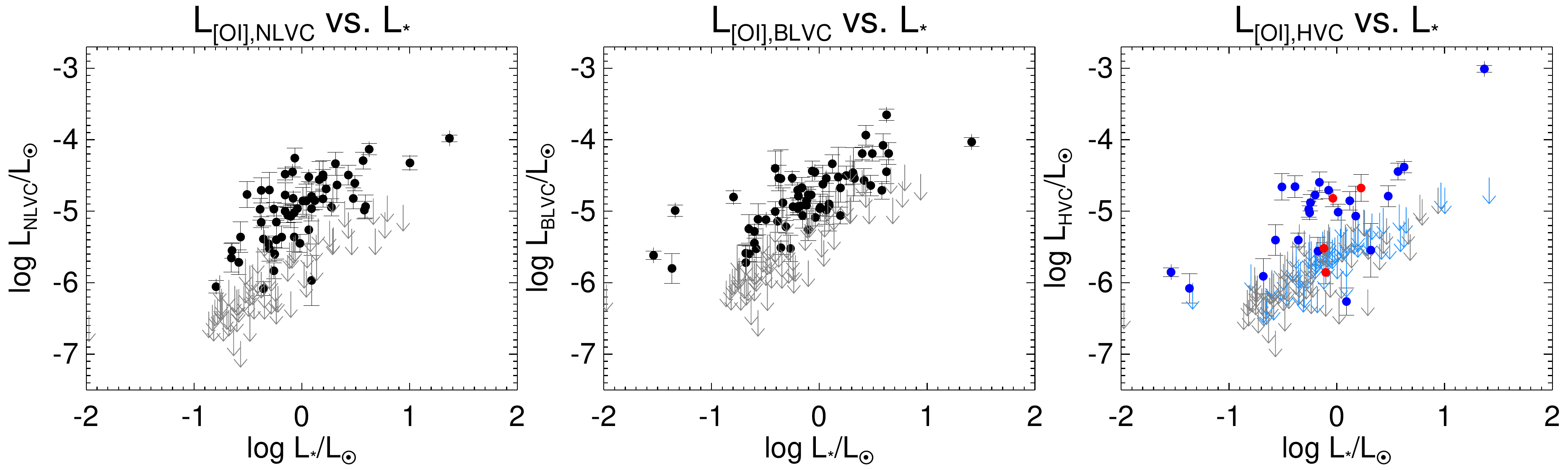} 
    \includegraphics[width=17cm]{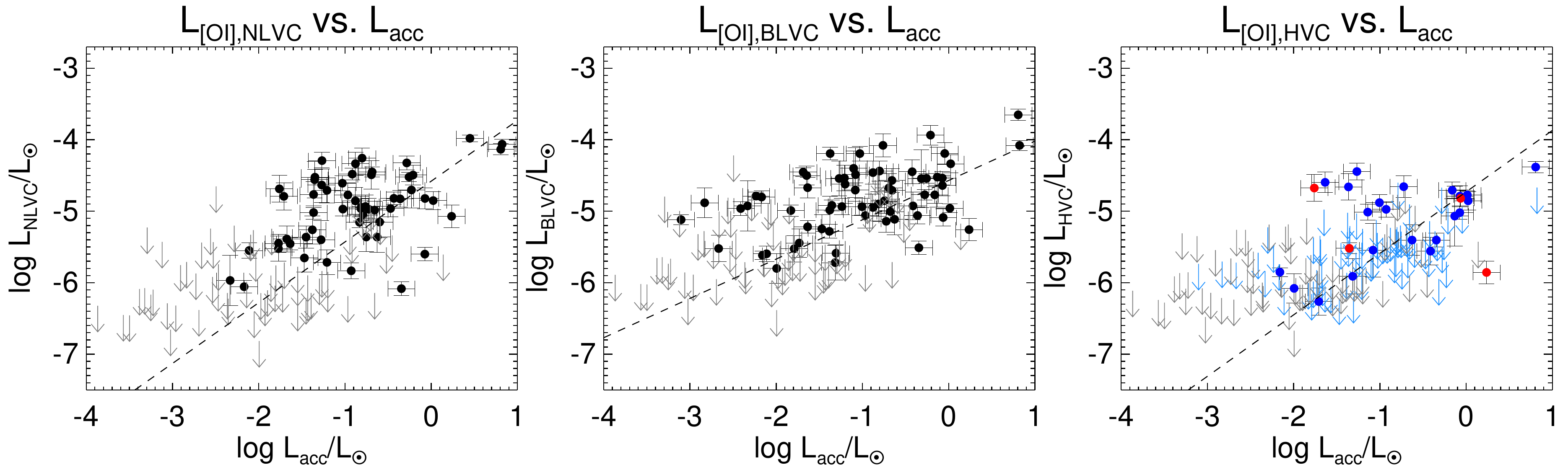} 
    \includegraphics[width=17cm]{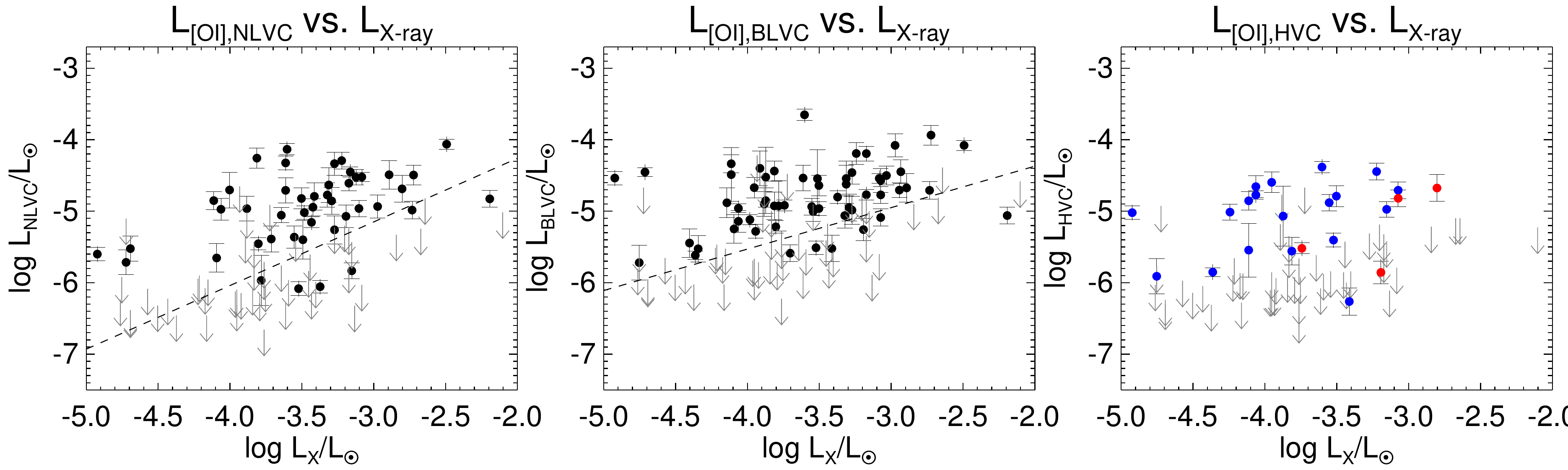} 
    \includegraphics[width=17cm]{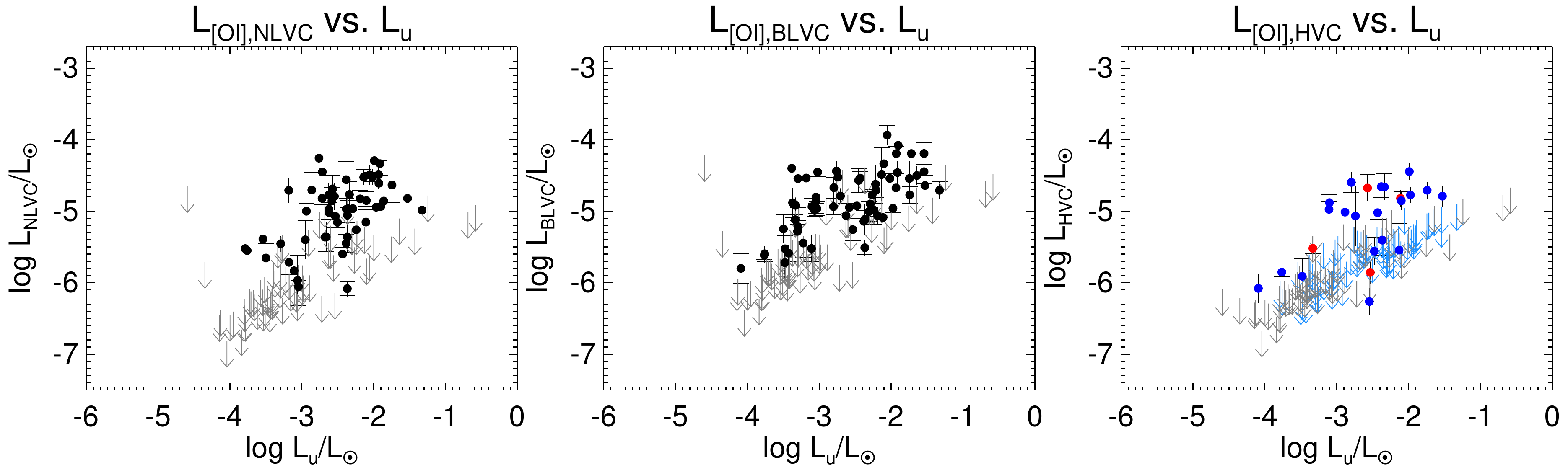} 
    \caption{First row: Luminosity of each [OI]$\lambda$6300 component versus stellar luminosity. %\\ 
    Second row: Luminosity of each [OI]$\lambda$6300 component versus accretion luminosity (derived 
from UV excess or H$\alpha$ profiles, in that order of preference). %\\ 
    Third row: Luminosity of each [OI]$\lambda$6300 component versus X-ray luminosity. %\\ 
    Fourth row: Luminosity of each [OI]$\lambda$6300 component versus the CFHT $u$-band luminosity. %\\ 
Gray arrows represent upper limits for [OI]$\lambda$6300 line luminosities when no emission in 
[OI]$\lambda$6300 was detected. 
In the HVC plots (right columns): light blue arrows represent upper limits for HVC luminosities 
when only low-velocity [OI]$\lambda$6300 emission was detected; blue filled circles represent 
blueshifted HVCs while red filled circles represent redshifted HVCs (systems that did not present 
a blueshifted component). 
    }\label{fig:oi_lum_lbol}
\end{figure*}

\end{appendix}

\end{document}